\documentclass[aps,pra,twocolumn,showpacs,10pt]{revtex4-1}

\usepackage{amsmath,amssymb}
\usepackage{graphicx}

\graphicspath{ {.} {./pics/} }

\newcommand{\rme}{\mathrm e}
\newcommand{\rmi}{\mathrm i}
\newcommand{\rmd}{\mathrm d}

\renewcommand{\Re}{\mathop{{\rm Re}}}

\begin{document}

\title{Dynamic Stark effect, light emission, and entanglement generation in a laser-driven quantum optical system}

\author{D. Pagel}
\email{pagel@physik.uni-greifswald.de}
\author{A. Alvermann}
\author{H. Fehske}
\affiliation{Institut f\"ur Physik, Ernst-Moritz-Arndt-Universit\"at, 17487 Greifswald, Germany}

\begin{abstract}

We calculate the emission spectra, the Glauber $g^{(2)}$ function, and the entanglement of formation for two-level emitters coupled to a single cavity mode and subject to an external laser excitation.
To evaluate these quantities we couple the system to environmental degrees of freedom, which leads to dissipative dynamics.
Because of the periodic time dependence of the system Hamiltonian, the coefficients of the Markovian master equation are constant only if Floquet states are used as the computational basis.
Studying the emission spectra, we show that the dynamic Stark effect first appears in second order of the laser intensity.
For the Glauber function, we find clearly distinguished parameter regimes of super- and sub-Poissonian light emission and explain the additional features appearing for finite laser intensity in terms of the quasienergy spectrum of the driven emitter-cavity system.
Finally, we analyze the temperature and emitter-cavity-coupling regimes where entanglement among the emitters is generated and show that the laser excitation leads to a decrease of entanglement.

\end{abstract}


\maketitle

\section{\label{sec:intro}Introduction}
The theoretical modeling of light-matter interaction becomes increasingly important when, with a view to quantum information applications~\cite{NC10}, the generation of nonclassical and entangled states of light~\cite{MW95} is investigated in the field of quantum optics~\cite{Car99}.
The Dicke Hamiltonian~\cite{Dic54} of two-level emitters interacting with a cavity photon mode is a generic model in this respect.
Many studies of the Dicke model focus on the superradiant phase transition~\cite{HL73,DK74, Hak84}.
Because most of the atom-field interactions in these studies only involve highly populated modes of the electric field, a semiclassical treatment that approximates the electric field as a $c$-number is sufficient.
The quantized Dicke Hamiltonian becomes important in cavity-quantum electrodynamics~\cite{WVEB06}, where many field modes contribute and where the light-matter-coupling constant is enhanced by the cavity.

Another important phenomenon arising from light-matter interaction is the Stark effect~\cite{HW04}, i.e., the splitting and shifting of atomic spectral lines in electric fields.
The Stark effect in constant fields can be observed for the Dicke Hamiltonian, whose eigenvalues shift with the emitter-cavity-coupling strength.
The dynamic Stark effect can be realized in such an optical system when it is driven by a laser.
In the dipole approximation this situation can be described by an explicit time-dependent contribution to the Dicke Hamiltonian.
Because the coupling to the external field is periodic with the laser frequency, the solutions of the Schr\"{o}dinger equation follow from Floquet theory.
For a single emitter with atom-cavity coupling and laser driving in the rotating-wave approximation, the Floquet states and quasienergies can be calculated analytically~\cite{AGC92}.
Shifted Rabi splittings are observed as a consequence of the dynamic Stark effect.
Interestingly, recent developments in the field of cavity quantum electrodynamics~\cite{WSBFHMKGS04, EFFSPV07, BGAANB09, FDLMGRSHM10, NDHMHSGRZHSM10} allow achieving the strong and ultrastrong light-matter-coupling regimes experimentally.
Hence, the full Dicke Hamiltonian including the contribution from the external laser has to be tackled to describe the dynamical properties.
This is the main purpose of the present study.

In more detail, we first analyze the dynamic Stark effect through evaluation of the emission spectrum~\cite{MW95,Car99} of the laser-driven Dicke system.
Because an analytical solution is not available in this case, we quantify the laser-intensity-dependent shifts of the emission peaks numerically.
Second, we calculate the Glauber function~\cite{Gla63}, which allows us to identify regimes where nonclassical light~\cite{MW95} is emitted.
Thereby, we provide a physical picture to interpret the features of the Glauber function appearing for finite laser intensity.
This generalizes our previous results without external drive~\cite{PAF15} and related studies~\cite{AT91, CBC05, DLGCC09, RLSH12, RSH13, SRDSHS13, GRSDSS13}.
Third, we consider the generation of entanglement~\cite{EPR35, Sch35} between two emitters, which is important for quantum optical applications~\cite{NC10}.
To this end, we quantify the generated bipartite entanglement by the so-called entanglement of formation.

To analyze the emission properties of such a coupled light-matter system, we use the full input-output formalism~\cite{CG84, GC85, Gra89, SG96pra} that explicitly takes the environmental degrees of freedom into account.
The resulting input-output relations connect expectation values of output operators to those of the system operators, which requires knowledge about the reduced (open) system dynamics~\cite{Wei12, BP02}.
For weak system-environment coupling the dissipative evolution of the system is described by Markovian master equations~\cite{BP02,ALZ06,SB08,BGB11,Sch14}.
In view of the periodicity of the system Hamiltonian, Floquet states can be used as the computational basis.
The resulting Floquet master equation~\cite{BBGSSW91, GH98} is Markovian and has time-independent coefficients.
Solution of the master equation requires the computation of the Floquet states, i.e., the solution of the corresponding Schr\"odinger equation without coupling to the environment.

The paper is organized as follows.
In Sec.~\ref{sec:Stark} we discuss the dynamic Stark effect for a single laser-driven emitter embedded in a cavity.
We start in Sec.~\ref{ssec:model} with the introduction of our model, continue in Sec.~\ref{ssec:formalism} with the formalism for the calculation of emission spectra, and finally present and analyze the results in Secs.~\ref{ssec:quasienergies} and~\ref{ssec:spectra}.
The statistics of the emitted photons and the emission of nonclassical light is studied for a single emitter in Sec.~\ref{sec:emission}, while the generation of entanglement among two emitters is considered in Sec.~\ref{sec:entangle}.
We conclude in Sec.~\ref{sec:conc}.
Further information on our theoretical approach is collected in the Appendixes:
In Appendix~\ref{app:TC} analytical results for the driven Tavis-Cummings model are summarized, details about the input-output approach for the description of the emission are given in Appendix~\ref{app:inout}, Appendix~\ref{app:Floquet} contains the derivation and a brief discussion of the Floquet master equation, and in Appendix~\ref{app:emitters} the emission spectra and Glauber function for a few emitters are presented and compared to the results for a single emitter.

\section{\label{sec:Stark}The dynamic Stark effect for a laser-driven Dicke system}
In this section we calculate emission spectra for laser-driven emitters in a cavity and analyze the shift of emission peaks with increasing laser intensity.
We first introduce the Dicke model and recapitulate the analytical results reported in Ref.~\cite{AGC92}.

\subsection{\label{ssec:model}Laser-driven emitters in a cavity}
The Dicke model~\cite{Dic54} (with $\hbar = 1$)
\begin{eqnarray}\label{HD}
  H_D &=& \omega_c a^\dagger a + \omega_x \sum_{j=1}^N \sigma_+^{(j)} \sigma_-^{(j)} + g \sum_{j=1}^N (a^\dagger \sigma_-^{(j)} + a \sigma_+^{(j)}) \nonumber\\
&& + g' \sum_{j=1}^N (a \sigma_-^{(j)} + a^\dagger \sigma_+^{(j)})
\end{eqnarray}
describes the interaction of $N$ two-level emitters with a single cavity mode.
The operator $a$ ($a^\dagger$) annihilates (creates) a cavity photon with frequency $\omega_c$.
Excitation and relaxation of the $j$th emitter with transition energy $\omega_x$ is provided by the spin operators $\sigma_+^{(j)}$ and $\sigma_-^{(j)}$, respectively.
The emitter-photon-coupling strength for the corotating (counterrotating) interaction terms is denoted by $g$ ($g'$).
Note that different coupling strengths for the corotating and counterrotating interaction terms can be realized experimentally~\cite{DEPC07, BAGPB14}.

The emitter-cavity system is excited by a laser with driving frequency $\omega_d$.
The interaction of the cavity mode with the laser field is described by the time-dependent Hamiltonian
\begin{equation}\label{HL}
  H_L(t) = \frac{\Omega}{2} (a \rme^{\rmi \omega_d t} + a^\dagger \rme^{-\rmi \omega_d t}) + \frac{\Omega'}{2} (a \rme^{-\rmi \omega_d t} + a^\dagger \rme^{\rmi \omega_d t}) \,.
\end{equation}
We allow for different photon-laser-coupling strengths (laser intensities) for the corotating ($\Omega$) and counterrotating ($\Omega'$) interaction terms.

The combined Hamiltonian
\begin{equation}\label{H}
  H(t) = H_D + H_L(t) \,,
\end{equation}
has a periodic time dependence $H(t) = H(t + T_d)$ with period $T_d = 2 \pi / \omega_d$.
Because of this discrete time-translation symmetry, solutions of the Schr\"{o}dinger equation are the Floquet states~\cite{Flo83}
\begin{equation}\label{floquet}
  | \psi_n(t) \rangle = \rme^{-\rmi \epsilon_n t} | \phi_n(t) \rangle \,.
\end{equation}
Here $\epsilon_n \in \mathbb{R}$ are quasienergies and $| \phi_n(t) \rangle = | \phi_n(t + T_d) \rangle$ is the time-periodic part of the state~\eqref{floquet}.
The quasienergies are unique up to multiples of $\omega_d$ and can therefore be mapped into the first quasienergy Brillouin zone, $-\omega_d / 2 \leq \epsilon_n < \omega_d / 2$.

Analytical results for the quasienergies of a single emitter ($N = 1$) at resonance ($\omega_c = \omega_x = \omega_d$) and in the rotating-wave approximation ($g' = \Omega' = 0$) were given in Ref.~\cite{AGC92} (see also Appendix~\ref{app:TC}).
The result without projection into the first quasienergy Brillouin zone is $\epsilon_n = \pm \sqrt{n} g \{1 - (\Omega / g)^2\}^{3/4}$.
The laser-induced dynamic Stark effect reduces the Jaynes-Cummings level splittings $\pm \sqrt{n} g$.
Taylor expansion of $\epsilon_n$ shows that this reduction is of order $\Omega^2$.

\subsection{\label{ssec:formalism}Input-output approach}
To evaluate the emission spectra of the laser-driven Dicke system beyond the rotating-wave approximation, we have to explicitly consider the coupling to environmental field modes.
We assume an interaction Hamiltonian of the form
\begin{equation}\label{HI}
  H_I = -\rmi X \sum_\alpha \lambda_\alpha (b_\alpha - b_\alpha^\dagger) \,,
\end{equation}
where $X = -\rmi (a - a^\dagger)$ is the field operator for the coupling of the cavity to the environment.
The operator $b_\alpha$ ($b_\alpha^\dagger$) annihilates (creates) environmental photons with frequencies $\omega_\alpha$, and the coupling constants are denoted by $\lambda_\alpha$.

The standard input-output formalism~\cite{CG84, GC85, Gra89, SG96pra, PAF15} with the interaction Hamiltonian~\eqref{HI} and the Floquet states~\eqref{floquet} as the computational basis leads us to the projected cavity-environment-coupling operator (see Appendix~\ref{app:inout})
\begin{multline}\label{X}
  \dot{X}_-(t) = -\rmi \sum_{m, n, \nu} (\epsilon_n - \epsilon_m + \nu \omega_d) \theta(\epsilon_n - \epsilon_m + \nu \omega_d) \\
    \times \sum_\mu | \psi_m(t) \rangle \langle \widetilde{\phi}_m(\mu - \nu) | X | \widetilde{\phi}_n(\mu) \rangle \langle \psi_n(t) | \,,
\end{multline}
where $\theta(\omega)$ is the Heaviside step function and the states $| \widetilde{\phi}_n(\nu) \rangle$ follow from Fourier expansion of the periodic states
\begin{equation}\label{fourier}
  | \phi_n(t) \rangle = \sum_{\nu=-\infty}^\infty \rme^{-\rmi \nu \omega_d t} | \widetilde{\phi}_n(\nu) \rangle \,.
\end{equation}
The output operator $\dot{X}_-(t)$ in Eq.~\eqref{X} is the projection of the field operator $X$, which couples the cavity and output field, onto transitions between Floquet states.
The corresponding matrix elements are sums over all Fourier modes with fixed mode number difference $\nu$ weighted with the respective transition energy $\epsilon_n - \epsilon_m + \nu \omega_d$.

The emission properties of the laser-driven Dicke system are characterized by correlation functions of $\dot{X}_-(t)$.
In particular, the emission spectrum is~\cite{Car99, DLGCC09}
\begin{multline}\label{S_def}
  S(\omega) = \frac{\gamma_c(\omega)}{\pi} \lim_{s \to \infty} \Re \Big\{ \frac{1}{T_d} \int_0^{T_d} \int_0^\infty \rme^{-\rmi \omega \tau} \\
    \times \langle \dot{X}_+(s + t + \tau) \dot{X}_-(s + t) \rangle \, \rmd\tau \, \rmd t \Big\} \,,
\end{multline}
where $\gamma_c(\omega)$ is the spectral function for the (cavity) environment and $\dot{X}_+ = \dot{X}_-^\dagger$.
In order to evaluate the emission spectrum~\eqref{S_def}, we have to calculate the long-time dynamics of the (system) operator $\dot{X}_-(t)$ or, equivalently, the evolution of the system density matrix $\rho(t)$.

The interaction with the (thermal) environment leads to an energy transfer between system and environment.
The dissipative dynamics of the system for weak coupling to the environment is described by a Markovian master equation
\begin{equation}
  \frac{\rmd}{\rmd t} \rho(t) = \mathcal{L}(t) \rho(t) \,,
\end{equation}
where $\mathcal{L}(t)$ is the generator of a quantum dynamical semigroup for $t \geq 0$.
We introduce the corresponding propagator
\begin{equation}\label{V}
  V(t, t') = T_\leftarrow \exp \bigg( \int_{t'}^t \mathcal{L}(\tau) \, \rmd\tau \bigg)
\end{equation}
($T_\leftarrow$ is the chronological time-ordering operator) that satisfies
\begin{equation}
  \frac{\partial}{\partial t} V(t, t') = \mathcal{L}(t) V(t, t') \,.
\end{equation}
Using the Floquet states as the computational basis, the generator $\mathcal{L}(t)$ becomes time-independent~\cite{GH98, BP02}.
As a result, the off-diagonal matrix elements $\rho_{m,n}(t) = \langle \psi_m(t) | \rho(t) | \psi_n(t) \rangle$ decay exponentially, while the diagonal elements $\rho_{n,n}(t)$ are given as the solution of a Pauli master equation (see Appendix~\ref{app:Floquet}).
Hence, the stationary state is periodic at long times:
\begin{eqnarray}\label{rho_oo}
  \rho^{\infty}(t) &=& \sum_n \rho_{n,n}^\infty | \psi_n(t) \rangle \langle \psi_n(t) | = \sum_n \rho_{n,n}^\infty | \phi_n(t) \rangle \langle \phi_n(t) | \nonumber\\
    &=& \rho^{\infty}(t + T_d) \,,
\end{eqnarray}
where $\rho_{m,n}^\infty = \lim_{t \to \infty} \rho_{m,n}(t) = \rho_{n,n}^\infty \delta_{m,n}$ are constant.
The oscillating asymptotic behavior is accounted for by the time average in Eq.~\eqref{S_def}.
Using the relations~\eqref{V} and~\eqref{rho_oo}, Eq.~\eqref{S_def} for the emission spectrum becomes
\begin{multline}\label{S}
  S(\omega) = \frac{\gamma_c(\omega)}{\pi} \Re \Big\{ \frac{1}{T_d} \int_0^{T_d} \int_0^\infty \rme^{-\rmi \omega \tau} \\
  \times \mathop{\text{Tr}} \big[ \dot{X}_+ V(t + \tau, t) \dot{X}_- \rho^\infty(t) \big] \, \rmd\tau \, \rmd t \Big\} \,,
\end{multline}
where $V(t + \tau, t) \dot{X}_- \rho^\infty(t)$ is the propagation of $\dot{X}_- \rho^\infty(t)$ from time $t$ until time $t + \tau$.

\subsection{\label{ssec:quasienergies}Quasienergy spectrum}
Before we present the results for $S(\omega)$ obtained from numerical solution of Eq.~\eqref{S}, we discuss the eigenvalues of the Dicke Hamiltonian $H_D$ and their relation to the quasienergies of $H(t)$.
We refer to the eigenvalues $E_n$ of $H_D$ in Eq.~\eqref{HD} as system energies.
The Floquet state $| \psi_n(t) \rangle$ of $H(t) = H_D + H_L(t)$ is characterized by a quasienergy $\epsilon_n$ and a whole bunch of Fourier modes with mode numbers $\nu$.

It is already evident from the analytical result in Ref.~\cite{AGC92} that the quasienergies for weak laser intensity are the system energies projected into the first quasienergy Brillouin zone $-\omega_d / 2 \leq \epsilon_n < \omega_d / 2$.
To zeroth order in the laser-driving strength, the periodic part $| \phi_n(t) \rangle$ of each Floquet state $| \psi_n(t) \rangle$ has a single Fourier mode $\nu$.
This mode number follows from the projection condition $E_n = \epsilon_n + \nu \omega_d$.
Additional Fourier modes $\nu \pm 1$ (first sidebands) contribute already in first order in $\Omega$, whereas modifications of $\epsilon_n$ occur for higher orders of $\Omega$ only.
In the weak driving regime $\Omega \ll \omega_0, g$, it thus suffices to take the system energies and the first two sidebands into account.

\begin{figure}
  \includegraphics[width=0.49\linewidth]{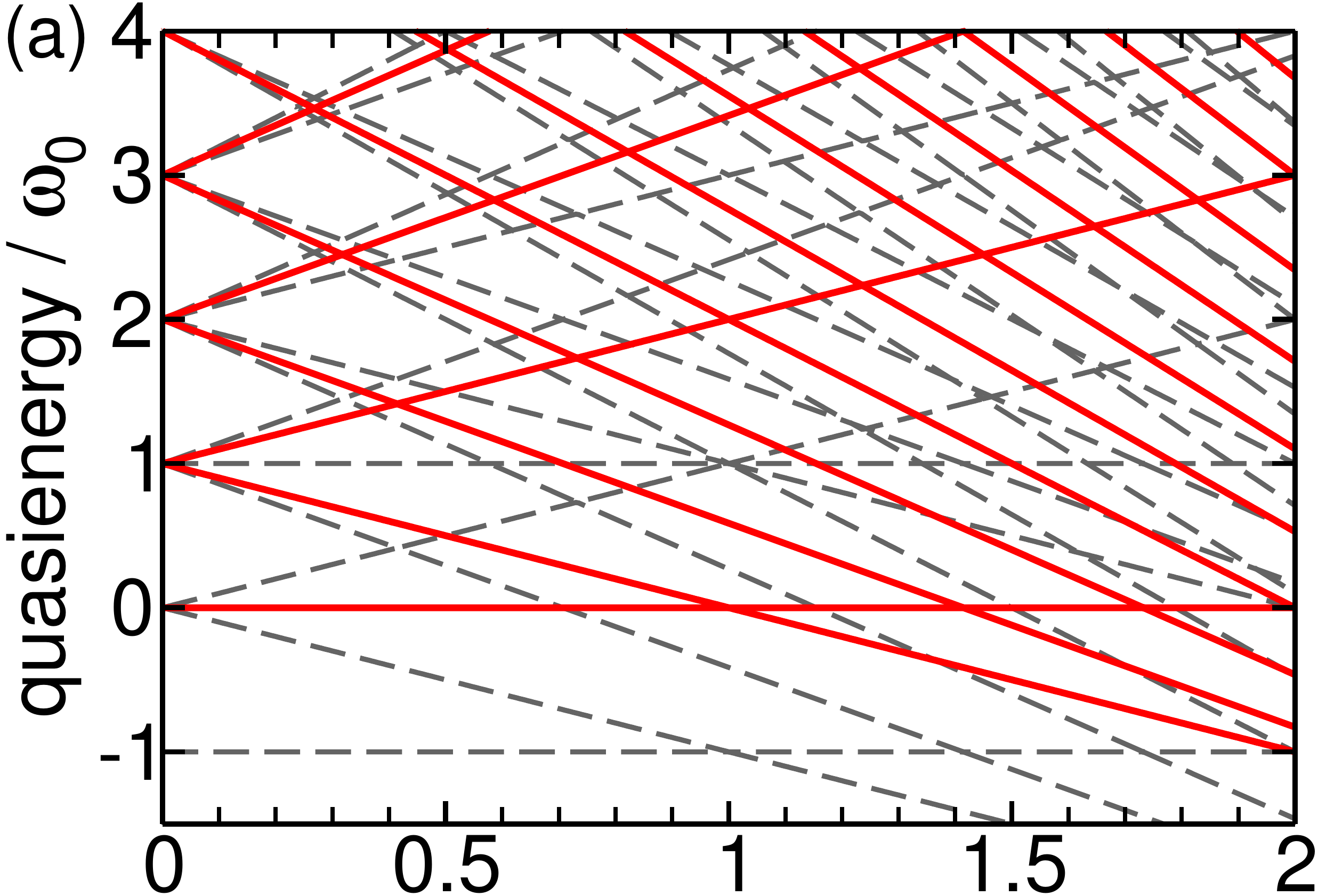}
  \includegraphics[width=0.49\linewidth]{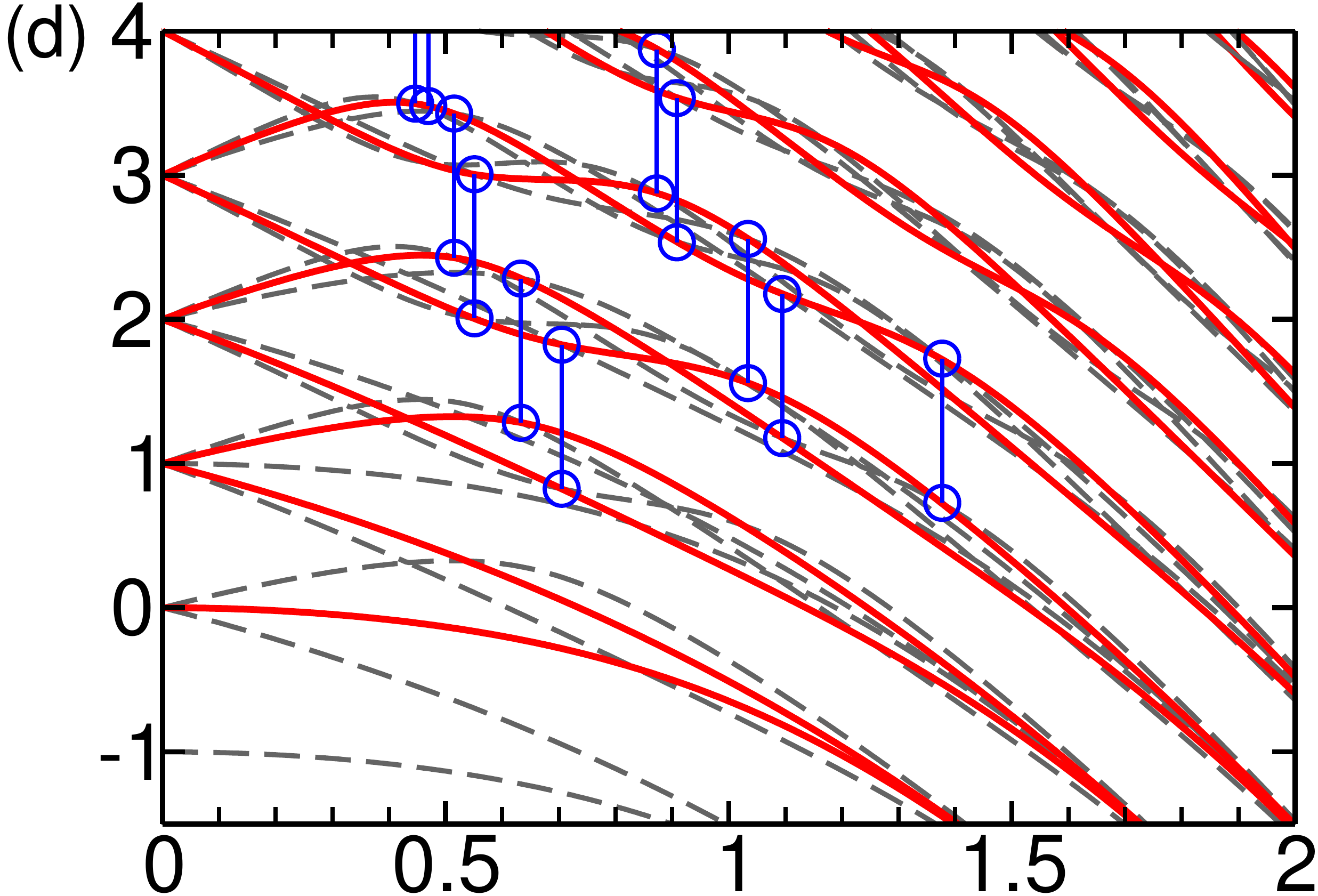}\\
  \includegraphics[width=0.49\linewidth]{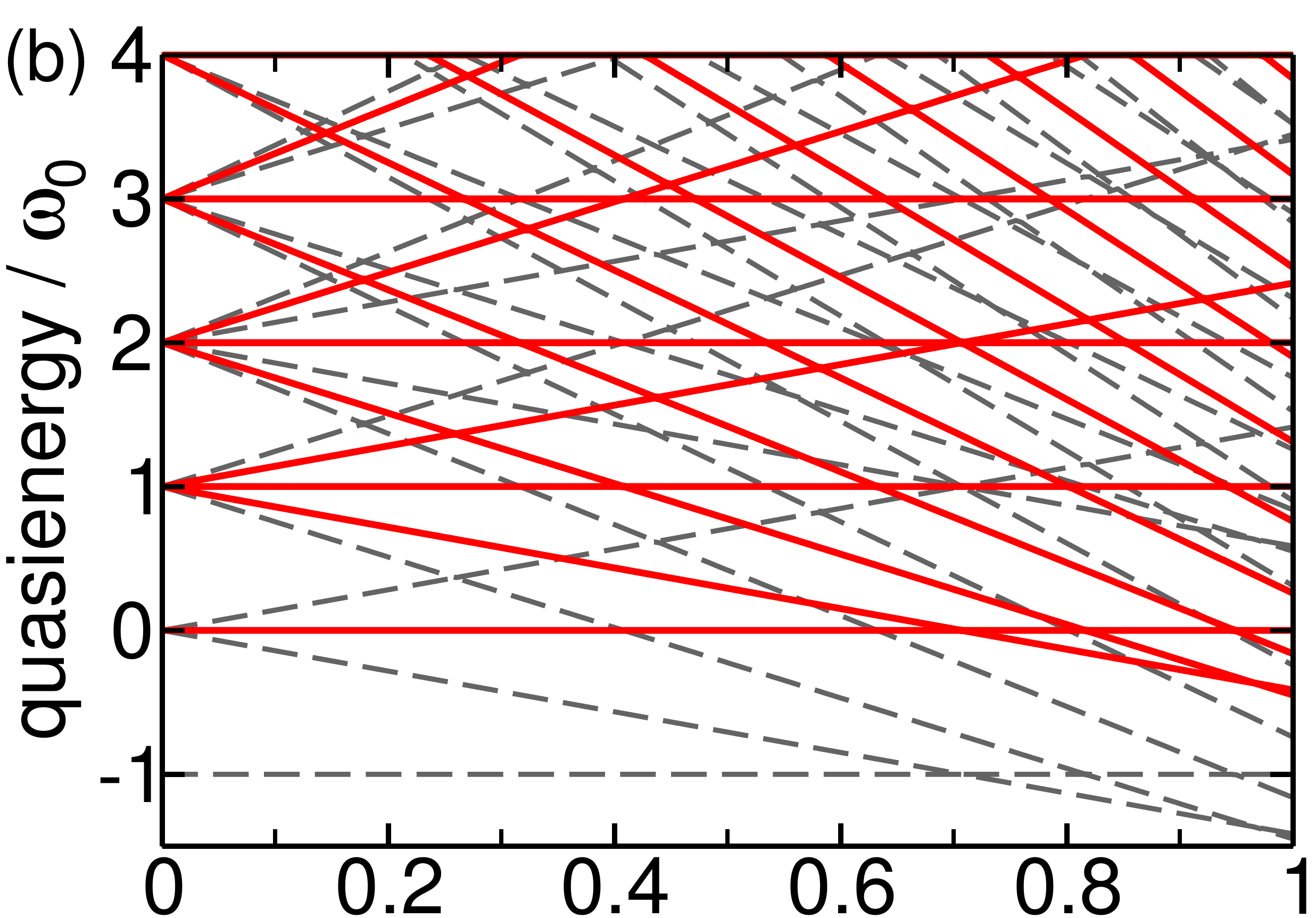}
  \includegraphics[width=0.49\linewidth]{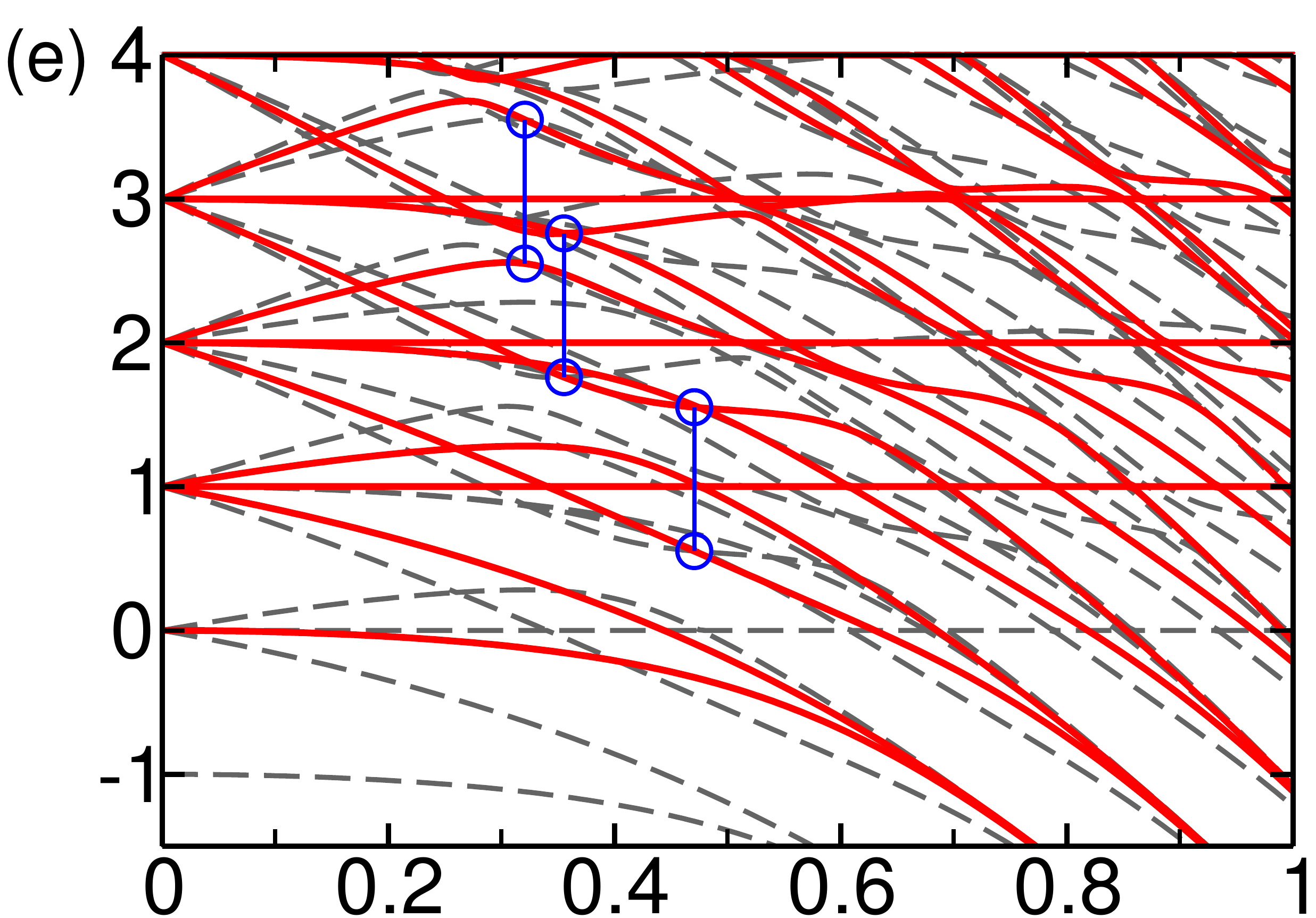}\\
  \includegraphics[width=0.49\linewidth]{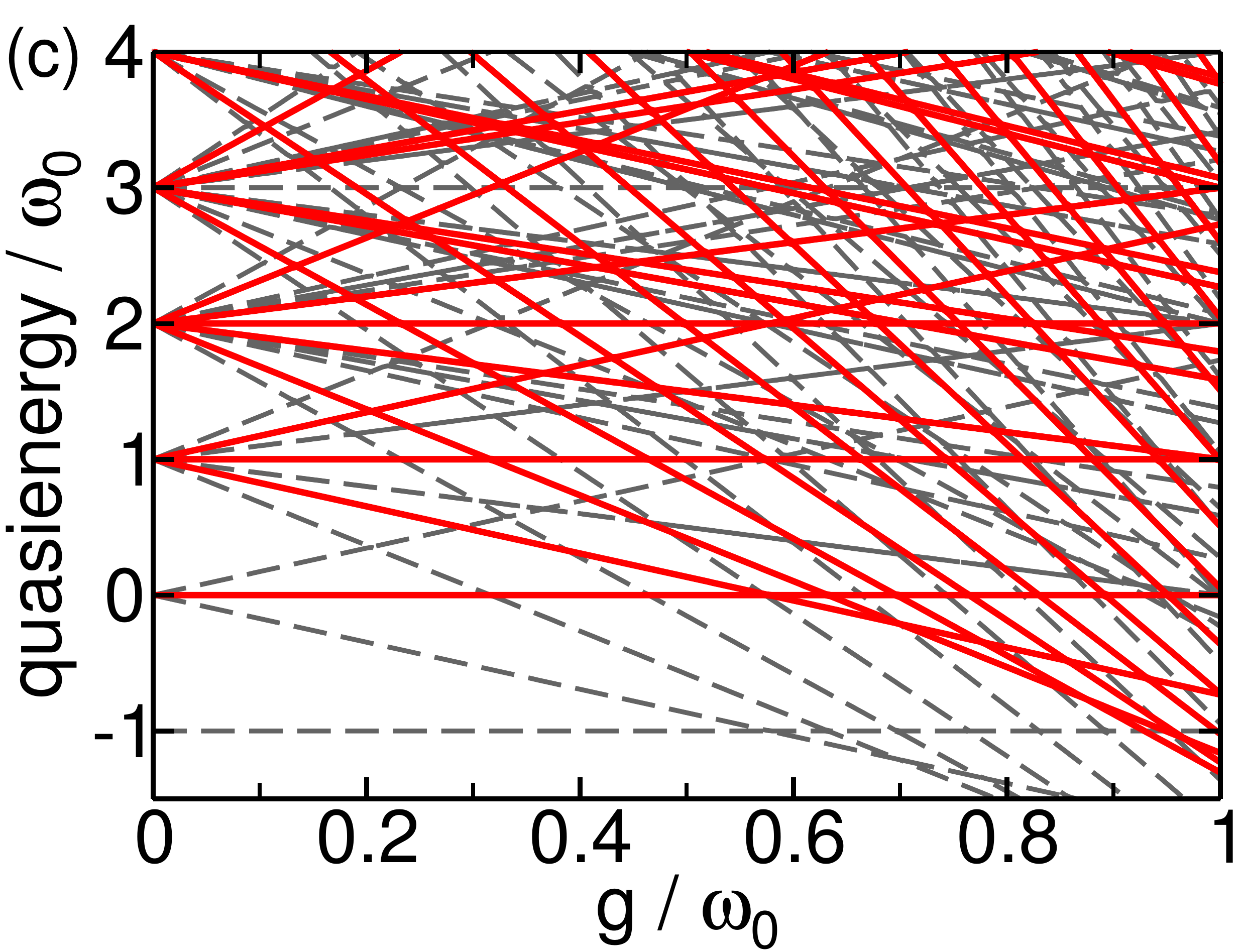}
  \includegraphics[width=0.49\linewidth]{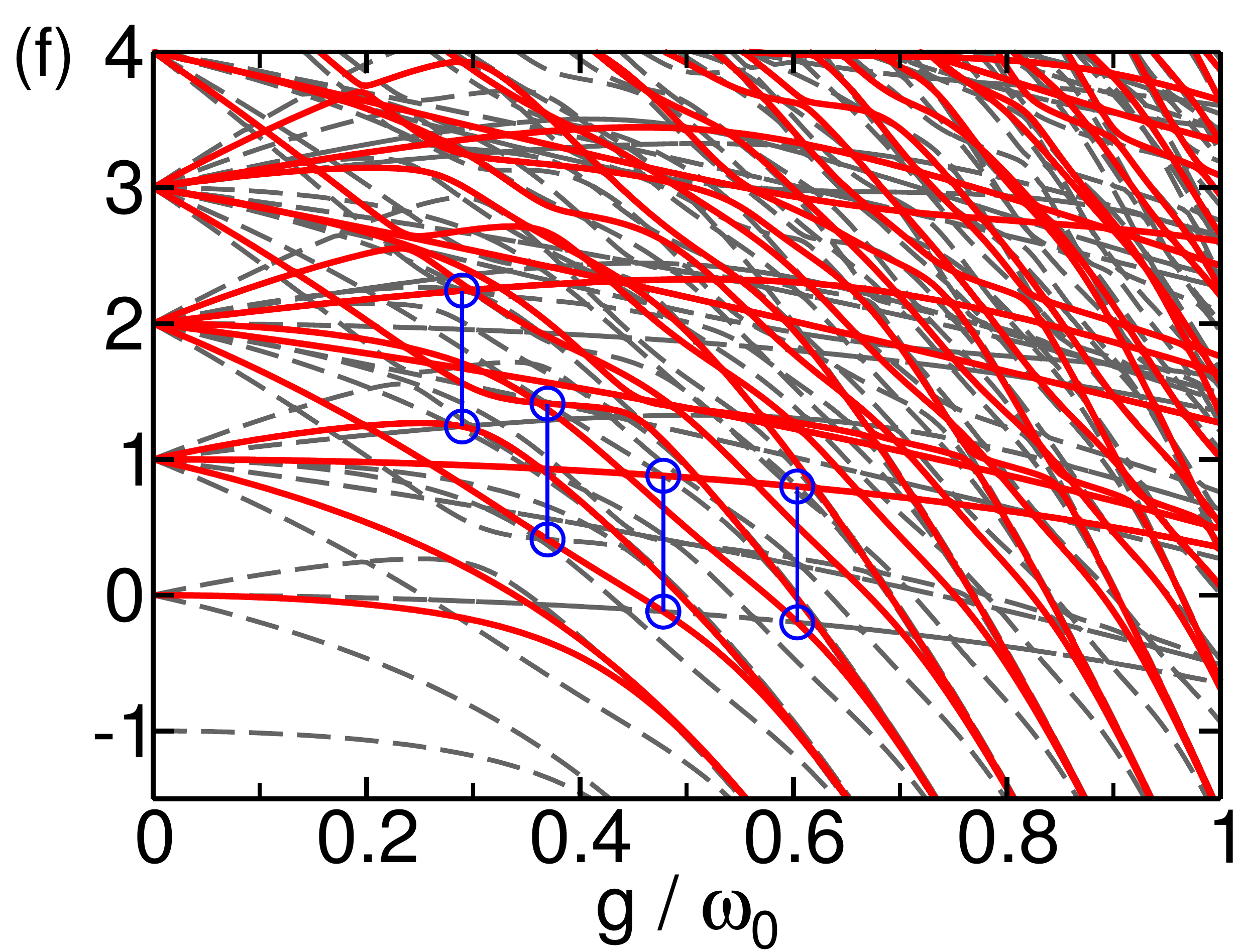}
  \caption{\label{fig:energy}Quasienergy spectra of $H(t)$ for (a)--(c) $g' = 0$ and (d)--(f) $g' = g$ as functions of the coupling strength $g$ for small laser intensity $\Omega \ll \omega_0, g$.
    Shown are the system energies $E_n = \epsilon_n + \nu \omega_d$ (red solid lines), the two sidebands $\epsilon_n + (\nu \pm 1) \omega_d$ (gray dashed lines), and crossings of system energies with sidebands (blue vertical lines) that are relevant for the Glauber function discussed in Sec.~\ref{sec:emission}.
    The results are shown for (a) and (d) $N = 1$, (b) and (e) $N = 2$, and (c) and (f) $N = 3$ emitters.}
\end{figure}

In Fig.~\ref{fig:energy} we plot the system energies $E_n = \epsilon_n + \nu \omega_d$ and the two sidebands $E_n \pm \omega_d = \epsilon_n + (\nu \pm 1) \omega_d$ as functions of the coupling strength $g$.
The values $E_n$ are obtained through numerical diagonalization of the Dicke Hamiltonian $H_D$.
Working at resonance $\omega_c = \omega_x = \omega_d = \omega_0$, the energies $M \omega_0$ of the uncoupled emitter-cavity system (with $g = 0$) are given by the total number $M = M_x + M_c$ of emitter ($M_x = 0, \dots, N$) and cavity ($M_c = 0, 1, \dots$) excitations.
In this sense, $M$ is the principal quantum number.
For $g' = 0$, we recover the well-known linear dispersions~\cite{TC68}, whereas, for $g' = g$, corrections arise from the coupling of states with different $M$.
These corrections increase if the number of emitters $N$ grows, because more and more states in the system energy spectrum are very close to each other.

\subsection{\label{ssec:spectra}Emission spectrum}
\begin{figure}
  \includegraphics[width=0.49\linewidth]{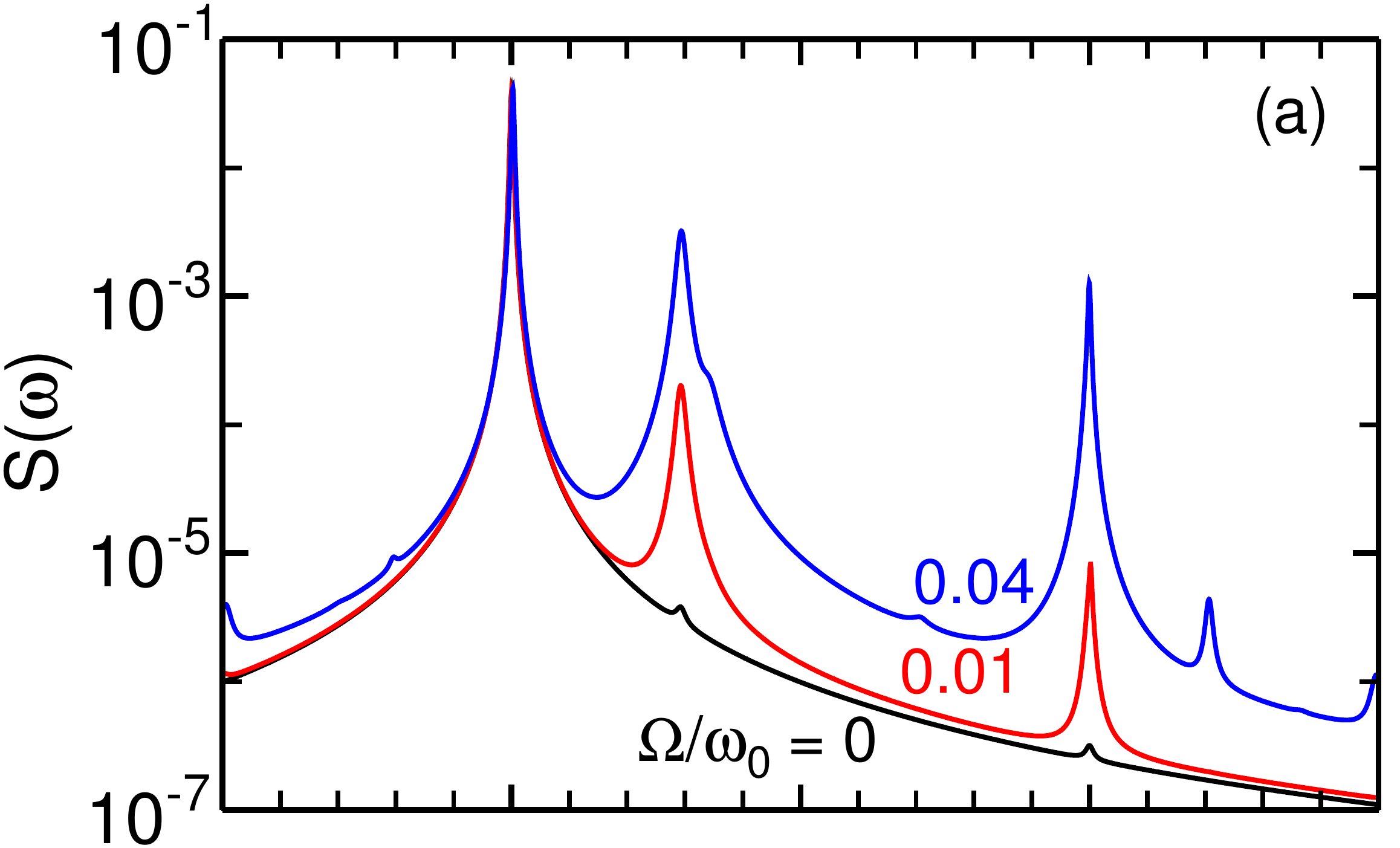}
  \includegraphics[width=0.49\linewidth]{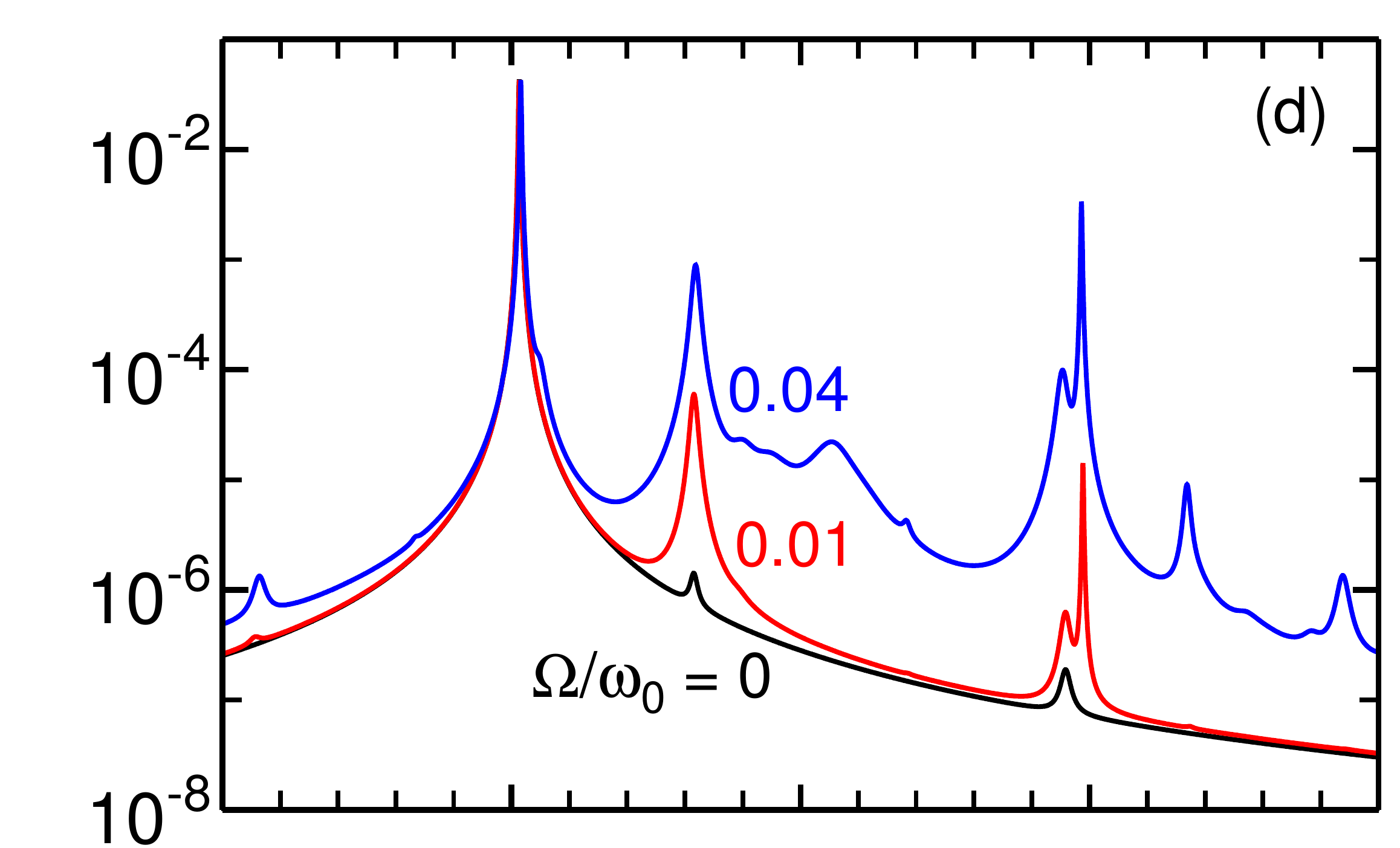}\\
  \includegraphics[width=0.49\linewidth]{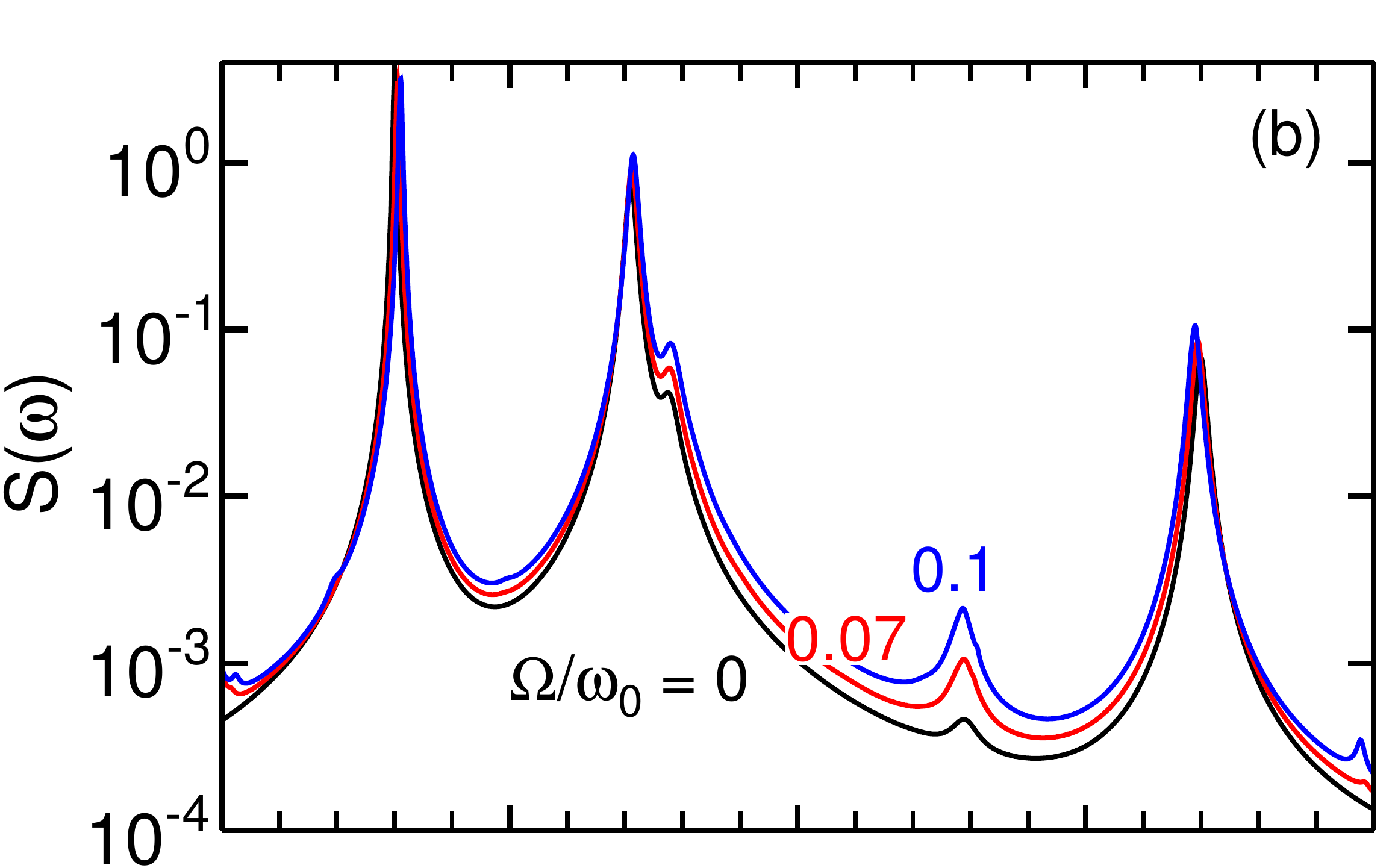}
  \includegraphics[width=0.49\linewidth]{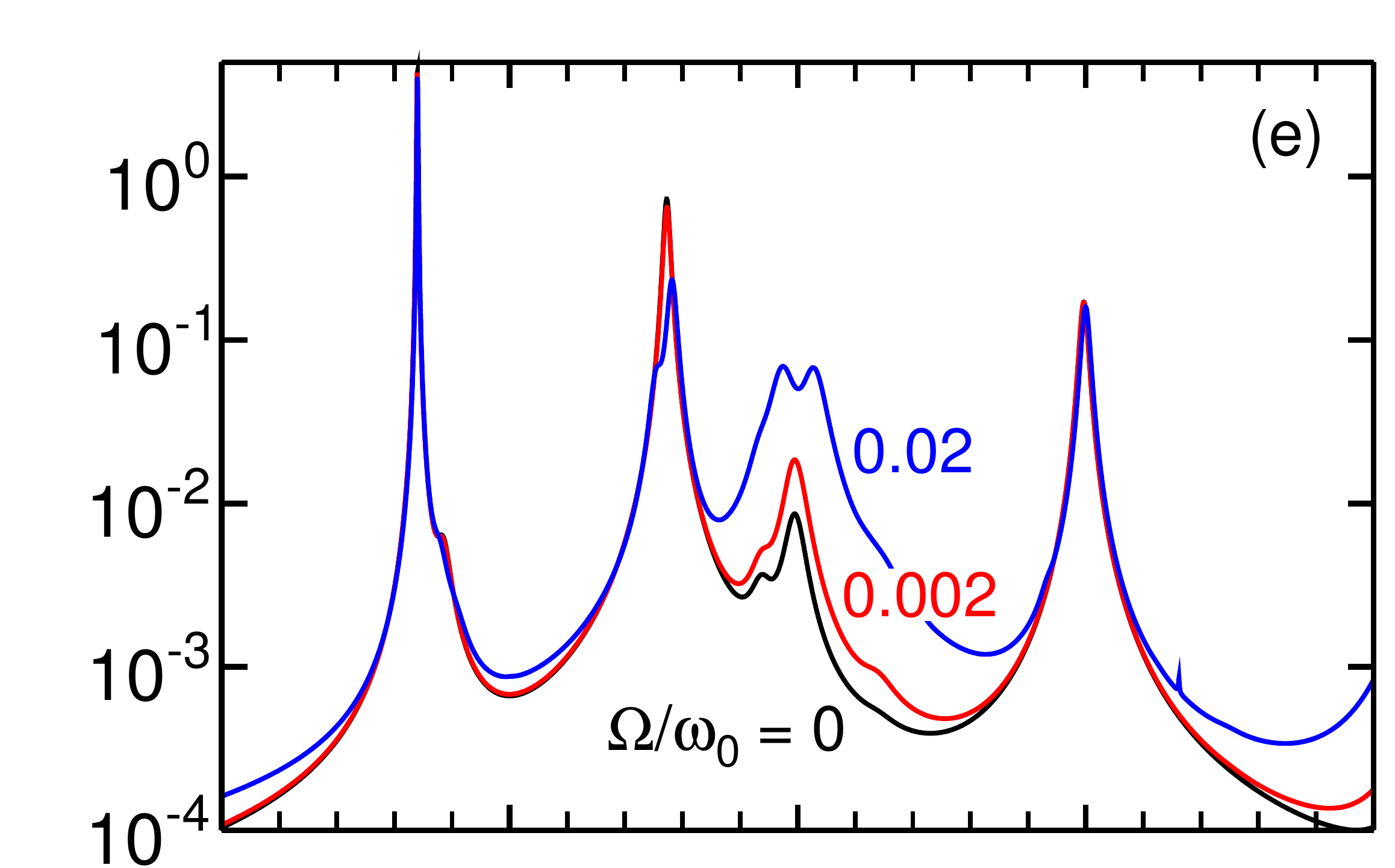}\\
  \includegraphics[width=0.49\linewidth]{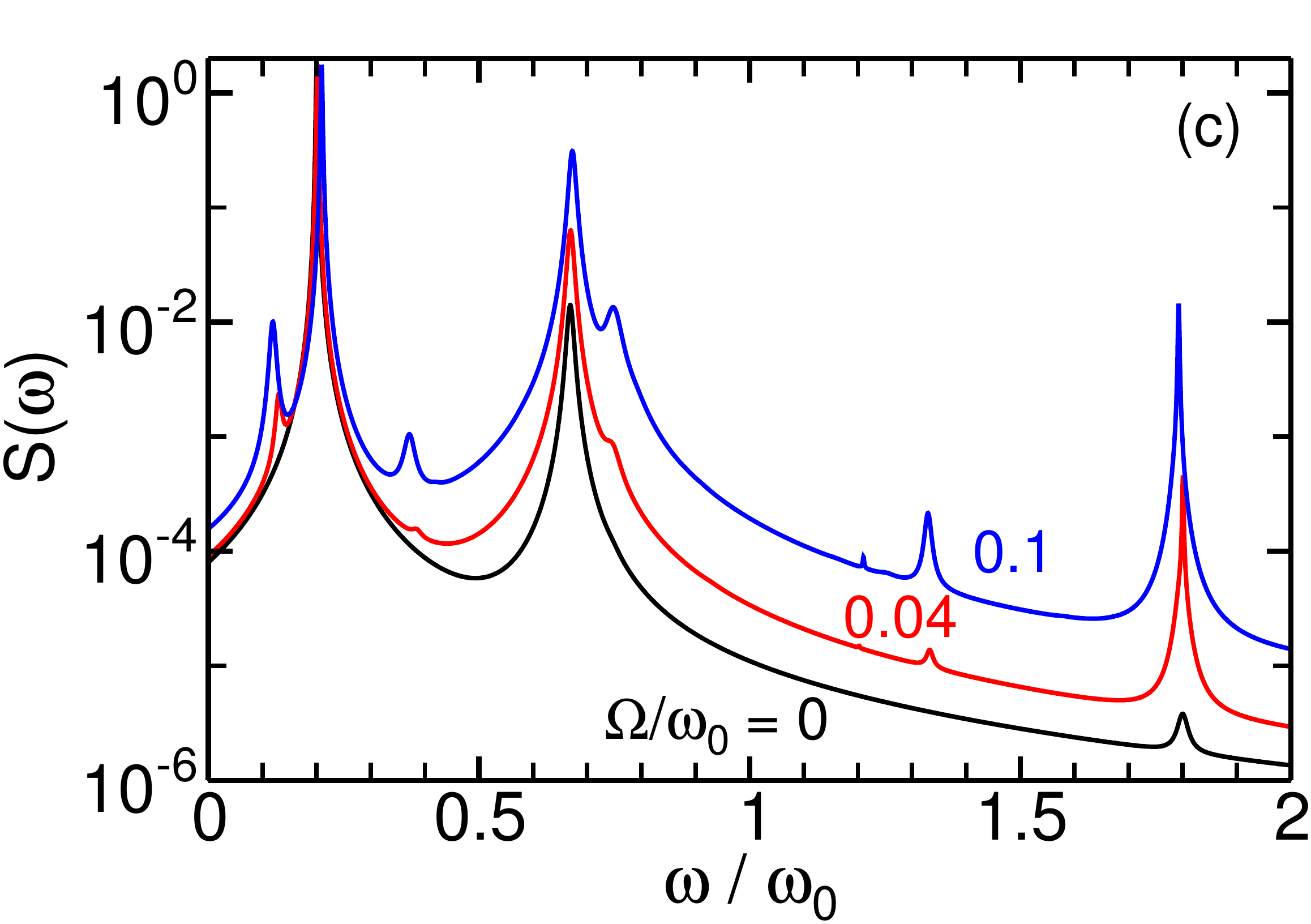}
  \includegraphics[width=0.49\linewidth]{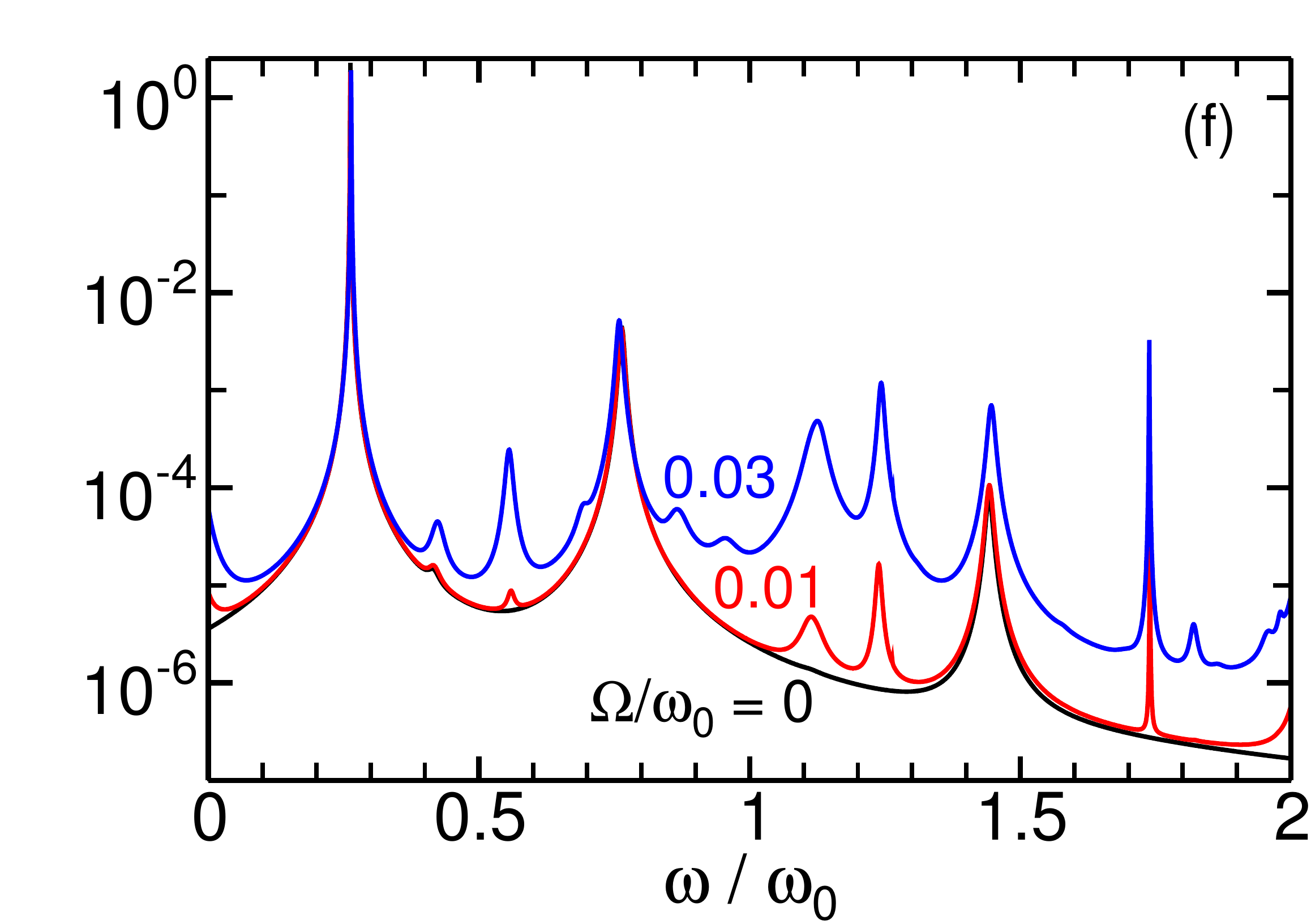}
  \caption{\label{fig:spct1}Emission spectra $S(\omega)$ for one emitter ($N = 1$) for different values of the laser intensity $\Omega$ as indicated in the plots.
    The left (right) column depicts the results for $g' = \Omega' = 0$ ($g' = g$ and $\Omega' = \Omega$).
    The emitter-cavity-coupling strength and the environment temperature are (a) and (d) $g = 0.5 \, \omega_0$ and $T = 0.07 \, \omega_0$, (b) and (e) $g = 0.7 \, \omega_0$ and $T = 0.23 \, \omega_0$, and (c) and (f) $g = 0.8 \, \omega_0$ and $T = 0.1 \, \omega_0$.}
\end{figure}

In Fig.~\ref{fig:spct1} we show the emission spectrum $S(\omega)$ calculated numerically from Eq.~\eqref{S}.
Technically, this requires the evaluation of (i) the Floquet states as the eigenstates of the one-cycle evolution operator, (ii) the (constant) coefficients of the master equation in the Floquet basis, (iii) the asymptotic state as the stationary solution for the diagonal density matrix elements, (iv) the output operator from Eq.~\eqref{X}, and finally (v) the spectrum~\eqref{S} as a sum of Lorentz peaks.
In these calculations, as well as in all the following ones, a maximal number of 50 cavity photons and 110 Fourier modes is used in the evaluation of the Floquet states, which is sufficient for the parameters used.
All results here and later are given at resonance $\omega_c = \omega_x = \omega_d = \omega_0$, and we compare the cases $g' = 0$ and $\Omega' = 0$ [Figs.~\ref{fig:spct1}(a)--\ref{fig:spct1}(c)] with $g' = g$ and $\Omega' = \Omega$ [Figs.~\ref{fig:spct1}(d)--\ref{fig:spct1}(f)].
The emission spectra in Fig.~\ref{fig:spct1}, as well as the Glauber functions discussed in Sec.~\ref{sec:emission}, are evaluated for a single emitter ($N = 1$).
The corresponding results for two and three emitters are given in Appendix~\ref{app:emitters}.

Of course, the emission spectra in Fig.~\ref{fig:spct1} for $\Omega = 0$ coincide with previous results~\cite{PAF15}:
For low temperatures $T \ll \omega_0$, the stationary thermal state is dominated by the ground state, and the spectrum in Figs.~\ref{fig:spct1}(a) and~\ref{fig:spct1}(d) exhibits a single peak marking the first possible transition into the ground state.
For finite laser intensity $\Omega > 0$,
the asymptotic stationary state has to be determined according to the temperature-dependent matrix elements of transitions between different Floquet states.
Because the laser excitation strongly affects these matrix elements if the corresponding transitions are in resonance with the laser frequency, the populations $\rho_{n,n}^\infty$ of higher excited states can be enhanced.
This leads to the increase of peak height for finite $\Omega$ in Figs.~\ref{fig:spct1}(a) and~\ref{fig:spct1}(d).

At first sight, according to the above arguments, increasing $\Omega$ should act in a similar manner as increasing the temperature.
However, because the resonance enhancement of transition matrix elements is not equal for all Floquet states, the asymptotic populations of Floquet states will no longer follow a thermal distribution.
It is thus no surprise that a high-energy spectral line may become stronger than a low-energy one [see Fig.~\ref{fig:spct1}(d)].
With increasing temperature, transitions involving higher excited states contribute to the emission spectrum in Figs.~\ref{fig:spct1}(b) and~\ref{fig:spct1}(e) already for $\Omega = 0$.
The changes of the spectral lines with the laser intensity now strongly depend on the choices of $g'$ and $\Omega'$.
The emission spectrum in Fig.~\ref{fig:spct1}(e) for $g' = g$ and $\Omega' = \Omega$ is much more sensitive to changes of $\Omega$ than the spectrum given in Fig.~\ref{fig:spct1}(b) for $g' = \Omega' = 0$.
This can be ascribed to the particular form of the quasienergy spectra displayed in Figs.~\ref{fig:energy}(a) and~\ref{fig:energy}(d):
Because only for $g' = g$ but not for $g' = 0$ pairs of system energies $E_n$ are very close to each other in the strong-coupling regime and the energy difference between neighboring pairs equals $\omega_0$, the number of resonant matrix element enhancements is increased for $g' = g$ relative to the case $g' = 0$.

The situation changes again if $g$ is increased from strong to ultrastrong coupling.
In addition to the markedly different behavior with modified laser intensity, a whole bunch of new spectral lines appears for finite $\Omega$ in Fig.~\ref{fig:spct1}(c) and~\ref{fig:spct1}(f).
The reason is that for finite laser intensity transitions not only between system energies $E_n = \epsilon_n + \nu \omega$ but also between their Fourier modes (e.g., their sidebands $\nu \pm 1$) are allowed.
For example, a transition from the ground to the first excited state has a negative transition energy and will not lead to a spectral line if $\Omega = 0$.
Nevertheless, an additional peak may occur for $\Omega > 0$ if the energy difference between the ground and the first excited state is less than the energy associated with the laser frequency, because the upper sideband of the ground state is then energetically higher than the first excited state.
Such processes lead, e.g., to the additional peak at $\omega \simeq 1.73 \omega_0$ in Fig.~\ref{fig:spct1}(f) that belongs to the transition from the ground- to the first-excited-state with energy difference $-0.27 \omega_0$.

\begin{figure}
  \includegraphics[width=0.49\linewidth]{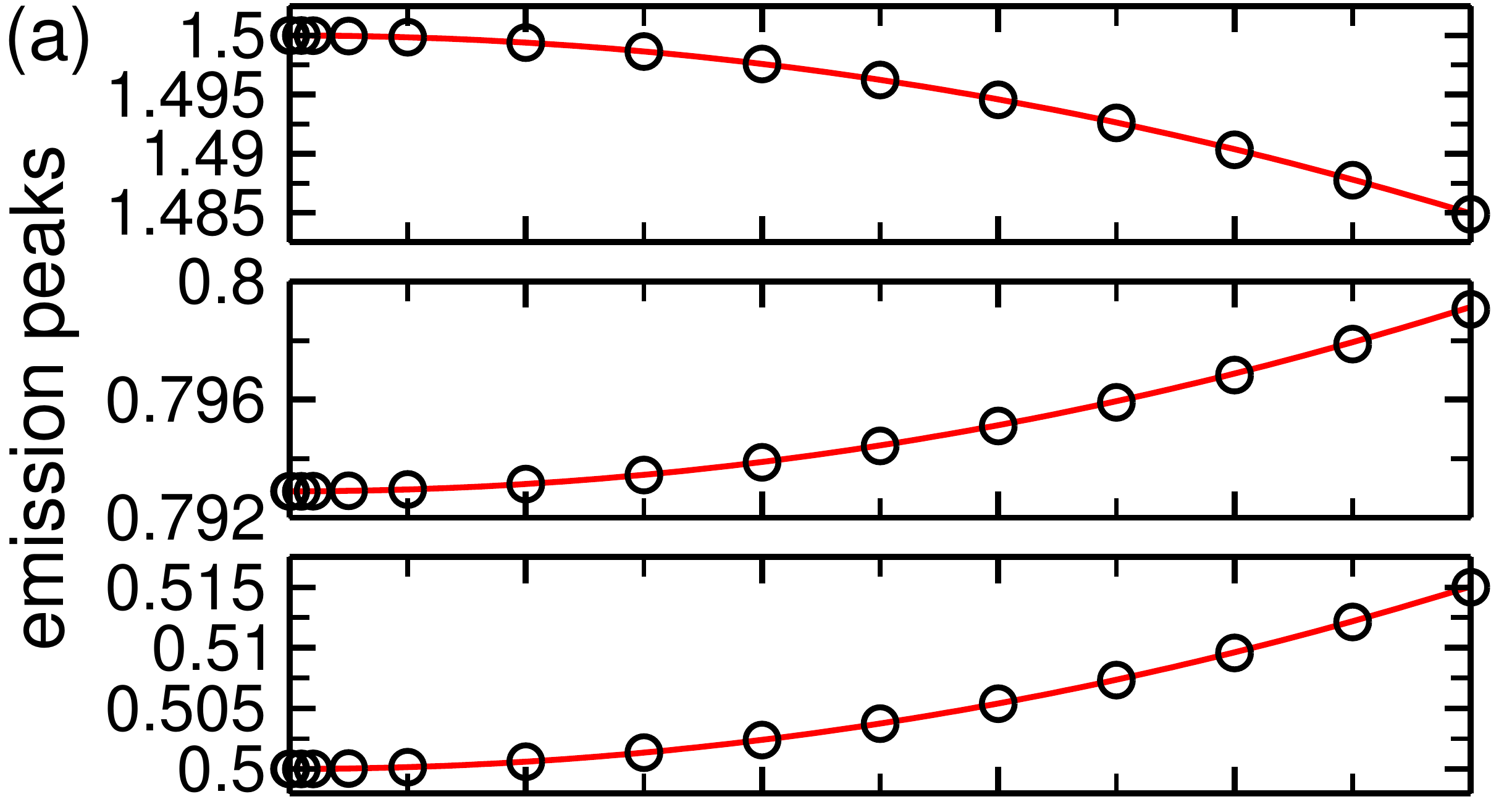}
  \includegraphics[width=0.49\linewidth]{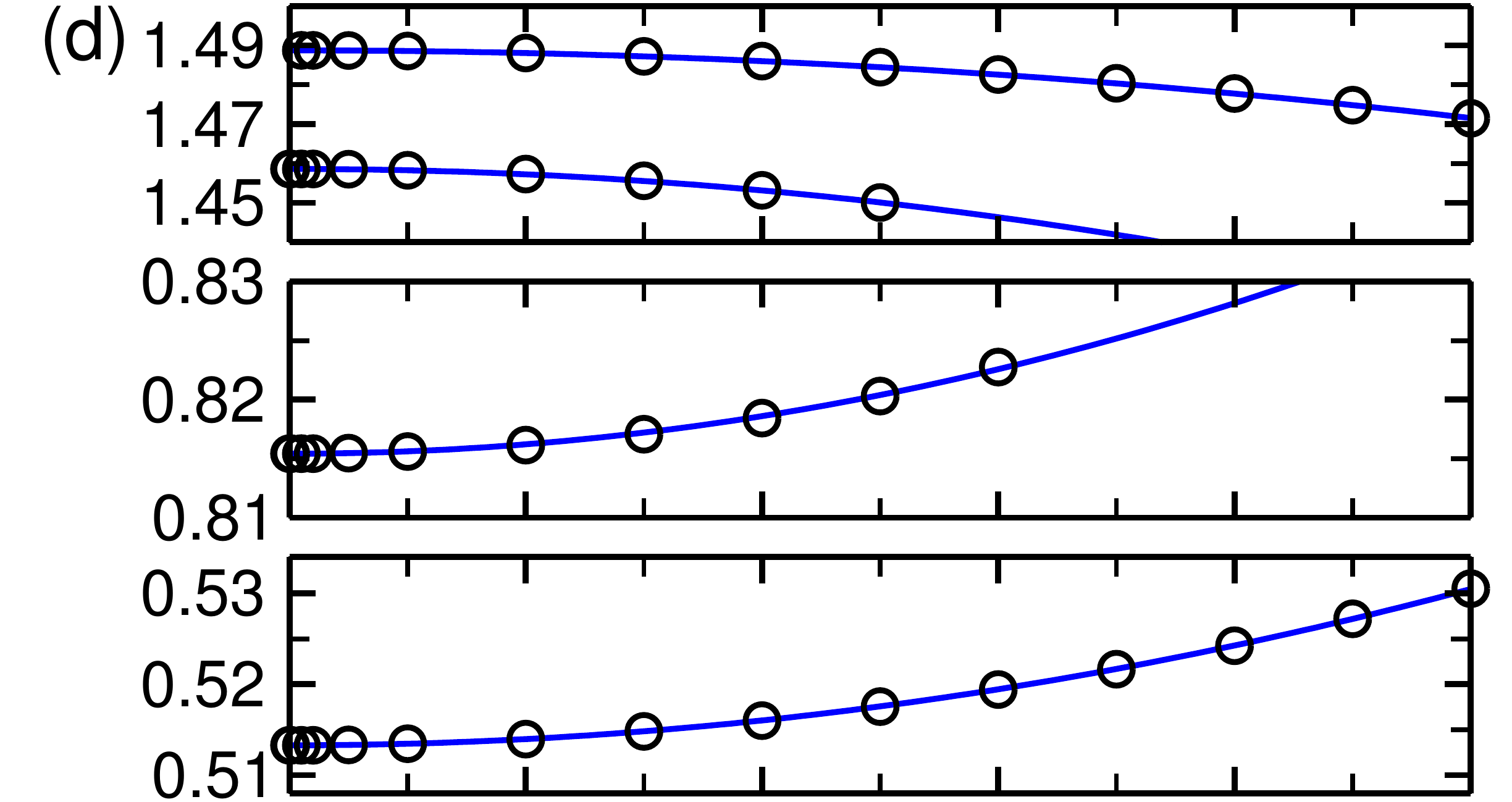}\\
  \includegraphics[width=0.49\linewidth]{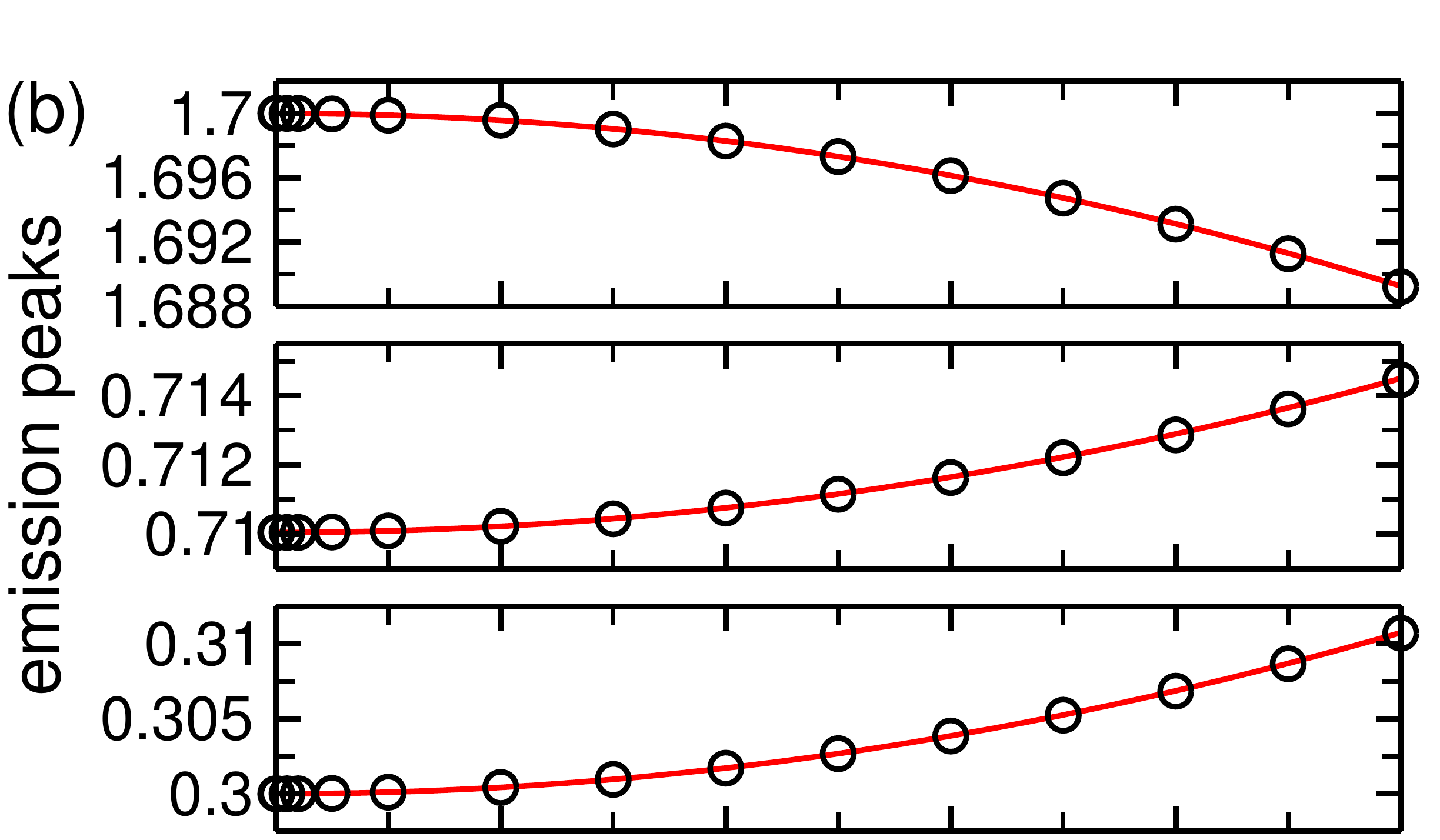}
  \includegraphics[width=0.49\linewidth]{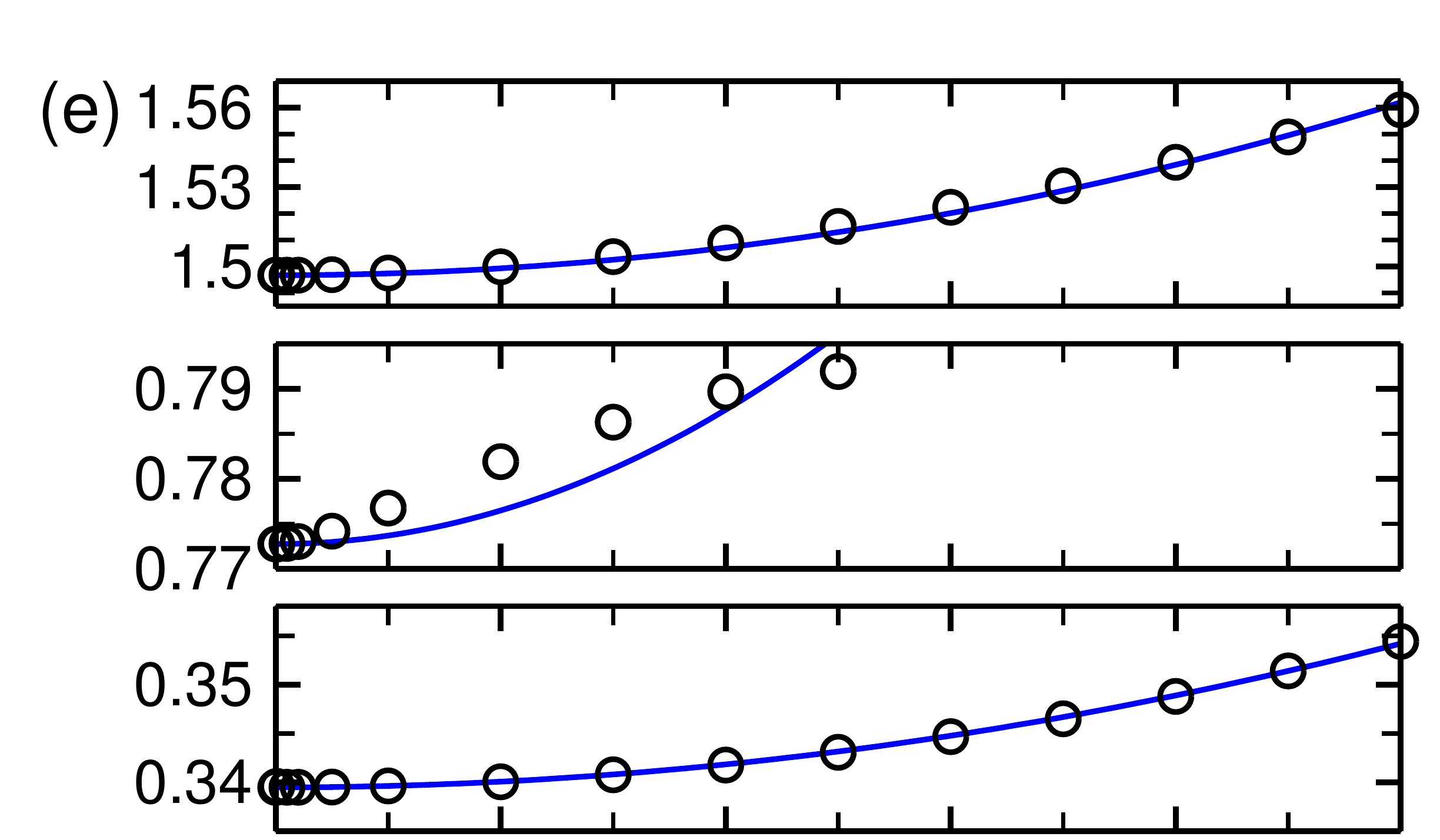}\\
  \includegraphics[width=0.49\linewidth]{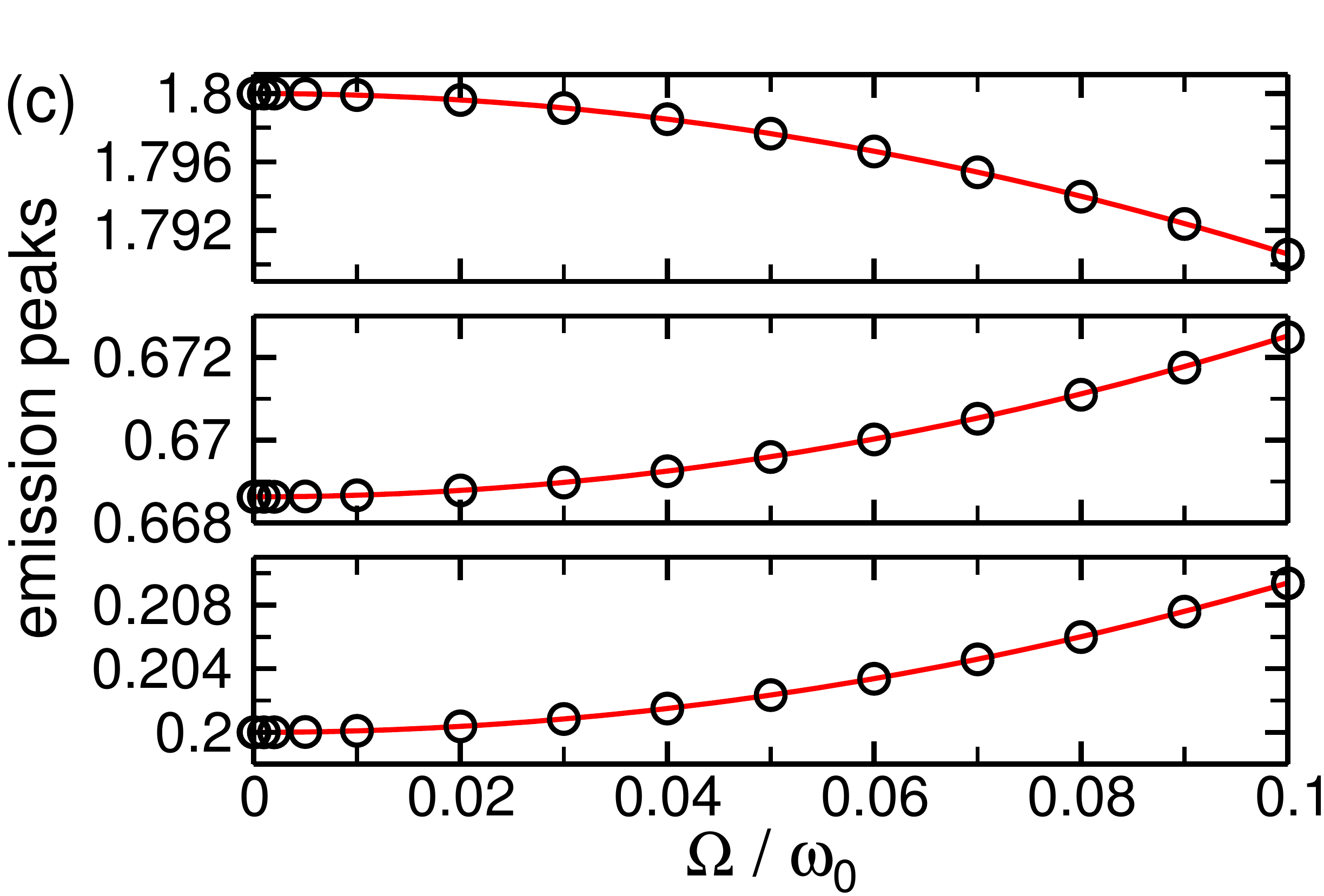}
  \includegraphics[width=0.49\linewidth]{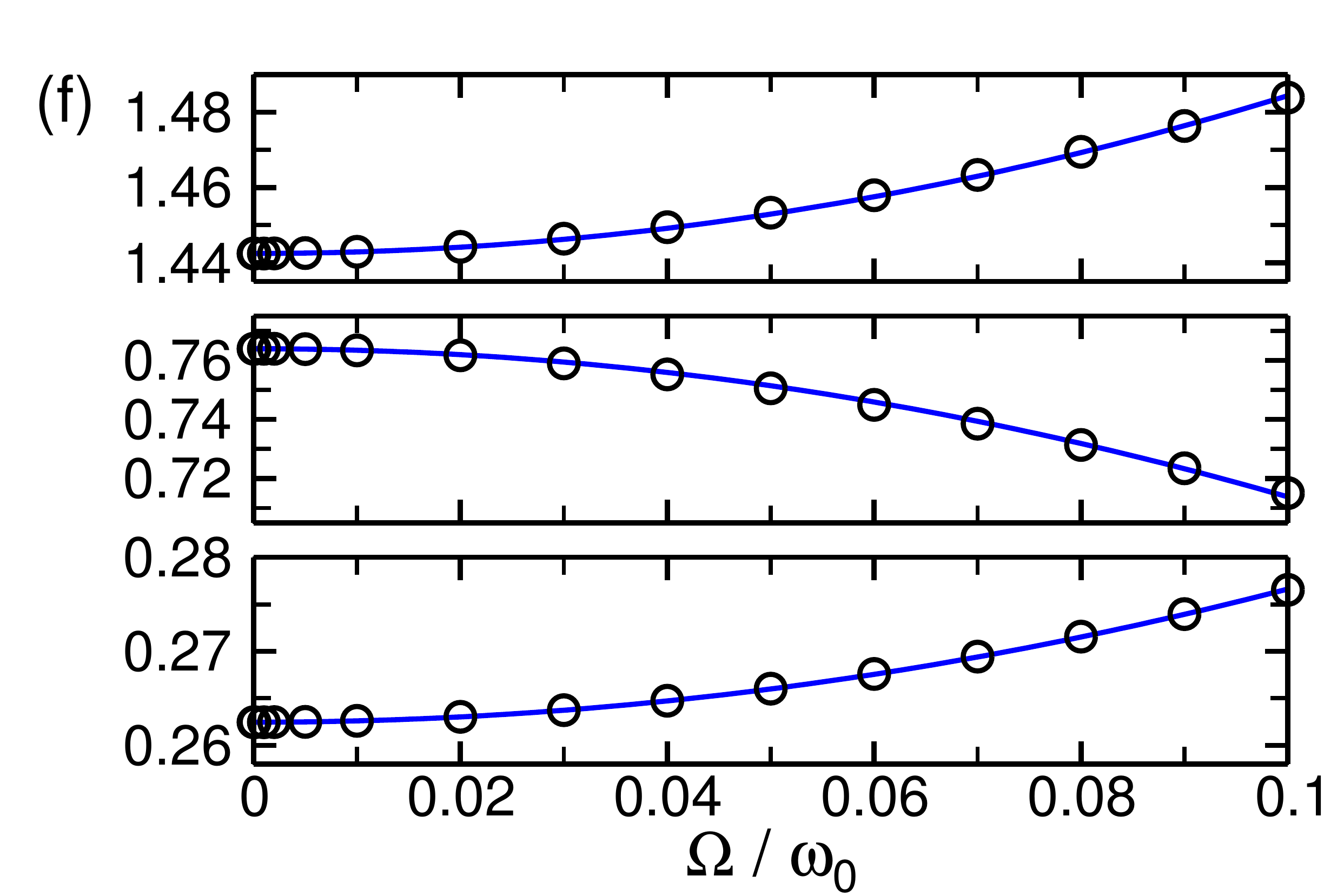}
  \caption{\label{fig:shft1}Shift of emission peaks as a function of the laser intensity $\Omega$ for one emitter.
    The left (right) column depicts the results for $g' = \Omega' = 0$ ($g' = g$ and $\Omega' = \Omega$).
    The other parameters are the same as those used in Fig.~\ref{fig:spct1}.
    The (red) lines in the left panels give the analytical result for the quasienergies~\cite{AGC92} (see Appendix~\ref{app:TC}) and the (blue) lines in the right panels depict fitted values proportional to $[1 - (\Omega / g)^2]^{3/4}$.}
\end{figure}

So far, the emission spectra in Fig.~\ref{fig:spct1} do not clearly show the expected dynamic Stark effect, i.e., a shift of the spectral lines with the laser intensity.
Therefore, we extract the position of selected spectral lines and plot their $\Omega$ dependence in Fig.~\ref{fig:shft1}.
The coupling strengths $g$ and environment temperatures $T$ are equal to that used in the calculation of the emission spectra.
The (red) lines in Fig.~\ref{fig:shft1} show the quasienergy transitions derived from the result in Appendix~\ref{app:TC} for $g' = 0$ and $\Omega' = 0$.
We find that our numerical data agree perfectly with the analytical calculation.
The results for $g' = g$ and $\Omega' = \Omega$ [Figs.~\ref{fig:shft1}(d)--\ref{fig:shft1}(f)] confirm that the dynamic Stark shift is proportional to $\Omega^2$ for small $\Omega \ll \omega_0, g$, with the apparent exception in Fig.~\ref{fig:shft1}(e), where the quadratic scaling is visible only for very small $\Omega$.
In addition, the fit proportional to $[1 - (\Omega/ g)^2]^{3/4}$ (blue lines in Fig.~\ref{fig:shft1}), which is based on the analytical result for $g' = \Omega' = 0$, almost perfectly agrees with our numerical data, but completely fails at describing the shift of a single spectral line in Fig.~\ref{fig:shft1}(e).
We expect that these deviations are caused by the interference of two (or even more) spectral lines whose height change with $\Omega$ such that not a single line is observed in Fig.~\ref{fig:shft1}(e).
This expectation is corroborated by the fact that all spectral lines in this frequency range disappear at $\Omega \simeq 0.06 \omega_0$, i.e., they cannot be observed for higher laser intensities.
Indeed, a closer look at the numerical data shows that in the range $0.77 \lesssim \omega / \omega_0 \lesssim 0.8$ a second spectral line appears with approximately half the weight of the plotted one.
What we observe in Fig.~\ref{fig:shft1}(e) is thus an avoided crossing between the two corresponding quasienergies.
With the exception of such anticrossings, we may therefore conclude that the quasienergies and hence the dynamic Stark effect are proportional to $[1 - (\Omega/ g)^2]^{3/4}$.

\section{\label{sec:emission}Statistics of emitted photons}
The possible generation of nonclassical light can be monitored by the second-order Glauber function~\cite{Gla63}
\begin{equation}\label{glauber}
  g^{(2)}(\tau) = \frac{1}{T_d} \int\limits_0^{T_d} \frac{\mathop{\text{Tr}} \big\{ \dot{X}_+ \dot{X}_- V(t + \tau, t) \dot{X}_- \rho^\infty(t) \dot{X}_+ \big\}}{\mathop{\text{Tr}} \big\{ \dot{X}_+ \dot{X}_- \overline{\rho}^\infty \big\}^2} \, \rmd t \,,
\end{equation}
where the state $\overline{\rho}^\infty$ in the denominator is the time-averaged stationary state
\begin{equation}
  \overline{\rho}^\infty = \frac{1}{T_d} \int_0^{T_d} \rho^\infty(t) \, \rmd t = \sum_{n, \nu} \rho_{n,n}^\infty | \widetilde{\phi}_n(\nu) \rangle \langle \widetilde{\phi}_n(\nu) | \,.
\end{equation}
The emitted photons have a super-Poissonian distribution if $g^{(2)}(0) > 1$, a Poissonian distribution if $g^{(2)}(0) = 1$, and a nonclassical sub-Poissonian distribution if $g^{(2)}(0) < 1$.
The value $g^{(2)}(0) = 2$ indicates thermal light emission.

\subsection{\label{ssec:glauber-0}Glauber function at zero time delay}
\begin{figure}
  \includegraphics[scale=0.3]{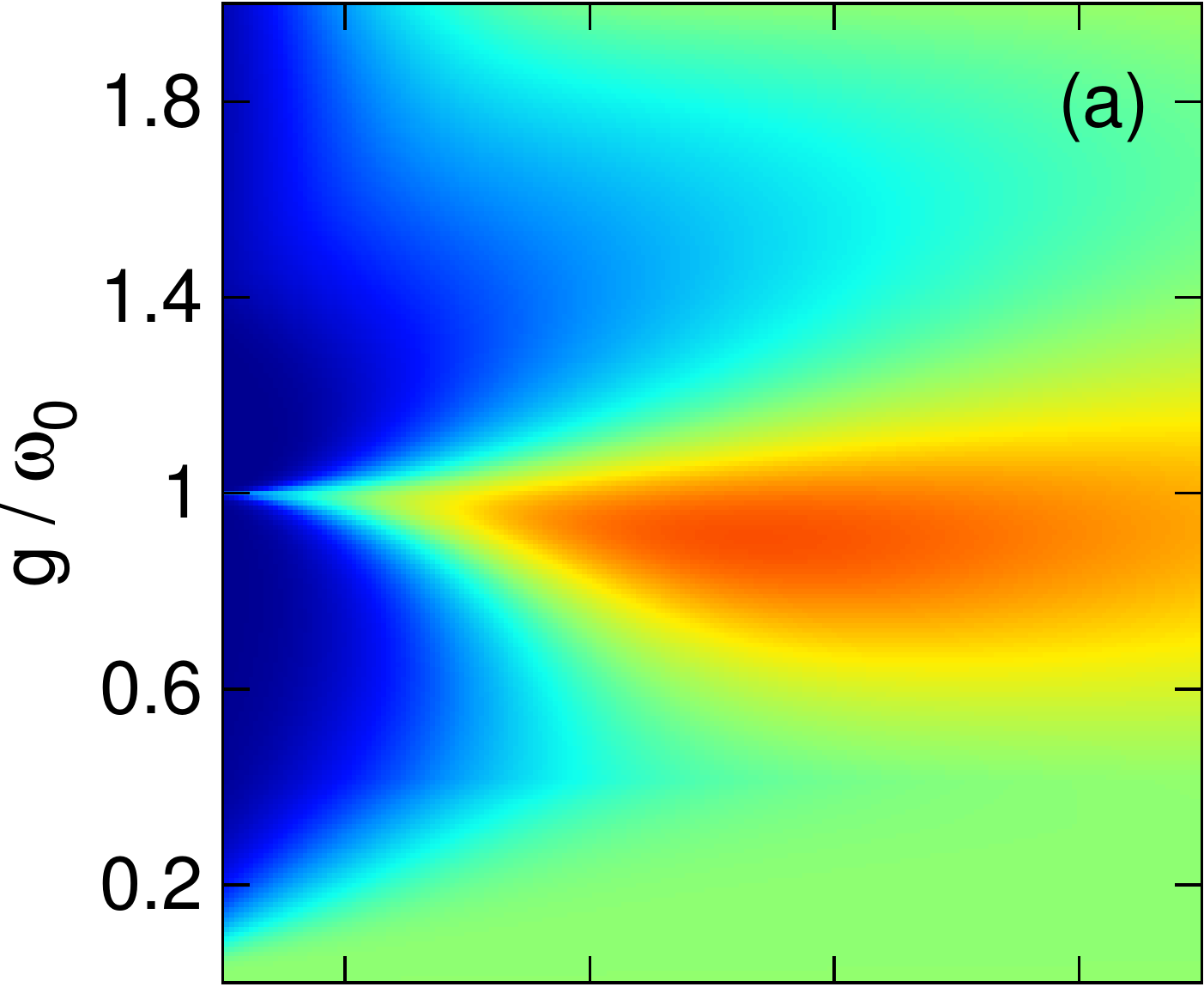}
  \includegraphics[scale=0.3]{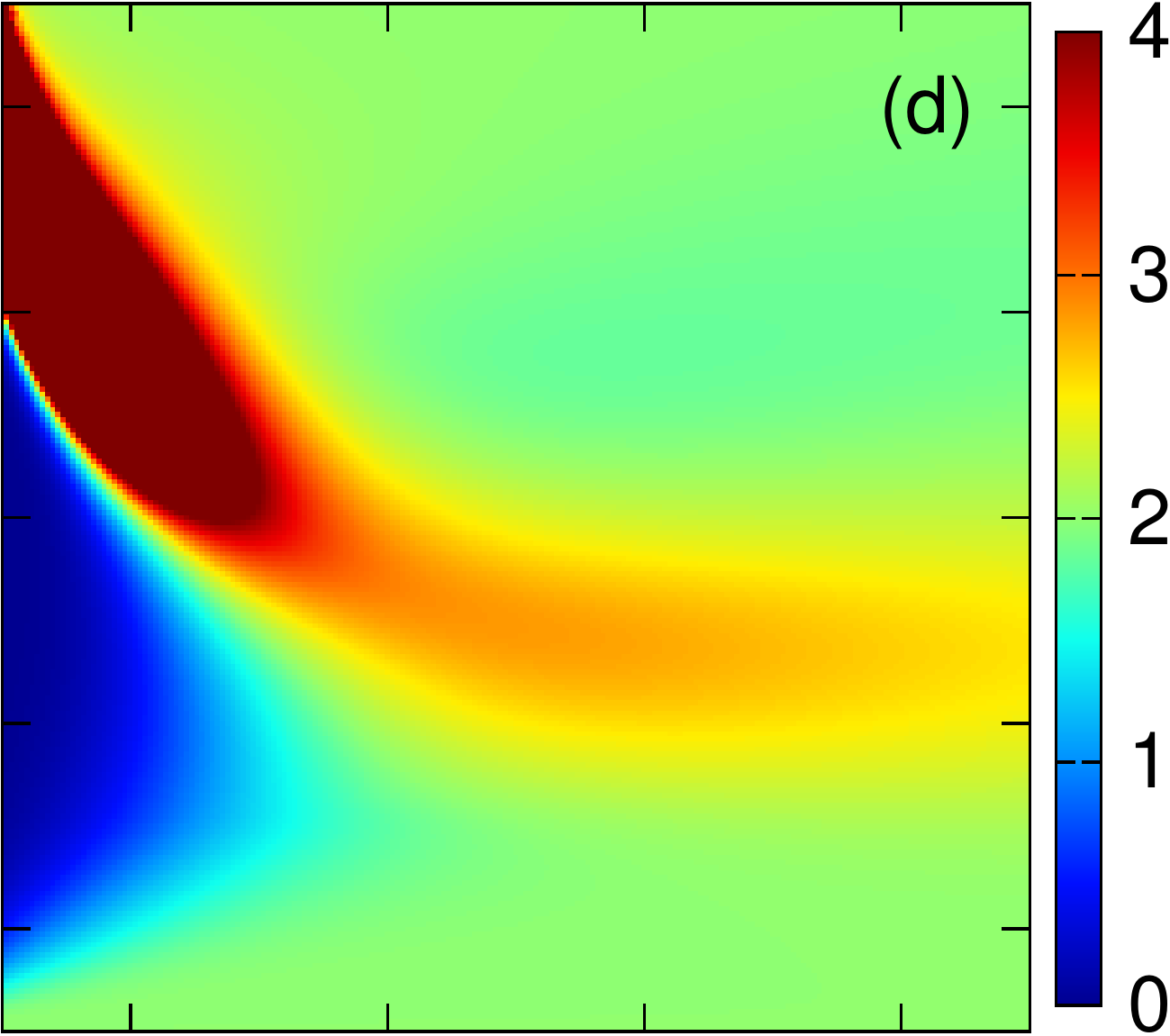}\\
  \includegraphics[scale=0.3]{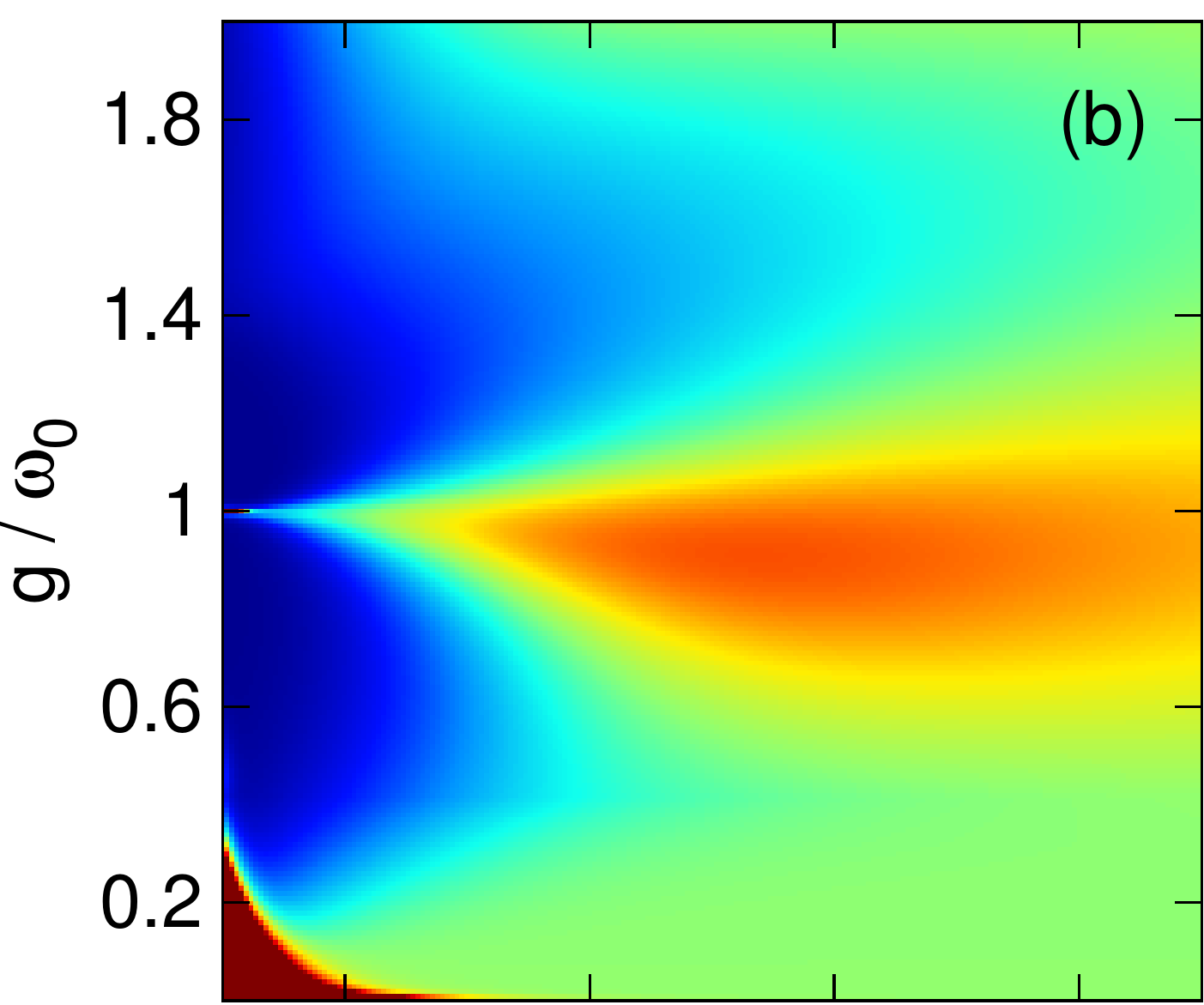}
  \includegraphics[scale=0.3]{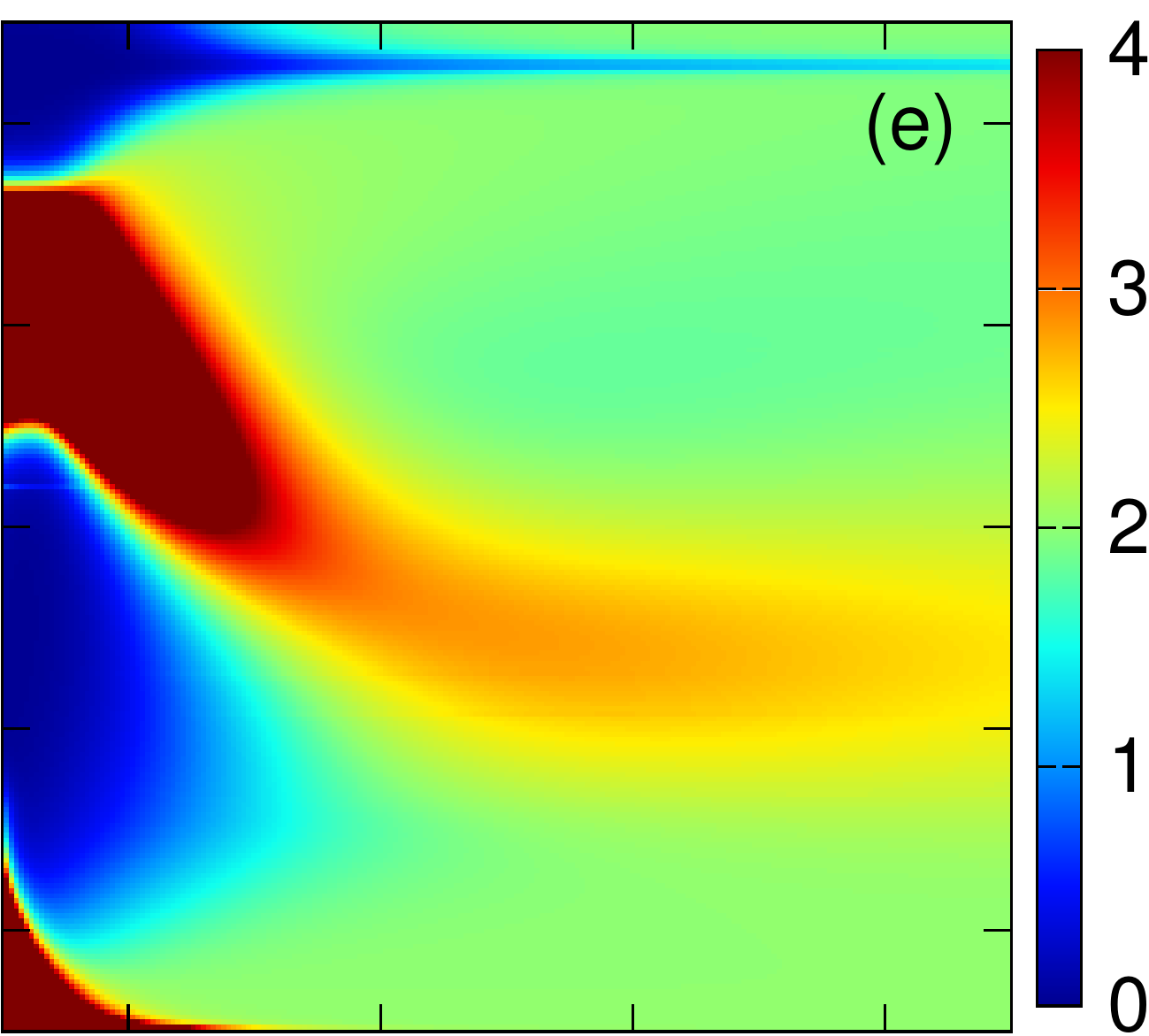}\\
  \includegraphics[scale=0.3]{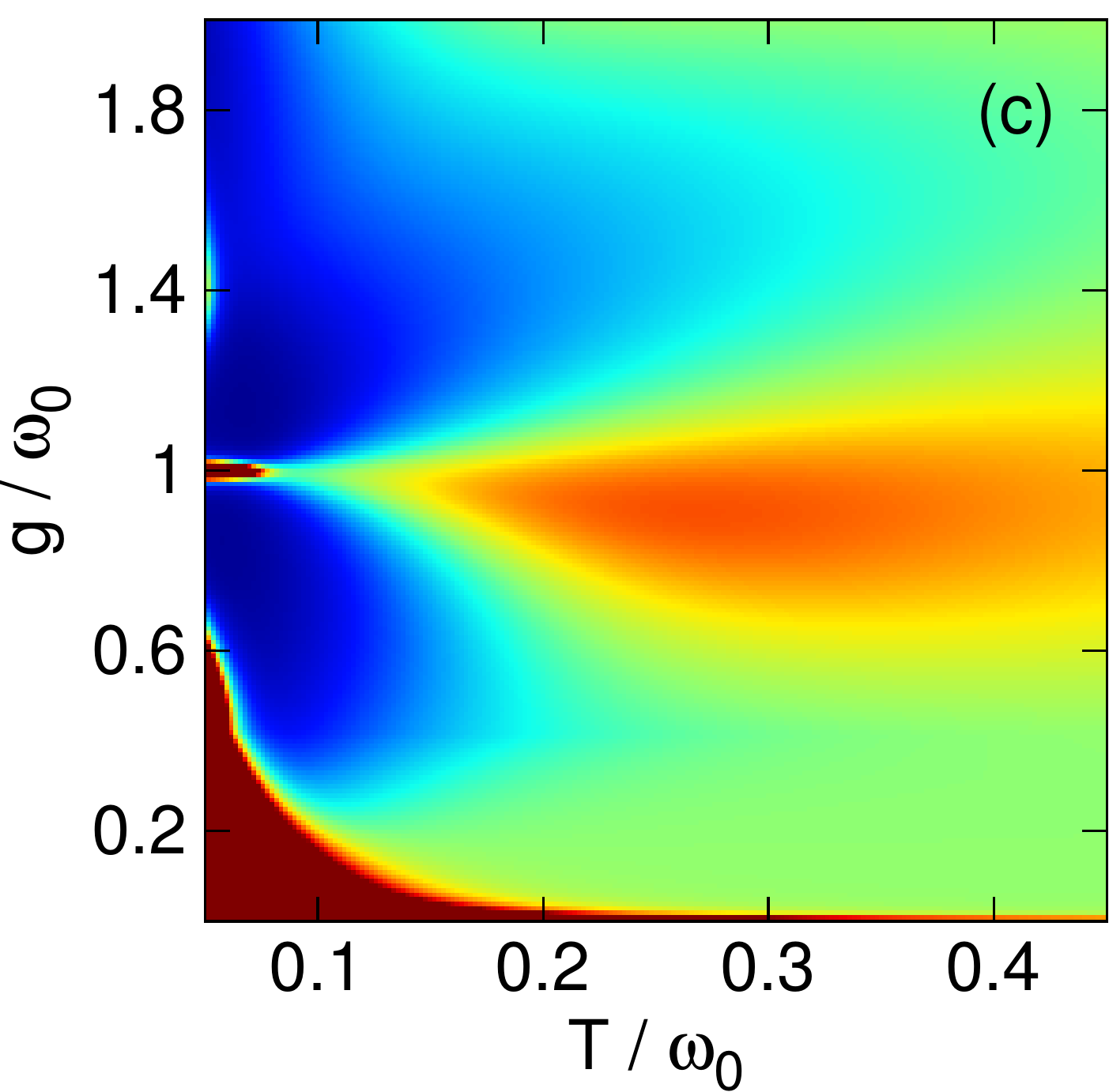}
  \includegraphics[scale=0.3]{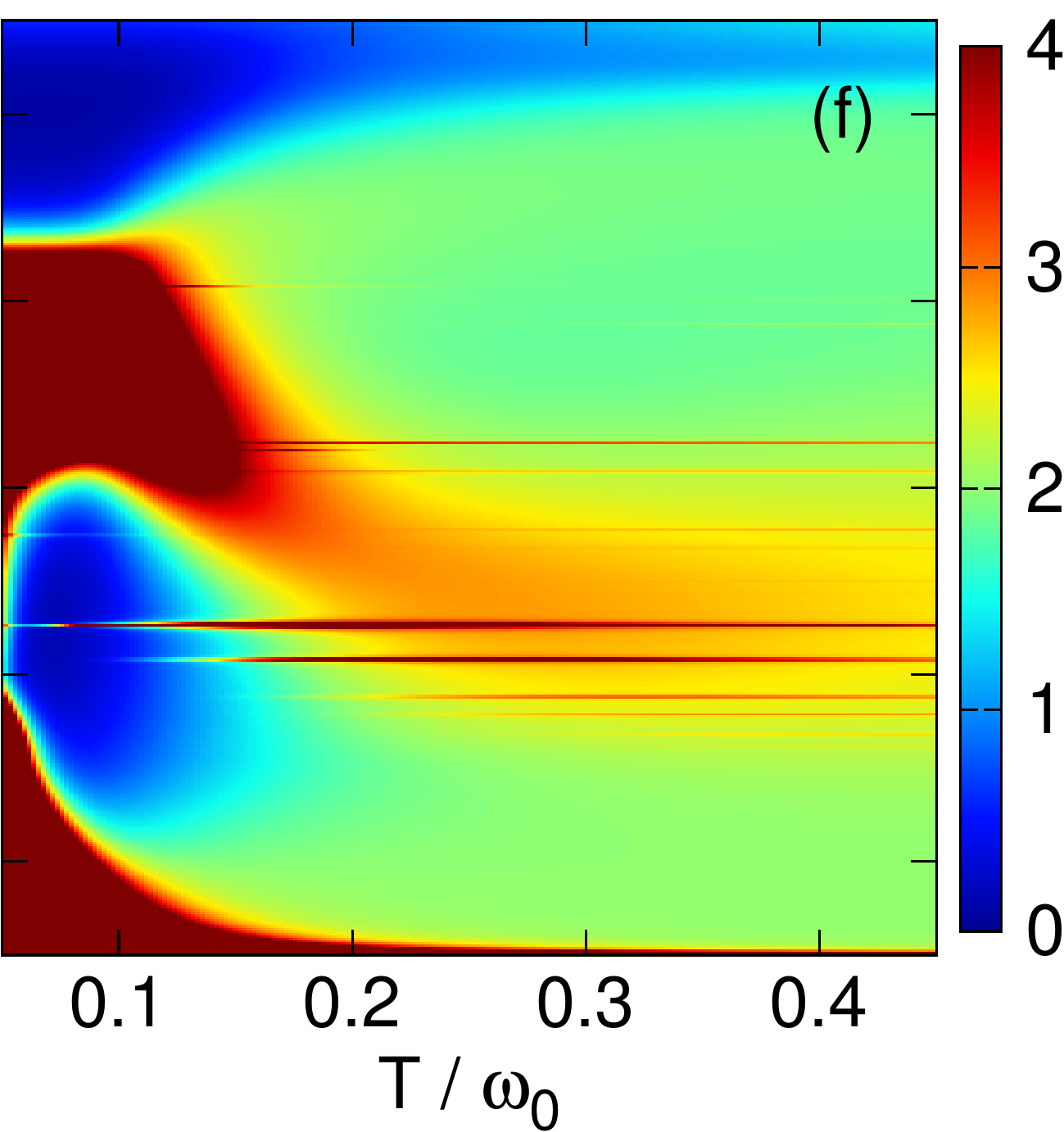}
  \caption{\label{fig:glauber1}Glauber function $g^{(2)}(0)$ at zero time delay for one emitter as a function of the environment temperature $T$ and the emitter-cavity-coupling strength $g$.
    The left (right) column depicts the results for $g' = \Omega' = 0$ ($g' = g$ and $\Omega' = \Omega$).
    The laser intensity is (a) and (d) $\Omega = 0$, (b) and (e) $\Omega = 10^{-4} \, \omega_0$, and (c) and (f) $\Omega = 10^{-3} \, \omega_0$.
    Note that all values $g^{(2)}(0) \geq 4$ are assigned the same [dark red (dark gray)] color in the density plots.}
\end{figure}

The Glauber function $g^{(2)}(0)$ for one emitter is shown for different laser intensities $\Omega$ in Fig.~\ref{fig:glauber1}.
For $\Omega = \Omega' = 0$~\cite{RSH13,PAF15}, the Glauber function depends on the system energies $E_n$ (red solid lines in Fig.~\ref{fig:energy}) and the corresponding eigenstates of $H_D$ in Eq.~\eqref{HD}.
The output operator $\dot{X}_-$ in Eq.~\eqref{X} allows for transitions between eigenstates where the change of the principal quantum number $M$ is $\Delta M = 1$ (dipole transitions).
At low temperatures, the denominator is dominated by the contribution from the transition $1 \to 0$, whereas the most important contribution to the numerator is given by the transition sequence $2 \to 1 \to 0$.
This leads to the triangular region with $g^{(2)}(0) < 1$ at low temperatures in Fig.~\ref{fig:glauber1}(d), with the emission of nonclassical light.
The triangular region lies below an elongated region with super-Poissonian photon statistics where $g^{(2)}(0) > 2$.
For $g' = 0$ in Fig.~\ref{fig:glauber1}(a) the super-Poissonian region is pushed back in favor of a second triangular region of sub-Poissonian light emission.

For finite $\Omega > 0$, the Glauber function $g^{(2)}(0)$ in Fig.~\ref{fig:glauber1} differs from these results in three aspects:
First, in Figs.~\ref{fig:glauber1}(b), \ref{fig:glauber1}(c), \ref{fig:glauber1}(e) and~\ref{fig:glauber1}(f), a triangular region of super-Poissonian photon statistics at low emitter-cavity-coupling strength $g$ and environment temperature $T$ is observed, indicating that the photon statistics becomes more classical.
Second, in Figs.~\ref{fig:glauber1}(e) and~\ref{fig:glauber1}(f) an elongated region of nonclassical light emission with sub-Poissonian photon statistics is formed at ultrastrong coupling.
Third, in Fig.~\ref{fig:glauber1}(f) additional horizontal lines with enhanced $g^{(2)}(0)$ appear for very specific emitter-cavity couplings $g$.
The second and third features are observed only for $g' = g$ in Figs.~\ref{fig:glauber1}(d)--\ref{fig:glauber1}(f) but not for $g' = 0$ in Figs.~\ref{fig:glauber1}(a)--\ref{fig:glauber1}(c).

To explain the above observations, we have to analyze the output operator~\eqref{X}, which involves the quasienergies shown in Fig.~\ref{fig:energy} and transitions between the corresponding Floquet states.
Because $\Omega \leq 10^{-3} \omega_0$ in Fig.~\ref{fig:glauber1}, we are in the regime of small laser intensity, where only the system energies $E_n = \epsilon_n + \nu \omega_d$ and the first two sidebands $E_n \pm \omega_d = \epsilon_n + (\nu \pm 1) \omega_d$ contribute.
In the denominator of $g^{(2)}(0)$ from Eq.~\eqref{glauber}, states that are connected by the action of a single output operator contribute.
Most relevant at low temperatures is the first excited state.
In contrast, for the numerator of $g^{(2)}(0)$, where each operator appears twice, states that are separated by two output operators contribute.
This difference will be of importance in the following discussion of Fig.~\ref{fig:glauber1}.

\subsubsection{First observation}
For $\Omega \ll g, \omega_0$, each Floquet state has a contribution from the system energy $E_n$ that does not depend on $\Omega$, and the corrections due to the two sidebands are linear in $\Omega$ (see Appendix~\ref{app:TC}).
Hence, the output operator $\dot{X}_-$ not only mediates dipole transitions with $\Delta M = 1$ but also transitions with $\Delta M = 0, 2$, which scale proportionally to $\Omega$.
As an exception, the transition $0 \to 0$ is forbidden.
Hence, at low temperatures $T$, the denominator of $g^{(2)}(0)$ is still dominated by the contribution from the transition $1 \to 0$ and remains (approximately) $\Omega$ independent.
The first correction arises from the transition $1 \to 1$ involving a sideband of $M = 1$.
This contribution scales proportionally to $\Omega^2$ because of the product $\dot{X}_+ \dot{X}_-$ in the denominator of Eq.~\eqref{glauber}.
Important for the numerator of Eq.~\eqref{glauber} is the transition sequence $1 \to 1 \to 0$.
Because of the starting point of this transition sequence, its contribution (proportional to $\Omega^2$) has to be multiplied with the stationary population of the state $M = 1$.
If the temperature $T$ is so small that the stationary population of the state $M = 2$ becomes comparable to that of the state $M = 1$ multiplied with $\Omega^2$, the contribution from the sequence $1 \to 1 \to 0$ will be comparable to that of $2 \to 1 \to 0$.
In that case, even a small laser intensity $\Omega \ll g, \omega_0$ will significantly increase the value of the Glauber function $g^{(2)}(0)$.
This increase of the numerator is the reason for the triangular region of highly classical light emission at low emitter-cavity coupling $g$ and low environment temperature $T$ in Figs.~\ref{fig:glauber1}(b), \ref{fig:glauber1}(c), \ref{fig:glauber1}(e), and~\ref{fig:glauber1}(f) and thus explains our first observation.

\subsubsection{Second observation}
The second observation, i.e., the decrease of the Glauber function $g^{(2)}(0)$ at ultrastrong emitter-cavity coupling, appears only for $g' = g$.
Our arguments in favor of this property thus have to involve the precise form of the quasienergy spectrum shown in Fig.~\ref{fig:energy}.
We already noted in Sec.~\ref{ssec:spectra} that only for $g' = g$, but not for $g' = 0$, pairs of system energies $E_n$ (red solid lines) are very close to each other if we are in the ultrastrong-coupling regime.
Hence, the dominant contributions to $g^{(2)}(0)$ are given by transitions between pairs of states.
For finite $\Omega$, an additional pair of sidebands below the lowest pair of system energies appears.
Corrections to the denominator (numerator) that scale proportionally to $\Omega^2$ thus involve transitions between these sidebands and the lowest (first excited) pair of system energies.
The relevant populations for $\Omega > 0$ compared to that for $\Omega = 0$ are thus shifted to the next lower-lying pair of states.
At low temperatures, this gain in state population may compensate for the decrease (proportional to $\Omega^2$) of the transition matrix element.
Then the denominator and numerator of $g^{(2)}(0)$ will be enhanced.
Nevertheless, the energy difference between neighboring pairs of states is constant, $\Delta E \simeq \omega_0$, and the expectation value in the denominator of the Glauber function is squared.
An equal increase of expectation values thus leads to a decreasing result and explains the elongated region of sub-Poissonian light emission at ultrastrong coupling in Figs.~\ref{fig:glauber1}(e) and~\ref{fig:glauber1}(f).
For $g' = 0$, the linearity of the dispersion relations prohibits an expectation value enhancement.
As a result, Figs.~\ref{fig:glauber1}(b) and~\ref{fig:glauber1}(c) do not show this region.

\subsubsection{Third observation}
The physics behind the third observation, i.e., the appearance of thin horizontal lines of highly classical light emission in Fig.~\ref{fig:glauber1}(f), is fundamentally different.
It can be verified numerically that the denominator does not change significantly if $g$ is varied across one of these lines.
The whole modification of $g^{(2)}(0)$ is carried by its numerator.
Hence, we have to search for special transition sequences $a \to b \to c$ between quasienergies to explain these strongly-$g$-dependent modifications.

Drawing vertical lines in Fig.~\ref{fig:energy} (blue lines) at those emitter-cavity-coupling strengths where the horizontal lines in Fig.~\ref{fig:glauber1}(f) for $g^{(2)}(0)$ appear, we realize that these couplings mark positions where system energies cross sidebands.
At these crossings, the energy difference between two system energies is $\omega_0$, i.e., $E_n - E_{n'} = \omega_0$ for some $n$, $n'$.
A transition between the corresponding states $| \psi_n \rangle$ and $| \psi_{n'} \rangle$ is in resonance with the laser driving.
The denominator of $g^{(2)}(0)$ remains unchanged because the resonant transitions already exist for $\Omega = 0$.
A new feature for finite $\Omega$ is the sideband below the lower system energy $E_{n'}$.
Denoting by $M$ and $M'$ the principal quantum numbers of the states $| \psi_n \rangle$ and $| \psi_{n'} \rangle$, respectively, we realize that this sideband allows for a transition sequence $M \to M' \to M'$.
The contribution from this resonant transition sequence leads to the enhancement of the numerator of $g^{(2)}(0)$ and hence to the sharp horizontal lines in Fig~\ref{fig:glauber1}(f).

Further inspection of Fig.~\ref{fig:energy} shows that there are additional crossings of system energies with sidebands at values of the emitter-cavity coupling $g$, where no horizontal lines in Fig.~\ref{fig:glauber1}(f) appear.
These additional crossings do not contribute to $g^{(2)}(0)$ due to a selection rule.
In particular, the resonant transition sequence $M \to M' \to M'$ (which belongs to one of these crossings) already contains the sideband transition $M' \to M'$ that enters with a scaling proportional to $\Omega^2$.
Hence, the remaining transition $M \to M'$ involves two system energies with the usual dipole selection rule $\Delta M = 1$.
Then only transitions marked with blue lines in Fig.~\ref{fig:energy} remain.

While we are now in the position to predict the thin lines in Fig.~\ref{fig:glauber1}(f) for specific values of the emitter-cavity-coupling strength, the temperature dependence along these lines remains open.
Inspection of Fig.~\ref{fig:energy} shows that the resonant transition sequences involve different but highly excited states.
The relevant population is that of the uppermost state in the sequence.
For very low temperatures, this state is not significantly populated.
Changes in $g^{(2)}(0)$ will therefore be visible for increasing temperature only.
The particular starting temperature depends on the specific transition sequence and will be larger if higher excited states are involved.
This prediction is confirmed when we compare Fig.~\ref{fig:energy} with the appearance of the horizontal lines in Fig.~\ref{fig:glauber1}(f).
Further increasing the environment temperature above the set-in threshold leads to enlarged contributions of many (nonresonant) transition sequences involving higher excited states.
Then the relative weight of the resonant transition sequences and their impact on $g^{(2)}(0)$ decreases.

\subsection{\label{ssec:glauber-t}Time-dependent Glauber function}
While the statistics of the emitted photons follows from the Glauber function $g^{(2)}(0)$ at zero time delay, the time-dependent function $g^{(2)}(t)$ determines their time-coincidence statistics.
In particular, photon bunching, i.e., the enhanced probability of observing two photons at equal times, is indicated by a nonpositive initial slope of $g^{(2)}(t)$ for $t = 0$.
Conversely, a positive initial slope of $g^{(2)}(t)$ proves photon antibunching, which is possible only for nonclassical light.

\begin{figure}
  \includegraphics[width=0.49\linewidth]{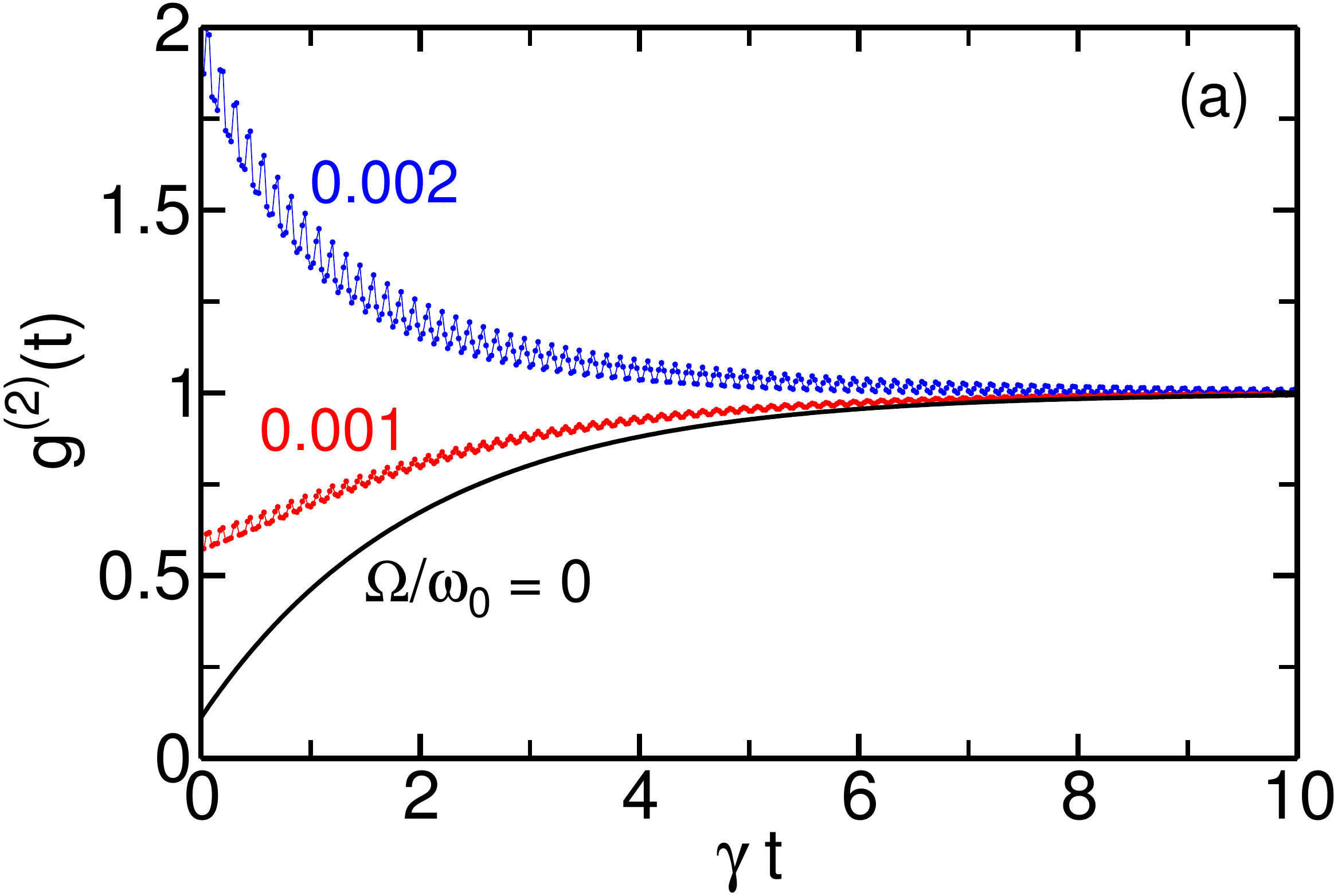}
  \includegraphics[width=0.49\linewidth]{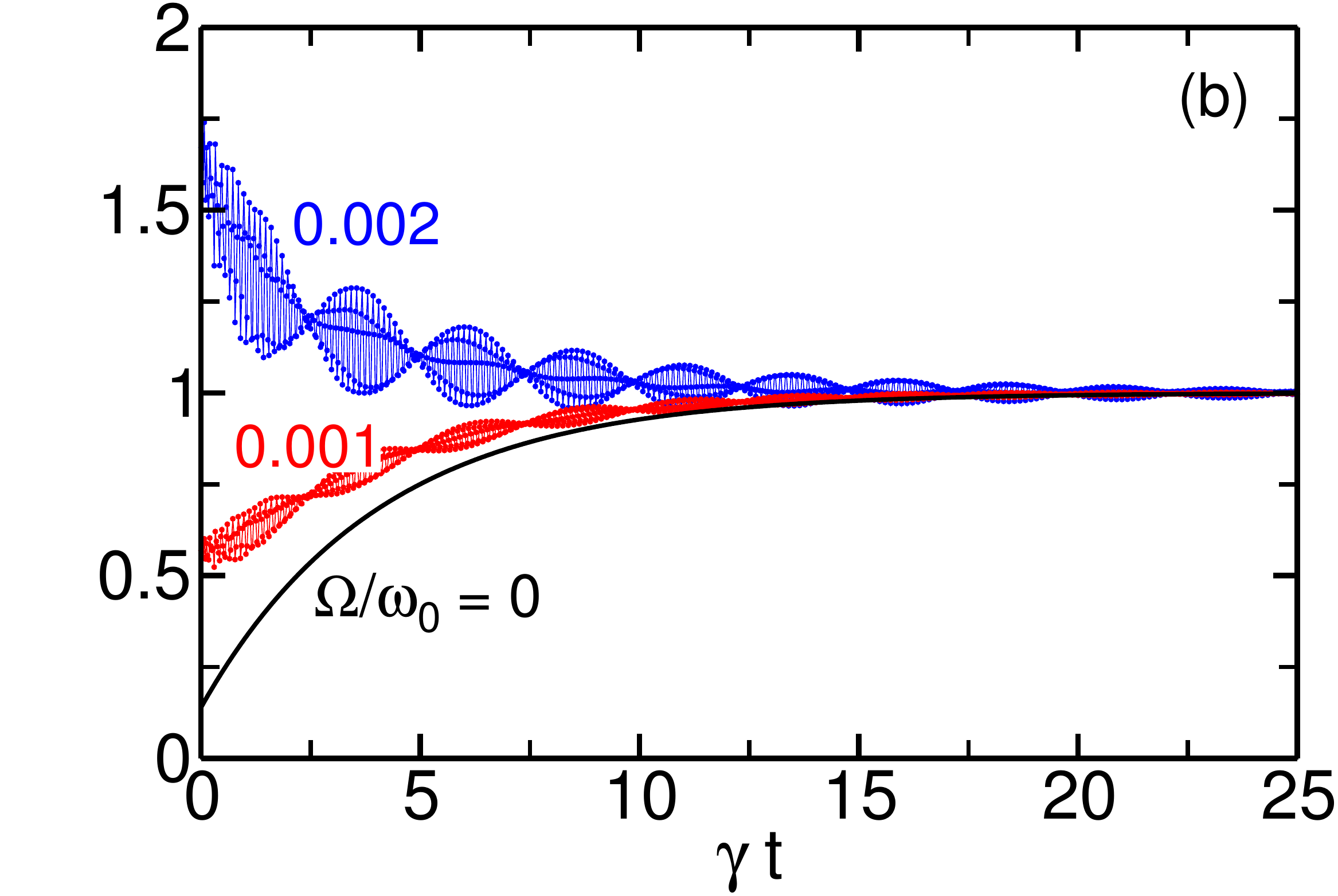}
  \caption{\label{fig:glaubert}Time-dependent Glauber function $g^{(2)}(t)$ for one emitter with $g = 0.5 \omega_0$ and $T = 0.07 \omega_0$.
    The case (a) $g' = 0$ and $\Omega' = 0$ is compared to (b) $g' = g$ and $\Omega' = \Omega$.}
\end{figure}

Figure~\ref{fig:glaubert} displays the time-dependent function $g^{(2)}(t)$ for the choice $g = 0.5 \omega_0$ and $T = 0.07 \omega_0$ for one emitter.
We see that an increasing laser intensity $\Omega$ induces oscillations in $g^{(2)}(t)$.
These oscillations may survive the long-time limit $t \to \infty$, where $\lim_{t \to \infty} g^{(2)}(t) = 1$ is fulfilled only for the average value of $g^{(2)}(t)$.
Increasing the laser intensity $\Omega$ changes the initial value $g^{(2)}(0)$ according to the results from the preceding section.
Interestingly, due to the oscillations, the overall behavior of $g^{(2)}(t)$ is not a monotonic function of $t$.
This indicates that photon antibunching, i.e., the positive initial slopes for $\Omega = 0.002 \omega_0$ in Fig.~\ref{fig:glaubert}, can occur even if the photon statistics is super-Poissonian [$g^{(2)}(0) > 1$].

\section{\label{sec:entangle}Generation of entanglement}
For a further classification of nonclassicality entanglement can be used.
Here we are interested in the generation of entanglement within the stationary state of two emitters inside the cavity.
This state follows as the partial trace of the averaged stationary density matrix $\overline{\rho}^\infty$ over the cavity degrees of freedom.
We get a bipartite emitter state of the form
\begin{equation}\label{state}
  \overline{\rho}_x^\infty = \begin{pmatrix} \rho_{11} & 0 & 0 & \rho_{14} \\ 0 & \rho_{22} & \rho_{23} & 0 \\ 0 & \rho_{32} & \rho_{33} & 0 \\ \rho_{41} & 0 & 0 & \rho_{44} \end{pmatrix} \,,
\end{equation}
which has nonzero elements only on its main diagonal and antidiagonal.
This structure of the density matrix follows from symmetry considerations within the $\text{SU}(2) \times \text{SU}(2) \times \text{U}(1)$ subalgebra of the full $\text{SU}(4)$ algebra of two quantum bits~\cite{Rau09} and was also observed for the Dicke model with $\Omega = 0$ (i.e., without time-dependent laser drive)~\cite{AE13}.
For all parameter combinations studied here, we numerically checked that we get the same structure of the reduced emitter state, even for finite laser intensity $\Omega > 0$.

For the bipartite state~\eqref{state}, the concurrence~\cite{HW97} is
\begin{equation}
  C(\overline{\rho}_x^\infty) = 2 \max \big\{ 0, |\rho_{14}| - \sqrt{\rho_{22} \rho_{33}}, |\rho_{23}| - \sqrt{\rho_{11} \rho_{44}} \big\} \,,
\end{equation}
with $0 \leq C \leq 1$.
The concurrence is an entanglement measure for two quantum bits.
A more basic measure, the entanglement of formation (EOF) that quantifies the resources needed to create a given entangled state, can be constructed from $C$~\cite{Woo98}.
Defining $\eta = (1 + \sqrt{1 - C^2}) / 2$, the EOF of the bipartite emitter state $\overline{\rho}_x^\infty$ becomes
\begin{equation}
  C_\text{EOF}(\overline{\rho}_x^\infty) = -\eta \log_2 \eta - (1 - \eta) \log_2 (1 - \eta) \,.
\end{equation}
The EOF is zero for separable states, finite for entangled ones, and approaches one for maximally entangled states.

\begin{figure}
  \includegraphics[scale=0.3]{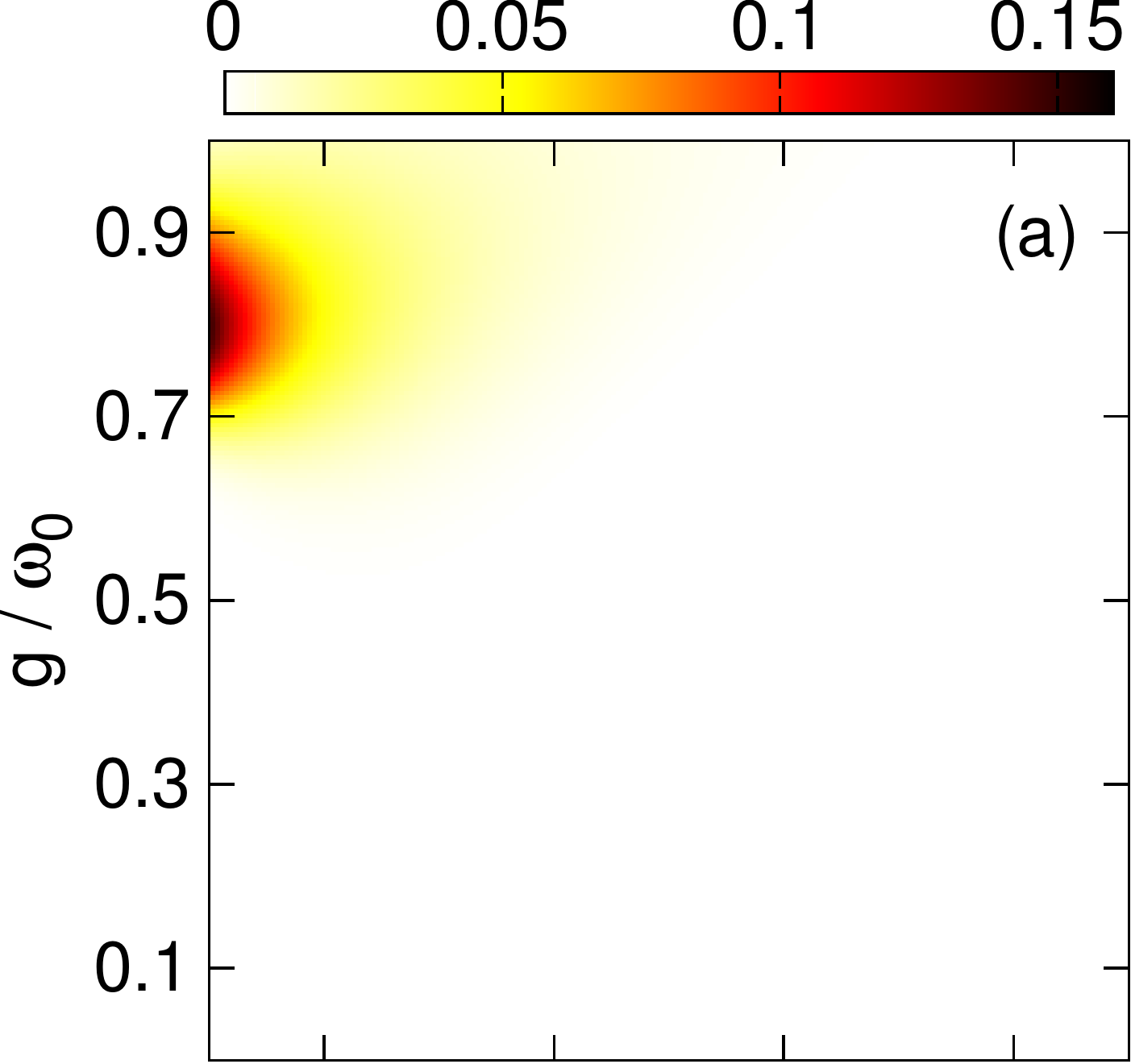}
  \includegraphics[scale=0.3]{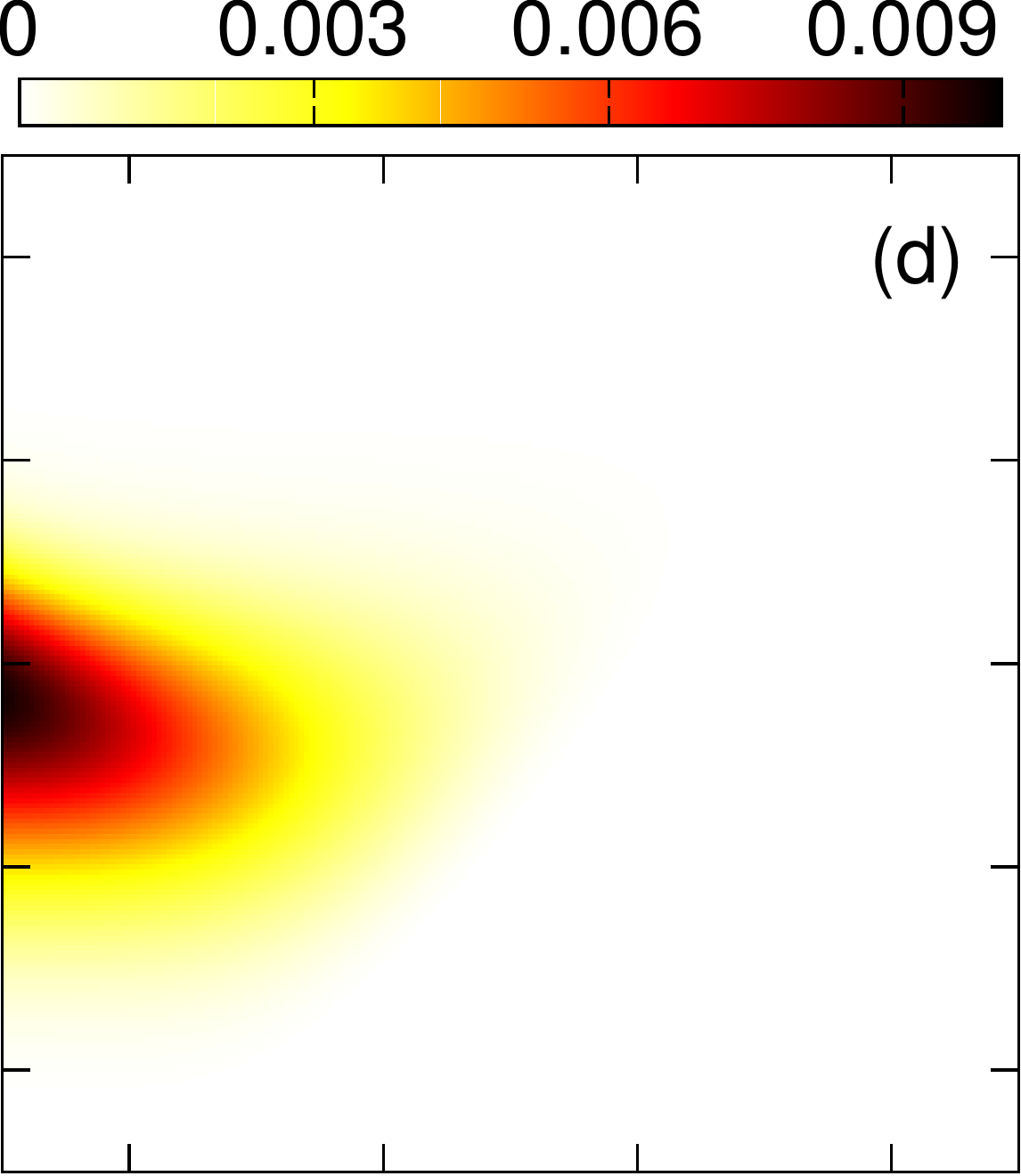}\\
  \includegraphics[scale=0.3]{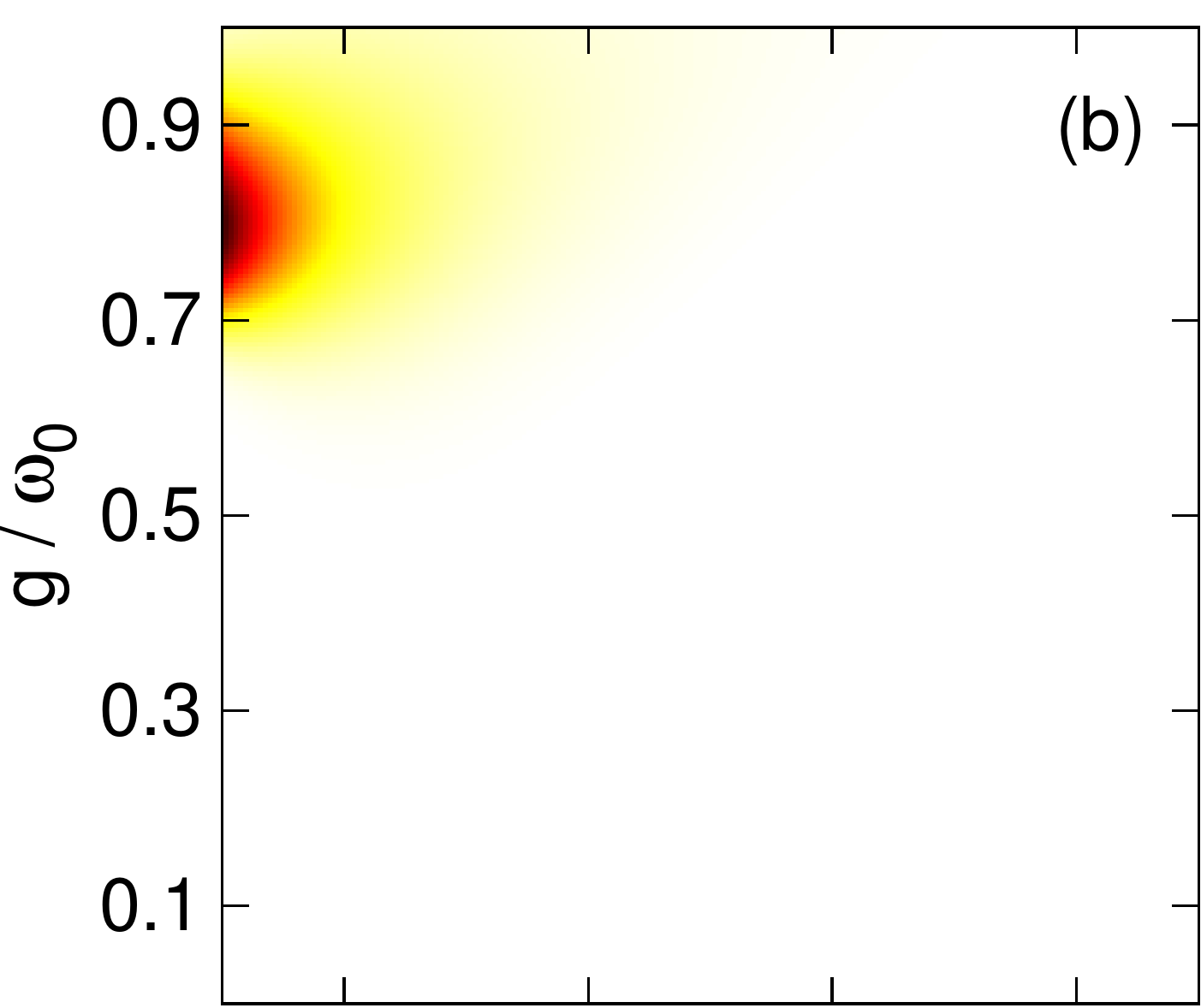}
  \includegraphics[scale=0.3]{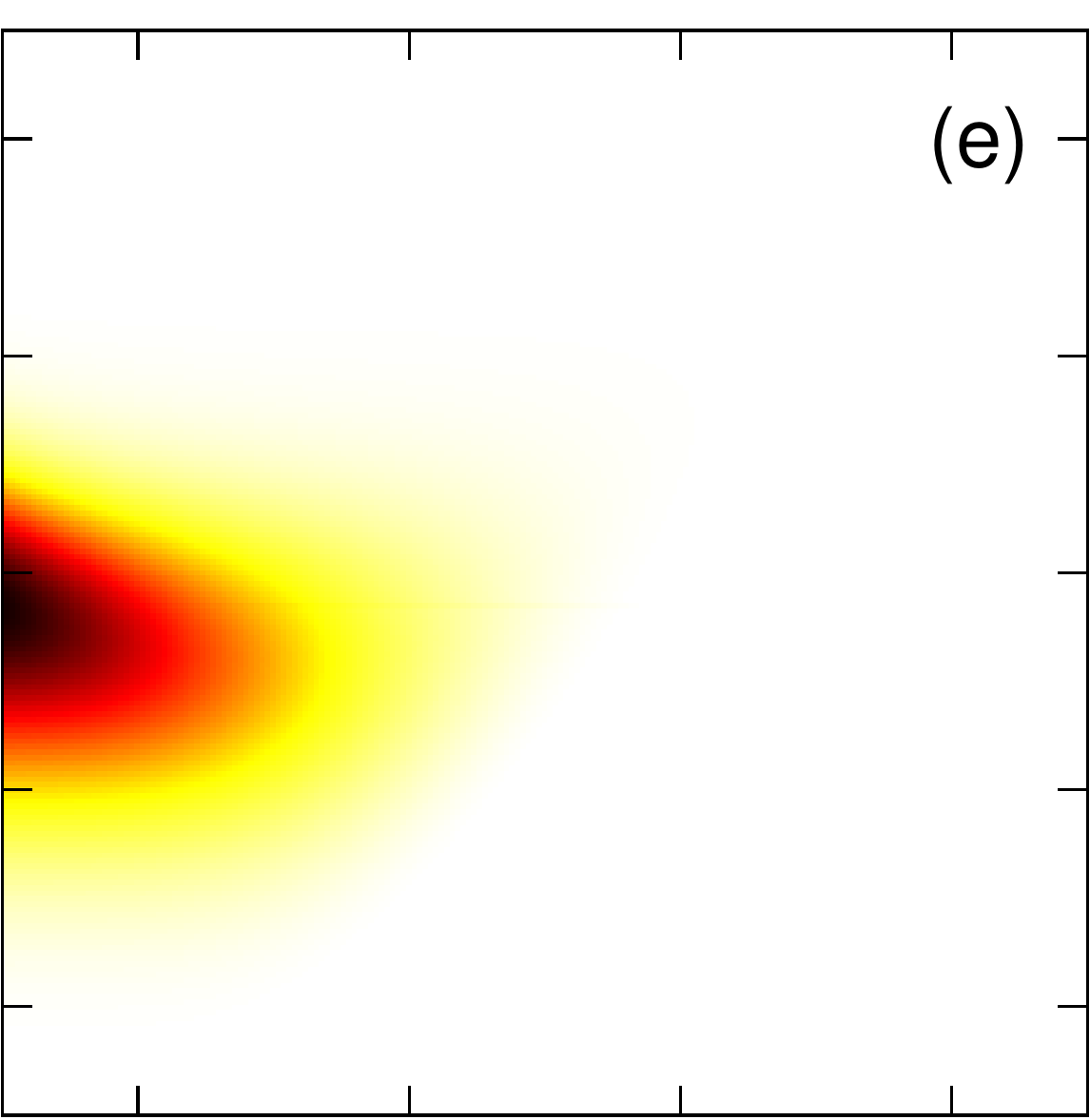}\\
  \includegraphics[scale=0.3]{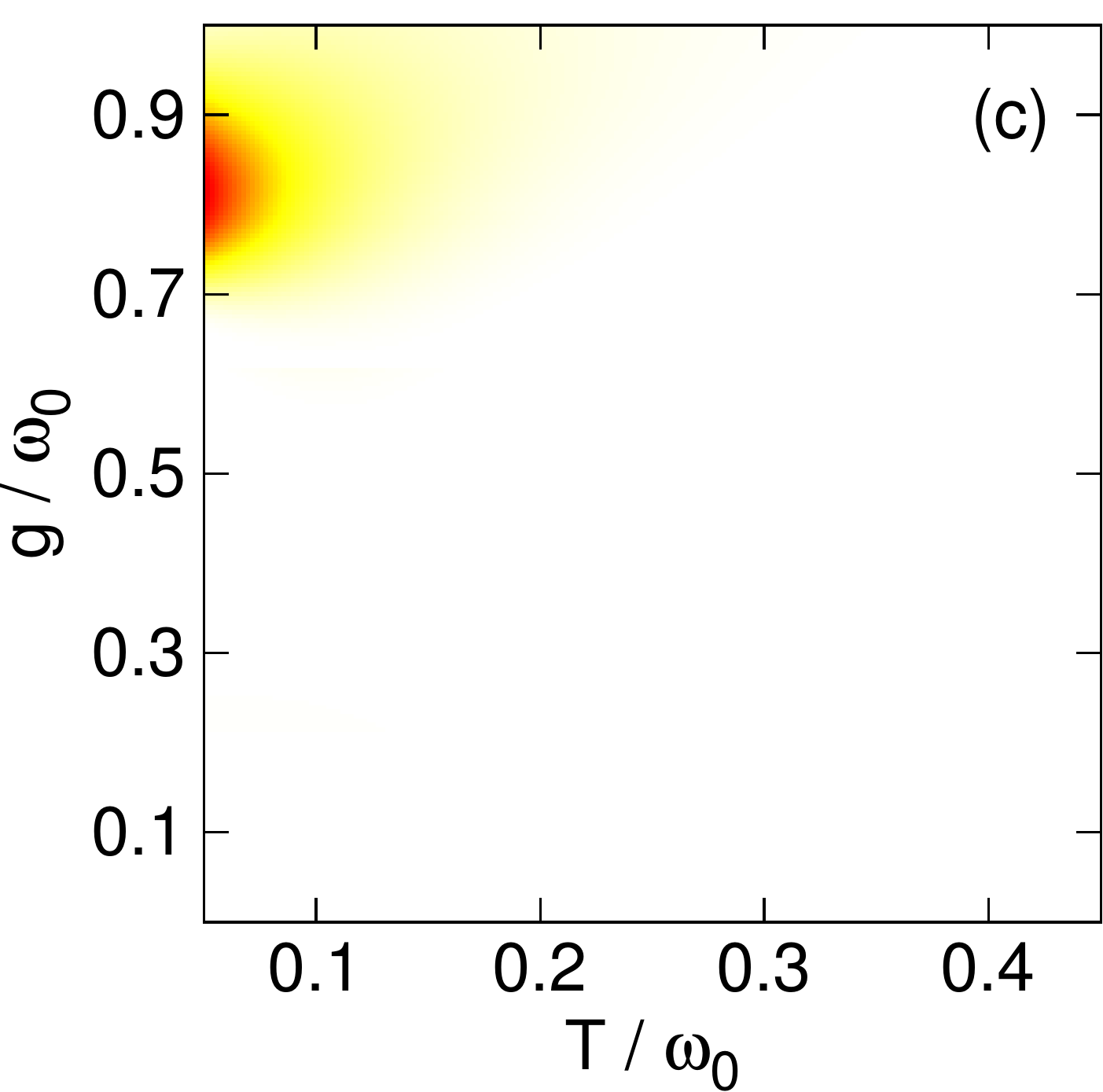}
  \includegraphics[scale=0.3]{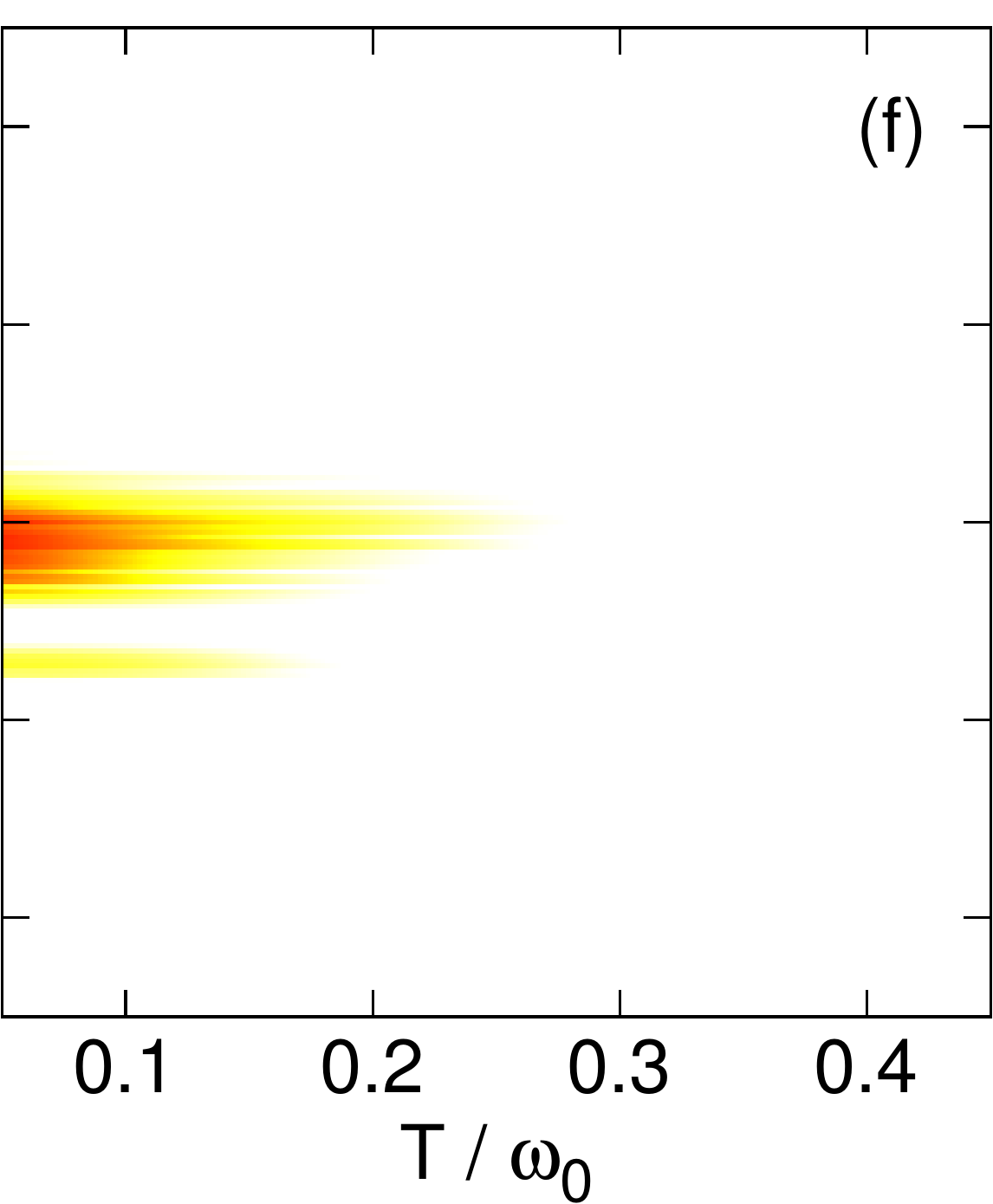}
  \caption{\label{fig:entangle}The EOF for two emitters as a function of the environment temperature $T$ and the emitter-cavity-coupling strength $g$.
    The left (right) column depicts the results for $g' = \Omega' = 0$ ($g' = g$ and $\Omega' = \Omega$).
    The laser intensity is (a) and (d) $\Omega = 0$, (b) and (e) $\Omega = 10^{-3} \, \omega_0$, and (c) and (f) $\Omega = 10^{-1} \, \omega_0$.}
\end{figure}

Figure~\ref{fig:entangle} depicts the EOF for two emitters as a function of the environment temperature $T$ and the emitter-cavity-coupling strength $g$ for different laser intensities $\Omega$.
Comparison with the Glauber function for two emitters (Fig.~\ref{fig:glauber2} in Appendix~\ref{app:emitters}) shows that entanglement between the two emitters is generated in parameter regions where nonclassical light is emitted.
Nevertheless, the opposite is not true, i.e., not in all regions with sub-Poissonian light statistics will the emitters be significantly entangled.
For example, at low environment temperatures and emitter-cavity coupling $0.3 \lesssim g / \omega_0 \lesssim 0.6$, entanglement is generated only for the full Hamiltonian including corotating and counterrotating terms, while the statistics of emitted photons is sub-Poissonian also in the rotating-wave approximation.
Hence, the generation of entanglement is more specific than the emission of nonclassical light.

Inspection of the magnitudes in Fig.~\ref{fig:entangle} reveals that the amount of generated entanglement is higher for $g' = \Omega' = 0$ than for $g' = g$ and $\Omega' = \Omega$.
In addition, the EOF is less sensitive to the laser intensity $\Omega$ than the Glauber $g^{(2)}(0)$ function.
The reason for this behavior is that $C_\text{EOF}$ is a property of the stationary state and does not involve details of the output operator $\dot{X}_-$.
While the stationary state merely depends on the quasienergies whose shift (dynamic Stark effect) is second order in $\Omega$, the output operator involves new transitions that already appear in first order of $\Omega$.

\section{\label{sec:conc}Conclusions}
Analyzing the properties of emitted light in a laser-driven quantum system, we have discussed the dynamic Stark effect, the photon statistics, and the generation of bipartite entanglement.
Thereby, the resonant case studied here requires the use of the full input-output formalism and the full master equation in the Floquet basis.
Essential is the careful distinction of the transitions between different Floquet states and their Fourier modes.

Without laser excitation, the stationary state of the emitter-cavity system weakly coupled to an environment at temperature $T$ is the thermal state.
This state is modified if the laser intensity is finite.
For $\Omega \ll g, \omega_0$, the populations of the Floquet states are thermally distributed.
Increasing $\Omega$, the dynamic Stark shift of the quasienergies leads to modified thermal populations.
In addition, the energy impact from the laser causes further population transfers.
The stationary state is thus no longer thermal.
Nevertheless, according to our results for the entanglement between two emitters (which depends only on the stationary populations), significant changes of the stationary state appear for $\Omega \gtrsim 10^{-1} \omega_0$.
The stationary emitter state and thus the generation of entanglement is quite robust against the laser excitation.

In contrast to the stationary state, the statistics of emitted photons is strongly influenced by the external laser.
The changes of the Glauber function for small laser intensity have to be explained by the specific form of the output operator that connects different Floquet states and includes transitions between their Fourier modes.
The contribution from these resonant transition sequences, in combination with a possible population enhancement due to an energetically lower starting point of these sequences, is responsible for the tremendous effect on the photon statistics.
Hence, changing the driving strength can significantly modify the statistics of the emitted photons.
In particular, thermal light emission at weak emitter-cavity coupling is replaced by the emission of photons with highly super-Poissonian statistics.

We showed that the shift of the quasienergies $\epsilon_n$ first appears in second order of the laser-driving strength $\Omega$, which is the dynamic Stark effect.
This result is in accordance with an analytical calculation of the quasienergies for a single emitter with interaction terms in the rotating-wave approximation.
Calculating the emission spectra for an emitter beyond the rotating-wave approximation, we verified the validity of the general proportionality $\epsilon_n \propto (1 - \Omega^2 / g^2)^{3/4}$, where $g$ is the emitter-cavity-coupling strength.
Exceptions to this rule arise at quasienergy anticrossings.
Laser-intensity-dependent spectroscopic measurements of the vacuum Rabi splitting might provide both the above proportionality of the quasienergies and the positions of avoided crossings.

The particular combination of Floquet theory, input-output theory, and the Floquet master equation was used here for the evaluation of the emission properties of a system beyond the rotating-wave approximation.
While the use of this method is restricted to low-dimensional systems because of the additional summations over Fourier modes, it includes the regimes of strong and ultrastrong light-matter interaction even if the Hamiltonian has a periodic time dependence.
This makes the study of nonclassical and entangled light emission in realistic quantum optical systems possible and hence a prediction of corresponding experimental outcomes.
For this task, we focused on the Glauber function as a particular combination of second- and first-order cumulants.
Future work should address the full counting statistics that includes all cumulants of the emitted photons.
This requires an extension of our approach with the concept of measurements at multiple times to evaluate the cumulant generating function.

\begin{acknowledgments}

This work was supported by Deutsche Forschungsgemeinschaft through SFB 652 (project B5) and SFB/TR 24 (project B10).

\end{acknowledgments}

\appendix

\section{\label{app:TC}The driven Tavis-Cummings model}
In this appendix we outline the calculation of the quasienergies and steady states of the driven Tavis-Cummings system~\cite{TC68, AGC92}.
Hence, we consider the emitter-cavity system in the rotating-wave approximation ($g' = 0$ and $\Omega' = 0$) for a single emitter ($N = 1$) at resonance ($\omega_c = \omega_x = \omega_d = \omega_0$) described by the Hamiltonian
\begin{eqnarray}
  H_\text{TC}(t) &=& \omega_0 a^\dagger a + \omega_0 \sigma_+ \sigma_- + g (a^\dagger \sigma_- + a \sigma_+) \nonumber\\
    && + \frac{\Omega}{2} (a \rme^{\rmi \omega_0 t} + a^\dagger \rme^{-\rmi \omega_0 t}) \,.
\end{eqnarray}

To solve the time-dependent Schr\"{o}dinger equation
\begin{equation}
  \rmi \frac{\partial}{\partial t} | \psi(t) \rangle = H_\text{TC}(t) | \psi(t) \rangle \,,
\end{equation}
we introduce the operator
\begin{equation}
  V(t) = \rme^{-\rmi \omega_0 (a^\dagger a + \sigma_+ \sigma_-) t} \,,
\end{equation}
which transforms the system to a frame rotating with the laser frequency $\omega_0$.
Introduction of rotating states $| \psi'(t) \rangle = V^\dagger(t) | \psi(t) \rangle$, with
\begin{equation}
  \rmi \frac{\partial}{\partial t} | \psi'(t) \rangle = H_\text{TC}' | \psi'(t) \rangle \,,
\end{equation}
yields the transformed Hamiltonian
\begin{eqnarray}
  H_\text{TC}' &=& V^\dagger(t) H_\text{TC}(t) V(t) - \omega_0 a^\dagger a - \omega_0 \sigma_+ \sigma_- \nonumber\\
    &=& g (a \sigma_+ + a^\dagger \sigma_-) + \frac{\Omega}{2} (a + a^\dagger) \,.
\end{eqnarray}
Because $H_\text{TC}'$ is time-independent, its eigenstates follow from diagonalization.
We obtain
\begin{equation}
  | \psi_0' \rangle = | \eta, 0; 0 \rangle | M \rangle \,,
\end{equation}
corresponding to the eigenenergy $E_0 = 0$, and
\begin{equation}
  | \psi_{n,\pm}' \rangle = \frac{1}{\sqrt{2}} \Big( | \eta, \alpha_{n,\pm}; n-1 \rangle | P \rangle \pm | \eta, \alpha_{n,\pm}; n \rangle | M \rangle \Big) \,,
\end{equation}
corresponding to $E_{n,\pm} = \pm \sqrt{n} g \{1 - (\Omega / g)^2\}^{3/4}$ for $n \geq 1$.
Here $| \eta, \alpha; n \rangle = D(\alpha) Q(\eta) | n \rangle$ are squeezed, displaced oscillator states with the displacement operator
\begin{equation}
  D(\alpha) = \exp \{ \alpha a^\dagger - \alpha^* a \}
\end{equation}
and the squeezing operator
\begin{equation}
  Q(\eta) = \exp \Big\{ \frac{1}{2} \big[ \eta (a^\dagger)^2 - \eta^* a^2 \big] \Big\} \,.
\end{equation}
The squeezing and displacement parameters expressed in terms of the ratio $\kappa = \Omega / g$ of laser intensity to emitter-cavity-coupling strength are
\begin{equation}
  \rme^{2 \eta} = \frac{1}{\sqrt{1 - \kappa^2}} \,, \qquad
  \alpha_{n,\pm} = \mp \sqrt{n} \kappa \,.
\end{equation}
The states
\begin{eqnarray}
  | M \rangle &=& \frac{1}{\sqrt{2}} \Big( \sqrt{1 + \sqrt{1 - \kappa^2}} | - \rangle - \sqrt{1 - \sqrt{1 - \kappa^2}} | + \rangle \Big) \,, \nonumber\\
  | P \rangle &=& \frac{1}{\sqrt{2}} \Big( \sqrt{1 + \sqrt{1 - \kappa^2}} | + \rangle - \sqrt{1 - \sqrt{1 - \kappa^2}} | - \rangle \Big) \nonumber\\
\end{eqnarray}
are defined in terms of the ground ($\sigma_+ \sigma_- | - \rangle = 0$) and excited ($\sigma_+ \sigma_- | + \rangle = | + \rangle$) emitter states.
We note that $| M \rangle$ and $| P \rangle$ are normalized but not orthogonal.

For weak driving $\kappa \ll 1$, the truncation of the eigenstates and eigenenergies of $H_\text{TC}'$ gives to lowest order in $\kappa$,
\begin{eqnarray}
  | \psi_0' \rangle &=& | 0, - \rangle - \frac{\kappa}{2} | 0, + \rangle \,, \nonumber\\
  | \psi_{n,\pm}' \rangle &=& \frac{1}{\sqrt{2}} \Big( | n-1, + \rangle \pm | n, - \rangle \Big) \nonumber\\
    && \pm \frac{\kappa}{\sqrt{2}} \Big[ \sqrt{n (n - 1)} | n-2, + \rangle \pm \Big( n - \frac{1}{2} \Big) | n-1, - \rangle \nonumber\\
    &&\quad - \Big( n + \frac{1}{2} \Big) | n, + \rangle \mp \sqrt{n (n + 1)} | n+1, - \rangle \Big] \,, \nonumber\\
\end{eqnarray}
corresponding to $E_0 = 0$ and $E_{n, \pm} = \pm \sqrt{n} g (1 - 3 \kappa^2 / 4)$, respectively.

\section{\label{app:inout}The input-output formalism}
We follow standard input-output theory~\cite{CG84, GC85, Gra89} to obtain a relation between input, intracavity, and output fields.
Assuming a coherent driving of the cavity with classical fields, the correlations of the output field are expressed as functions of intra-cavity correlations only~\cite{FS13}.
Using the Floquet states as the computational basis for the intra-cavity system dynamics, the output operator in Eq.~\eqref{X} is obtained.

The Hamiltonian for the interaction of the cavity with the environment is given in Eq.~\eqref{HI}.
This operator together with the free Hamiltonian $\sum_\alpha \omega_\alpha b_\alpha^\dagger b_\alpha$ leads to the equation of motion
\begin{equation}
  \dot{b}_\alpha = -\rmi \omega_\alpha b_\alpha + \lambda_\alpha X
\end{equation}
for the environmental photon operator.
The formal solution of this equation for $t_0 < t < t_1$ is
\begin{eqnarray}\label{b-alpha}
  b_\alpha(t) &=& \rme^{-\rmi \omega_\alpha (t - t_0)} b_\alpha(t_0) + \lambda_\alpha \int_{t_0}^t \rme^{-\rmi \omega_\alpha (t - t')} X(t') \, \rmd t' \nonumber\\
    &=& \rme^{-\rmi \omega_\alpha (t - t_1)} b_\alpha(t_1) - \lambda_\alpha \int_t^{t_1} \rme^{-\rmi \omega_\alpha (t - t')} X(t') \, \rmd t' \;. \nonumber\\
\end{eqnarray}
Defining the input (output) field operators
\begin{equation}\label{inoutfields}
  b_\text{in (out)}(t) = \sum_\alpha \lambda_\alpha \rme^{-\rmi \omega_\alpha (t - t_{0\,(1)})} b_\alpha(t_{0\,(1)})
\end{equation}
and inserting Eq.~\eqref{b-alpha}, the relation
\begin{equation}\label{inout_disc}
  b_\text{out}(t) = b_\text{in}(t) + \int_{t_0}^{t_1} \sum_\alpha \lambda_\alpha^2 \rme^{-\rmi \omega_\alpha (t - t')} X(t') \, \rmd t'
\end{equation}
is obtained.
Performing the thermodynamic limit, where summations over $\alpha$ can be replaced by frequency integrations~\cite{HR85, PAF13}, and assuming an Ohmic environment spectral function $\gamma(\omega) = \sum_\alpha \lambda_\alpha^2 \delta(\omega - \omega_\alpha) = \gamma \omega / \omega_0$ that is consistent with the Markov approximation, the input-output relation becomes~\cite{RLSH12}
\begin{equation}\label{inout_cont}
  b_\text{out}(t) = b_\text{in}(t) + 2 \pi \rmi \frac{\gamma}{\omega_0} \dot{X}_-(t) \,,
\end{equation}
where $\dot{X}_-$ is the positive-frequency part of the operator $\dot{X}$.

The definition~\eqref{inoutfields} of input and output fields explicitly contains the coupling constants $\lambda_\alpha$, which account for the energy-dependent coupling between the cavity and the environmental field modes.
With an Ohmic spectral function as in Ref.~\cite{RLSH12}, chosen for consistency with the Markovian master equation used to propagate the system density matrix, the weight of environmental field modes in Eq.~\eqref{inoutfields} is proportional to $\sqrt{\omega}$.
Equal weights for each environmental field mode [i.e., $\gamma(\omega) \equiv \gamma$] are recovered under the additional assumption of frequency-independent coupling constants~\cite{GC85}.
Note that in the relevant input-output relation~\eqref{inout_cont} the coupling constant $\gamma$ appears only in front of the operator $\dot X_-$ that enters the expression~\eqref{S} for the emission spectrum in the main text.
For zero system-environment coupling ($\gamma = 0$) we recover the identity $b_\text{out}(t) = b_\text{in}(t)$.
For finite coupling, the difference between the output and input fields is proportional to the coupling strength $\gamma$.

We consider a coherent driving of the cavity with a classical laser field added to the quantum vacuum in the input and output channels.
Then the modified system dynamics is described by the additional Hamiltonian~\eqref{HL}.
Because all normal-ordered cross correlations between the input and intracavity fields vanish, the correlations in the output channel can be expressed as functions of intracavity correlations only~\cite{FS13}.
For example, the time-resolved number of photons collected in the output channel is
\begin{equation}\label{Nout}
  N_\text{out}(t) = \langle b_\text{out}^\dagger(t) b_\text{out}(t) \rangle = 4 \pi^2 \frac{\gamma^2}{\omega_0^2} \langle \dot{X}_+(t) \dot{X}_-(t) \rangle \,,
\end{equation}
where $\dot{X}_+ = \dot{X}_-^\dagger$.

For the evaluation of the expectation value in Eq.~\eqref{Nout} a computational basis has to be chosen.
Because of the periodic time dependence of the system Hamiltonian~\eqref{H}, the operator $\dot{X}_-(t)$ can be expanded in the Floquet states~\eqref{floquet}.
The projection onto the positive-frequency components, inherent in $\dot{X}_-$, requires the expansion of the periodic parts of the Floquet states in Fourier modes [see Eq.~\eqref{fourier}].
The resulting spectral decomposition of the output operator is given in Eq.~\eqref{X}.

\section{\label{app:Floquet}The Floquet master equation}
We consider the dynamics of the system density matrix $\rho(t)$ for a time-dependent system Hamiltonian $H(t)$ in the limit of weak system-environment coupling.
The adiabatic approximation for slowly varying $H(t)$~\cite{CFP01} fails in the quantum optical domain, where the system Hamiltonian oscillates at optical frequencies~\cite{Car99}.
Nevertheless, the periodicity of $H(t)$ may then be used in a description where the Floquet states~\cite{Flo83} are the computational basis.
The resulting Floquet master equation~\cite{BBGSSW91, GH98} has constant coefficients.
We here recapitulate its derivation.

The total Hamiltonian is the sum of the time-dependent system part $H(t)$, the contribution from the reservoir $H_R$, and the interaction $H_I = X R$ in Eq.~\eqref{HI}, where $R = -\rmi \sum_\alpha \lambda_\alpha (b_\alpha - b_\alpha^\dagger)$ for abbreviation.
The dynamics of the density matrix $\rho_\text{tot}(t)$ of the total system is described by the von Neumann equation,
\begin{equation}
  \frac{\rmd}{\rmd t} \hat{\rho}_\text{tot}(t) = -\rmi [\hat{H}_I(t), \hat{\rho}_\text{tot}(t)]
\end{equation}
(operators in the interaction picture are marked with a caret).
The density operator in the interaction picture is
\begin{equation}
  \hat{\rho}_\text{tot}(t) = U_\text{tot}^\dagger(t, 0) \rho_\text{tot}(t) U_\text{tot}(t, 0) \,,
\end{equation}
where the time-evolution operator of the uncoupled system and environment is
\begin{equation}
  U_\text{tot}(t, s) = T_\leftarrow \exp \Big( -\rmi \int_s^t H(\tau) \, \rmd\tau \Big) \, \rme^{-\rmi H_R (t - s)}
\end{equation}
($T_\leftarrow$ denotes chronological time ordering).

For weak system-environment coupling, the Born and Markov approximations are performed~\cite{BP02}.
In particular, one sets $\rho_\text{tot}(t) = \rho(t) \rho_R$, assuming initial factorization $\rho_\text{tot}(0) = \rho(0) \rho_R$ and neglecting the backaction of the system onto the reservoir.
The constant reservoir state $\rho_R \propto \rme^{-H_R / T}$ is assumed to be a thermal state at temperature $T$ (which is measured in units of energies).
In addition, the density matrix $\rho(\tau)$ in integrals over retarded times $\tau \in [0, t]$ is replaced by $\rho(t)$ at the local time $t$.
Then the dissipative dynamics of the system density operator $\rho(t)$ is described by a Markovian master equation~\cite{Car99,BP02}
\begin{equation}\label{BMeom}
  \frac{\rmd}{\rmd t} \hat{\rho}(t) = \int_0^\infty \big[ \hat{X}(t - \tau) \hat{\rho}(t), \hat{X}(t) \big] C(\tau) \, \rmd\tau + \text{H.c.}
\end{equation}
(\text{H.c.}~means the Hermitian conjugation).
In Eq.~\eqref{BMeom},
\begin{equation}
  C(\tau) = \text{Tr}_R \big\{ \rme^{\rmi H_R \tau} R \, \rme^{-\rmi H_R \tau} R \rho_R \big\} = C(-\tau)^*
\end{equation}
is the reservoir correlation function, with $\text{Tr}_R\{ \cdot \}$ denoting the partial trace over the reservoir degrees of freedom.

Equation~\eqref{BMeom} is the standard Born-Markov equation of motion.
Solution of this master equation requires the choice of a computational basis.
As explained in Sec.~\ref{ssec:model}, the natural basis states for the description of a driven system with a time-periodic Hamiltonian are the Floquet states~\eqref{floquet}.
We find
\begin{eqnarray}
  X_{m,n}(t) &=& \langle \psi_m(0) | \hat{X}(t) | \psi_n(0) \rangle \nonumber\\
    &=& \rme^{-\rmi (\epsilon_n - \epsilon_m) t} \langle \phi_m(t) | X | \phi_n(t) \rangle \;.
\end{eqnarray}
Expanding the periodic part $| \phi_n(t) \rangle$ of the Floquet states in Fourier modes, we obtain
\begin{equation}
  X_{m,n}(t) = \sum_{\mu,\nu} \rme^{-\rmi (\epsilon_n - \epsilon_m) t} \rme^{-\rmi \nu \omega_d t} \langle \widetilde{\phi}_m(\mu - \nu) | X | \widetilde{\phi}_n(\mu) \rangle \;.
\end{equation}
Introducing the operator
\begin{multline}\label{transOp}
  X_{\omega, \nu} = \sum_{m, n} \sum_\mu \langle \widetilde{\phi}_m(\mu - \nu) | X | \widetilde{\phi}_n(\mu) \rangle \\
  \times | \psi_m(0) \rangle \langle \psi_n(0) | \delta_{\epsilon_n - \epsilon_m, \omega} \,,
\end{multline}
which is a projection of $X$ onto transitions between Floquet states $| \psi_m(t) \rangle$, $| \psi_n(t) \rangle$ with quasienergy difference $\omega = \epsilon_n - \epsilon_m$, yields
\begin{equation}\label{hS_t}
  \hat{X}(t) = \sum_{\omega, \nu} \rme^{-\rmi (\omega + \nu \omega_d) t} X_{\omega, \nu} \,.
\end{equation}
In addition, we introduce the even and odd Fourier transforms of the reservoir correlation function
\begin{eqnarray}
  \chi(\omega) &=& \int_{-\infty}^\infty C(\tau) \rme^{\rmi \omega \tau} \rmd\tau = \chi(\omega)^* \,, \\
  \xi(\omega) &=& \frac{1}{\rmi} \int_{-\infty}^\infty C(\tau) \mathop{\mathrm{sgn}}(\tau) \rme^{\rmi \omega \tau} \rmd\tau = \xi(\omega)^* \,,
\end{eqnarray}
which are given by
\begin{equation}
  \chi(\omega) = \begin{cases} \gamma(\omega) [n(\omega, T) + 1] \quad & \text{ if } \omega > 0 \\ \gamma(-\omega) n(-\omega, T) & \text{ if } \omega < 0 \end{cases}
\end{equation}
and
\begin{equation}
  \xi(\omega) = \begin{cases} \Re \Gamma(\omega + \rmi 0^+) [n(\omega, T) + 1]   \quad & \text{ if } \omega > 0 \\ -\Re \Gamma(-\omega + \rmi 0^+) n(-\omega, T) & \text{ if } \omega < 0 \end{cases}
\end{equation}
for a thermal reservoir with spectral function $\gamma(\omega)$ and its analytical continuation $\Gamma(\omega)$ into the upper half plane, where $\gamma(\omega) = \mp\Gamma(\pm\omega + \rmi 0^+)$.
The function $n(\omega, T)$ is the Bose-Einstein distribution
\begin{equation}
  n(\omega, T) = \frac{1}{\rme^{\beta \omega} - 1} \,.
\end{equation}
With these definitions and Eq.~\eqref{hS_t} we find
\begin{eqnarray}\label{BMFeom}
  \frac{\rmd}{\rmd t} \hat{\rho}(t) &=& \frac{1}{2} \sum_{\omega, \omega'} \sum_{\nu, \nu'} \big\{ \chi(\omega' + \nu' \omega_d) + \rmi \xi(\omega' + \nu' \omega_d) \big\} \nonumber\\
    && \times \rme^{\rmi (\omega - \omega') t} \, \rme^{\rmi (\nu - \nu') \omega_d t} \big[ X_{\omega', \nu'} \hat{\rho}(t), X_{\omega, \nu}^\dagger \big] + \text{H.c.} \nonumber\\
\end{eqnarray}
Equation~\eqref{BMFeom} is the Born-Markov master equation in the Floquet basis.
Because it is not of Lindblad type, it does not preserve the positivity of the density operator.

A master equation preserving positivity is obtained within the secular approximation, where all contributions with $\omega' \neq \omega$ and $\nu' \neq \nu$ are neglected.
This simplification is justified if the relaxation of the system is slow compared with all oscillations $\rme^{\pm \rmi (\omega - \omega') t}$ and $\rme^{\pm\rmi (\nu - \nu') \omega_d t}$.
The resulting Floquet master equation reads~\cite{BBGSSW91, GH98}
\begin{eqnarray}\label{master}
  \frac{\rmd}{\rmd t} \hat{\rho}(t) &=& -\rmi \frac{1}{2} \sum_{\omega, \nu} \xi(\omega + \nu \omega_d) \big[ X_{\omega, \nu}^\dagger X_{\omega, \nu}^{}, \hat{\rho}(t) \big] \nonumber\\
    && + \frac{1}{2} \sum_{\omega, \nu} \chi(\omega + \nu \omega_d) \Big\{ \big[ X_{\omega, \nu}^{} \hat{\rho}(t), X_{\omega, \nu}^\dagger \big] \nonumber\\
    && + \big[ X_{\omega, \nu}^{}, \hat{\rho}(t) X_{\omega, \nu}^\dagger \big] \Big\} \,.
\end{eqnarray}
It contains dissipative terms proportional to $\chi(\omega)$ and the Lamb-shift terms proportional to $\xi(\omega)$.
These reservoir-induced dissipation effects are included to lowest order in the system-reservoir-coupling strength.
Nevertheless, the periodic driving of the system as well as all intrasystem couplings is included to all orders.

The master equation~\eqref{master} splits into two equations of motion
\begin{eqnarray}\label{drho_nn}
  \frac{\rmd}{\rmd t} \rho_{n,n}(t) &=& \sum_{k \neq n} \sum_\nu \chi(\omega_{kn\nu}) |X_{n,k,\nu}|^2 \rho_{k,k}(t) \nonumber\\
    && - \sum_{k \neq n} \sum_\nu \chi(\omega_{nk\nu}) |X_{k,n,\nu}|^2 \rho_{n,n}(t) \,, \qquad
\end{eqnarray}
\begin{equation}\label{drho_mn}
  \frac{\rmd}{\rmd t} \rho_{m,n}(t) = -Z_{m,n} \rho_{m,n}(t) \qquad (m \neq n) \,,
\end{equation}
for the matrix elements $\rho_{m,n}(t) = \langle \psi_m(t) | \rho(t) | \psi_n(t) \rangle$ of the system density operator.
In these equations, $\omega_{kn\nu} = \epsilon_k - \epsilon_n + \nu \omega_d$,
\begin{equation}
  X_{n,k,\nu} = \sum_\mu \langle \widetilde{\phi}_n(\mu - \nu) | X | \widetilde{\phi}_k(\mu) \rangle \,,
\end{equation}
and
\begin{eqnarray}
  Z_{m,n} &=& \frac{1}{2} \sum_{k, \nu} \big[ \chi(\omega_{mk\nu}) + \rmi \xi(\omega_{mk\nu}) \big] |X_{k,m,\nu}|^2 \nonumber\\
    && + \frac{1}{2} \sum_{k, \nu} \big[ \chi(\omega_{nk\nu}) - \rmi \xi(\omega_{nk\nu}) \big] |X_{k,n,\nu}|^2 \nonumber\\
    && - \sum_\nu \chi(\nu \omega_d) X_{m,m,\nu} X_{n,n,\nu}^* \,.
\end{eqnarray}
Since
\begin{eqnarray}
  \Re Z_{m,n} &=& \frac{1}{2} \sum_\nu \chi(\nu \omega_d) \big| X_{m,m,\nu} - X_{n,n,\nu} \big|^2 \nonumber\\
    && + \frac{1}{2} \sum_{k \neq m, \nu} \chi(\omega_{mk\nu}) | X_{k,m,\nu} |^2 \nonumber\\
    && + \frac{1}{2} \sum_{k \neq n, \nu} \chi(\omega_{nk\nu}) | X_{k,n,\nu} |^2
\end{eqnarray}
is positive for all $m \neq n$, the general solution of Eq.~\eqref{drho_mn},
\begin{equation}
  \hat{\rho}_{m,n}(t) = \rme^{-Z_{m,n} t} \hat{\rho}_{m,n}(0) \qquad (m \neq n) \,,
\end{equation}
shows an exponential decay of the off-diagonal density matrix elements.

\section{\label{app:emitters}Emission properties of a few emitters}
In this appendix we present and discuss the emission spectra, the shift of spectral lines, and the Glauber function for two and three emitters.

\subsection{Emission spectra}
\begin{figure}
  \includegraphics[width=0.49\linewidth]{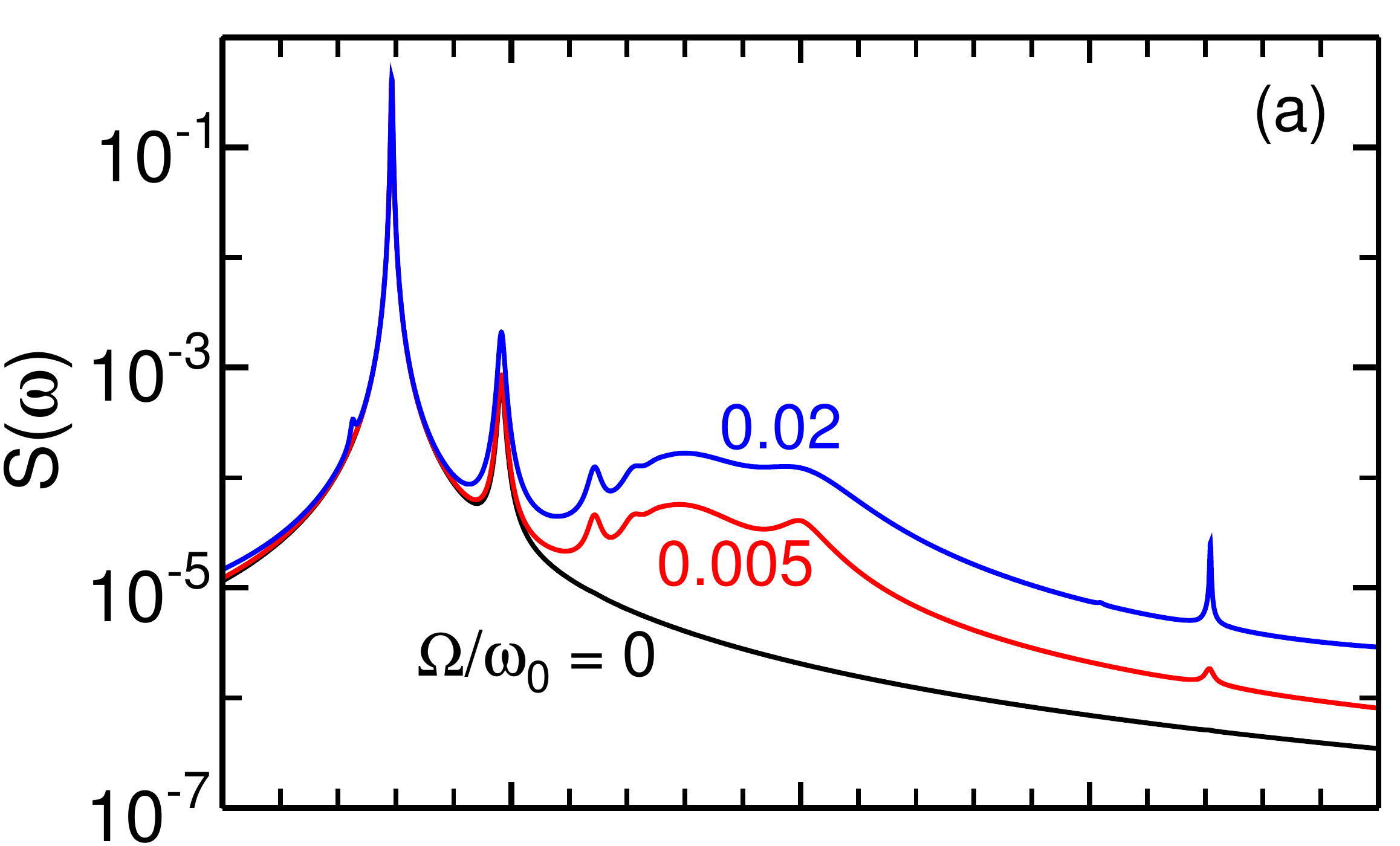}
  \includegraphics[width=0.49\linewidth]{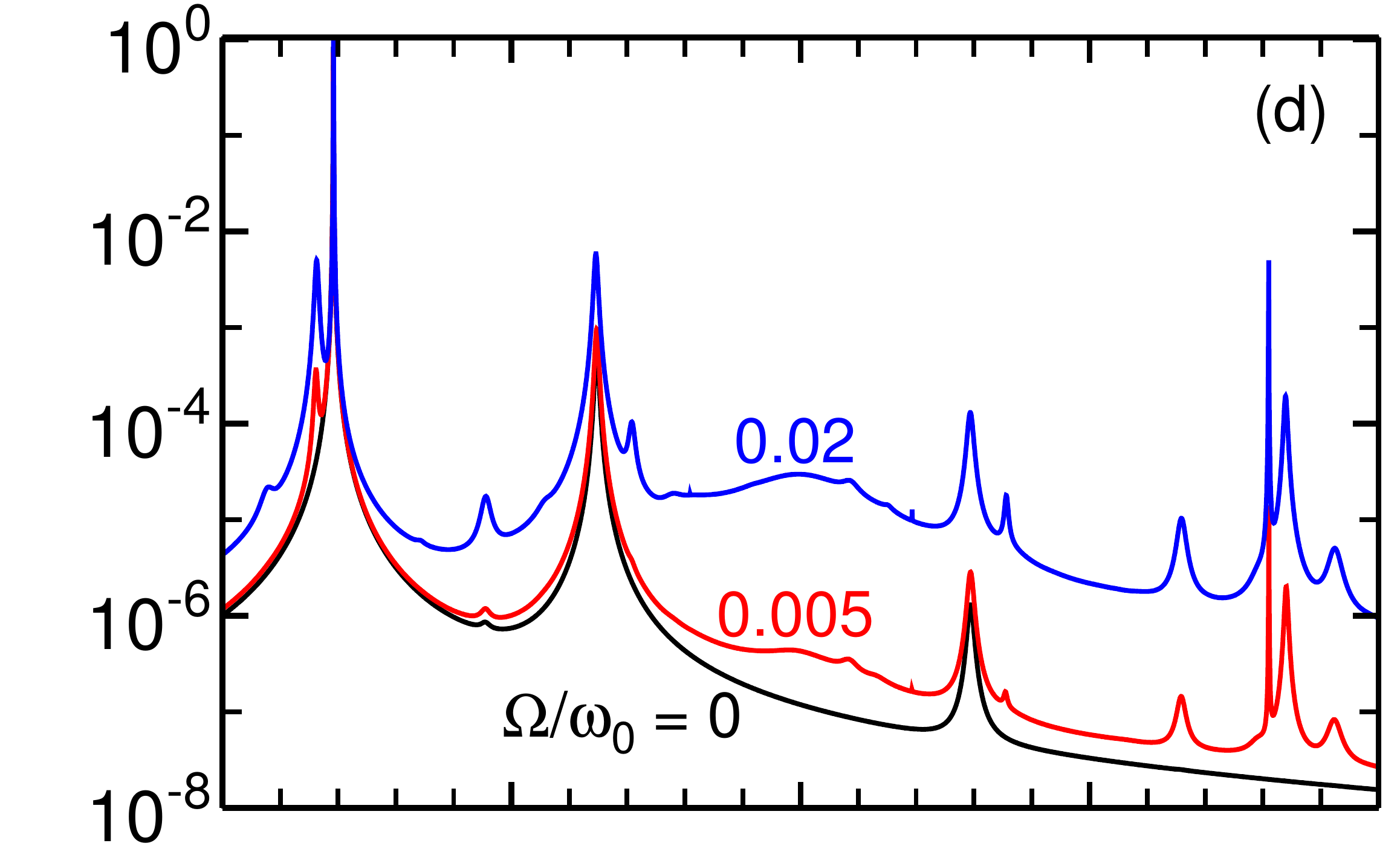}\\
  \includegraphics[width=0.49\linewidth]{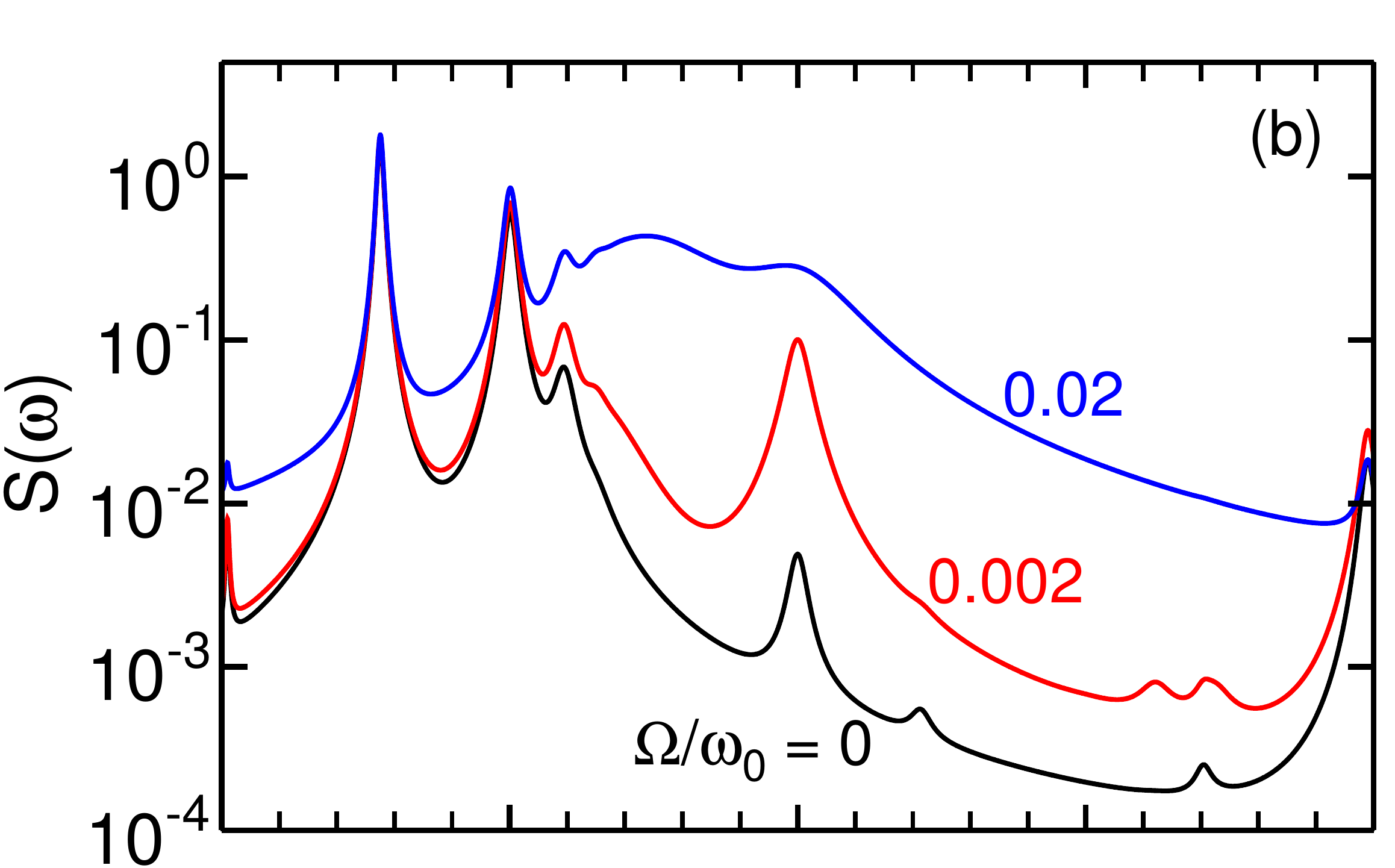}
  \includegraphics[width=0.49\linewidth]{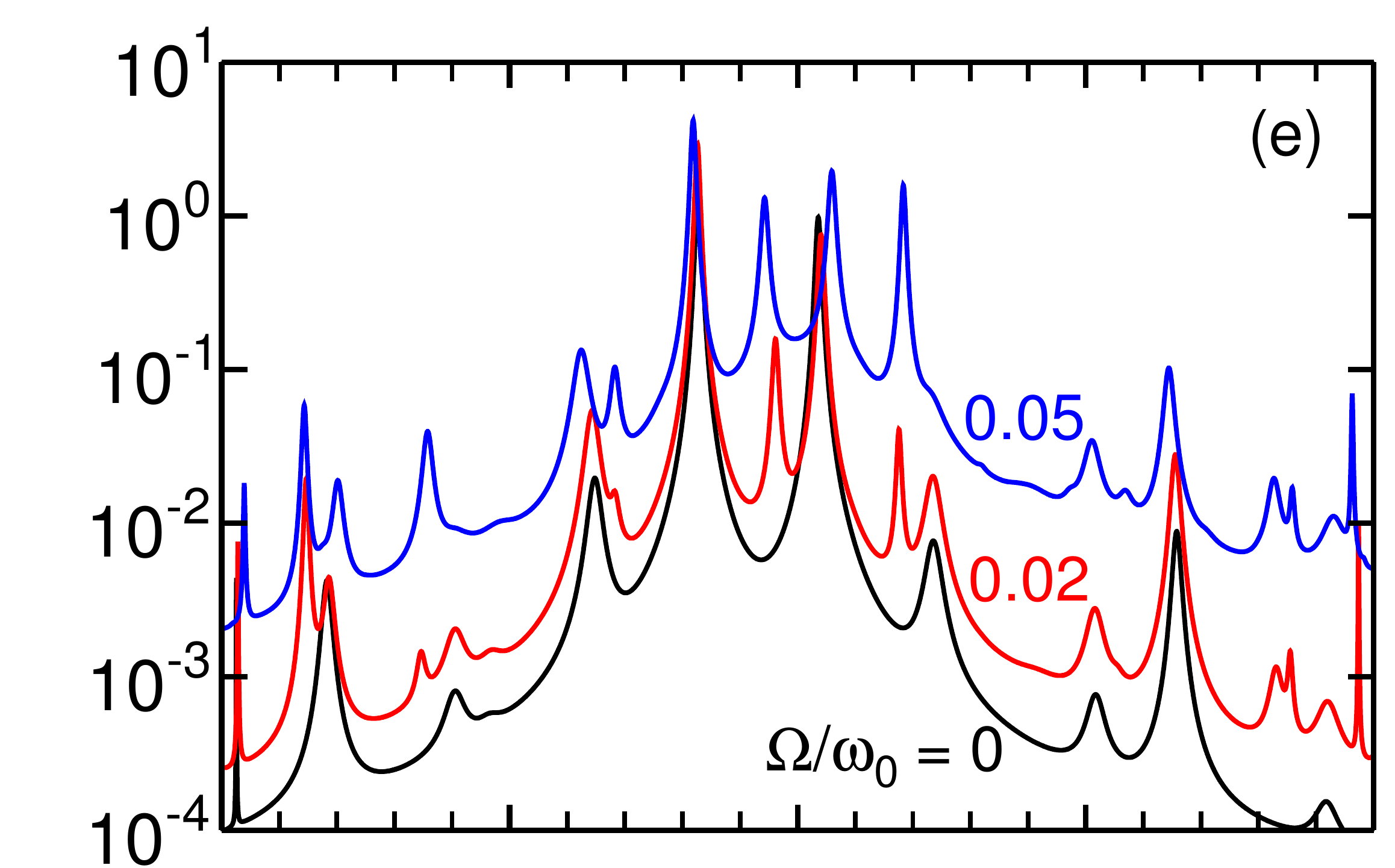}\\
  \includegraphics[width=0.49\linewidth]{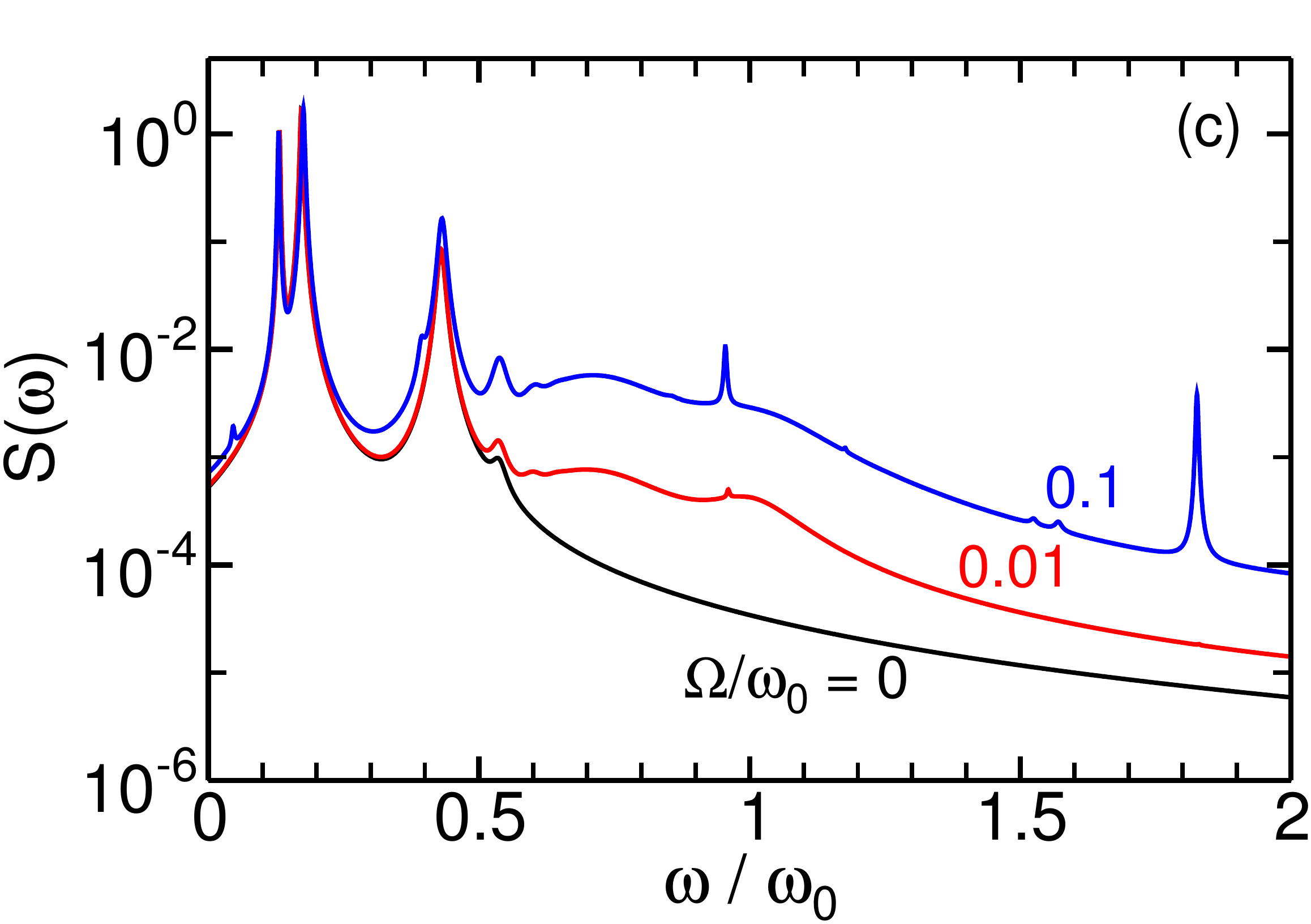}
  \includegraphics[width=0.49\linewidth]{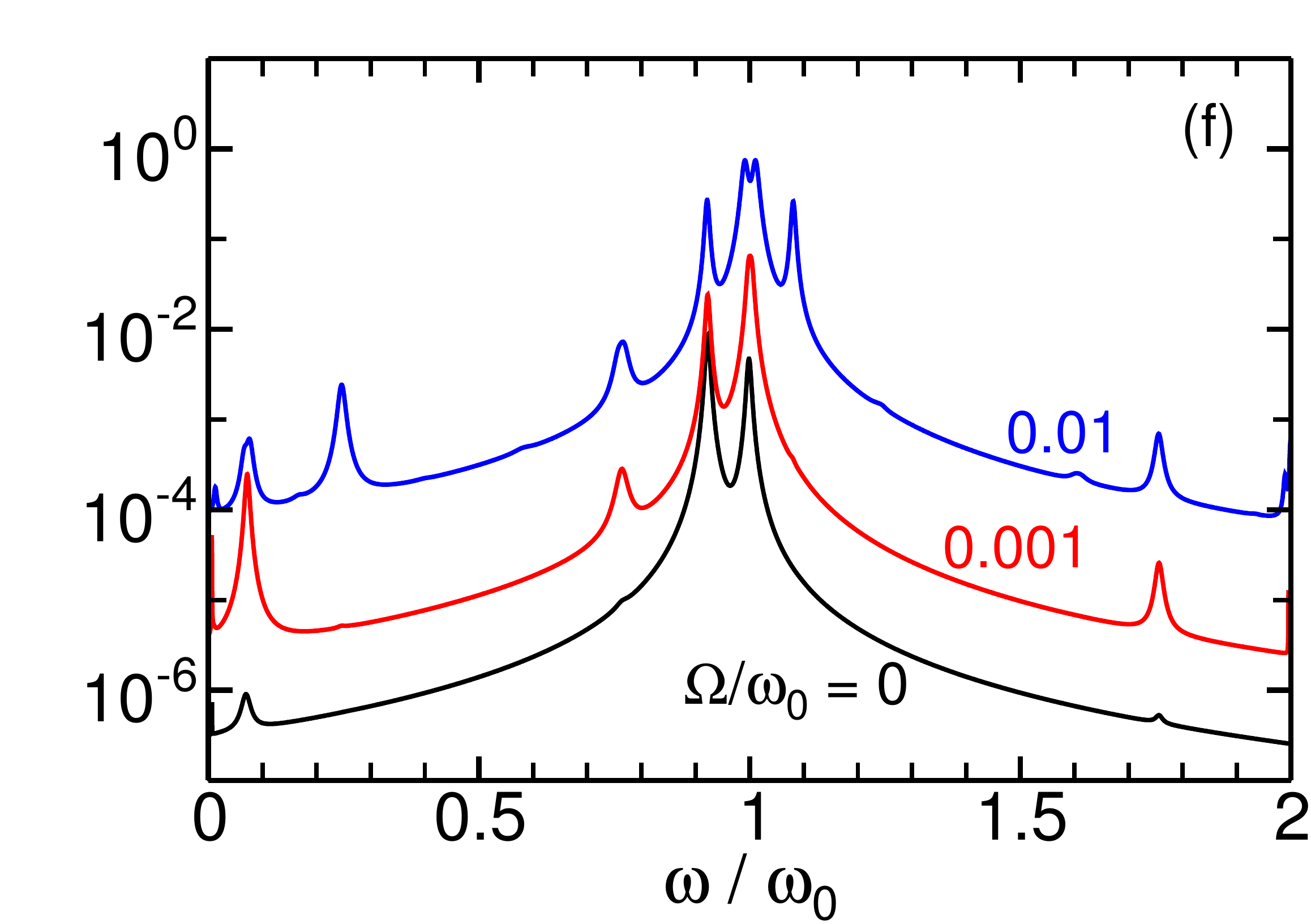}
  \caption{\label{fig:spct2}Emission spectra $S(\omega)$ for two emitters.
    The left (right) column depicts the results for $g' = \Omega' = 0$ ($g' = g$ and $\Omega' = \Omega$).
    The other parameters are the same as in Fig.~\ref{fig:spct1}.}
\end{figure}

\begin{figure}[b]
  \includegraphics[width=0.49\linewidth]{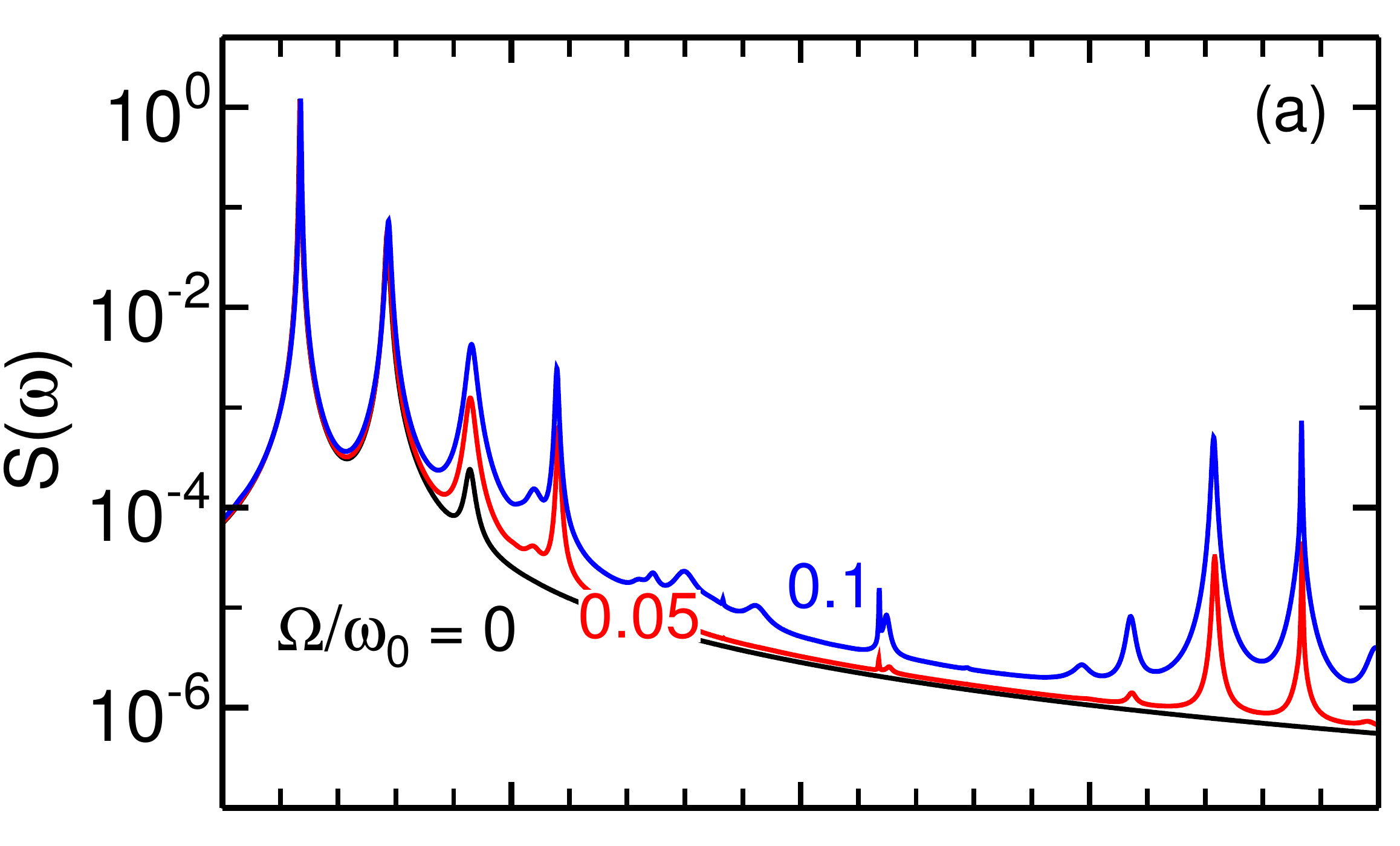}
  \includegraphics[width=0.49\linewidth]{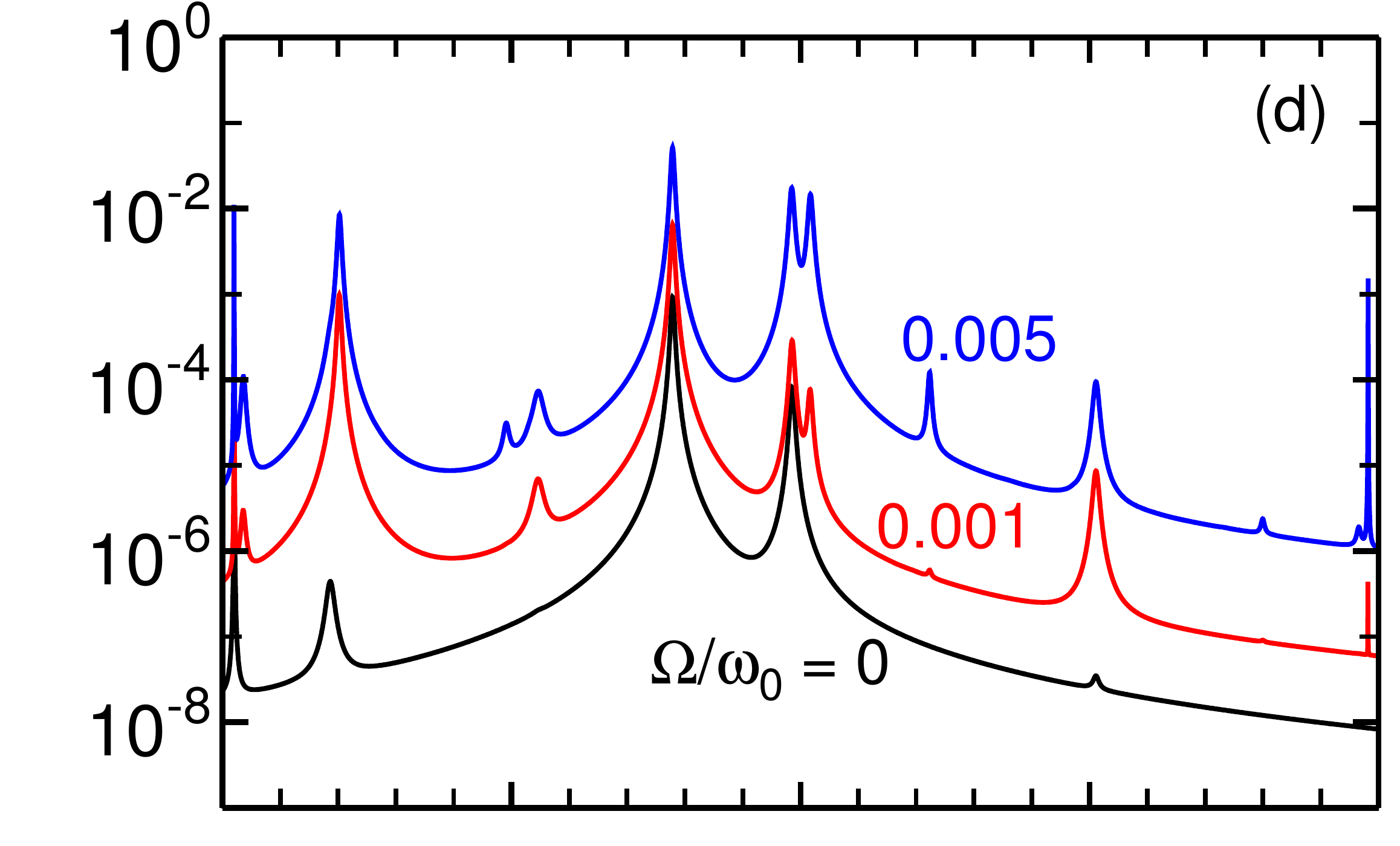}\\
  \includegraphics[width=0.49\linewidth]{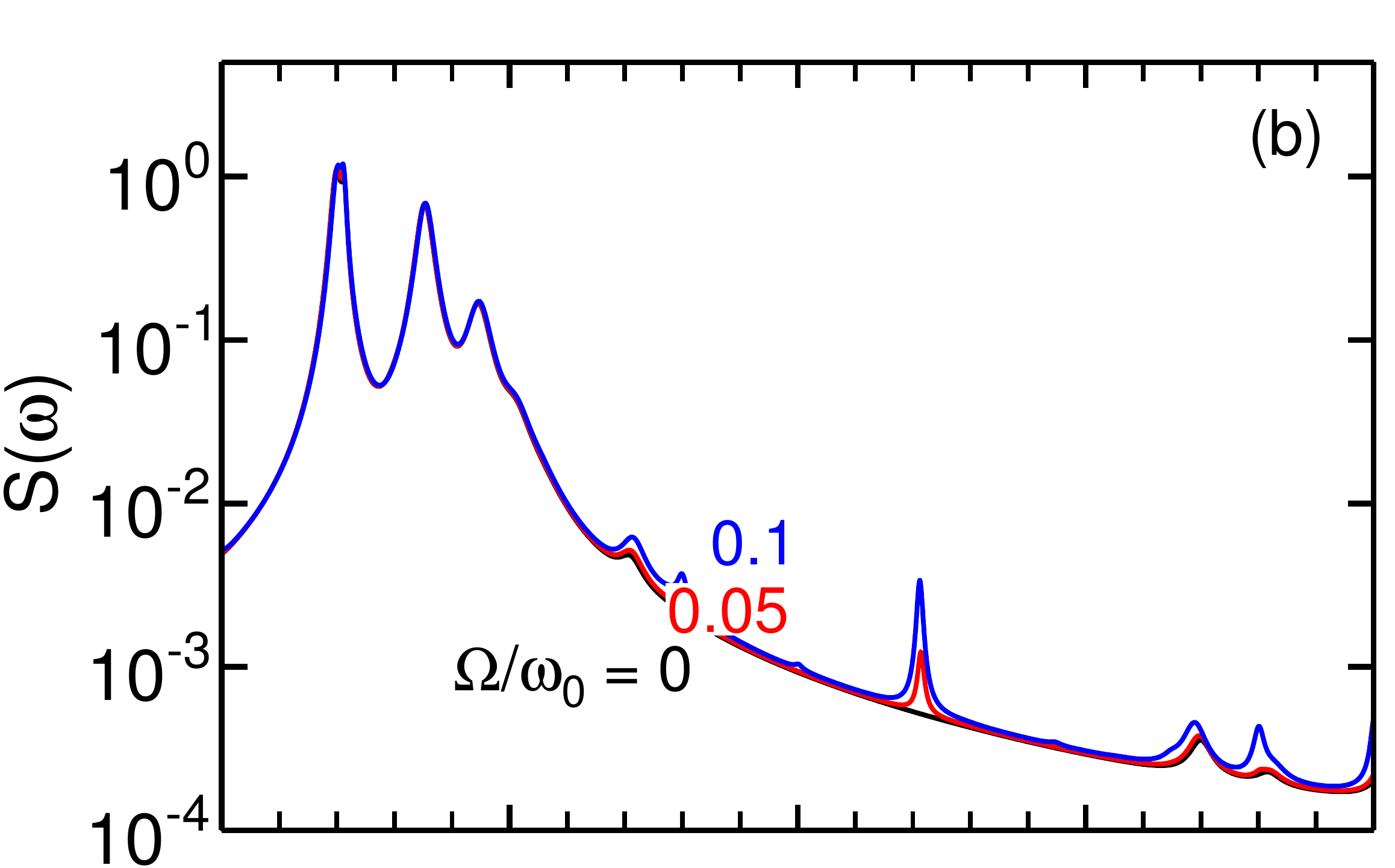}
  \includegraphics[width=0.49\linewidth]{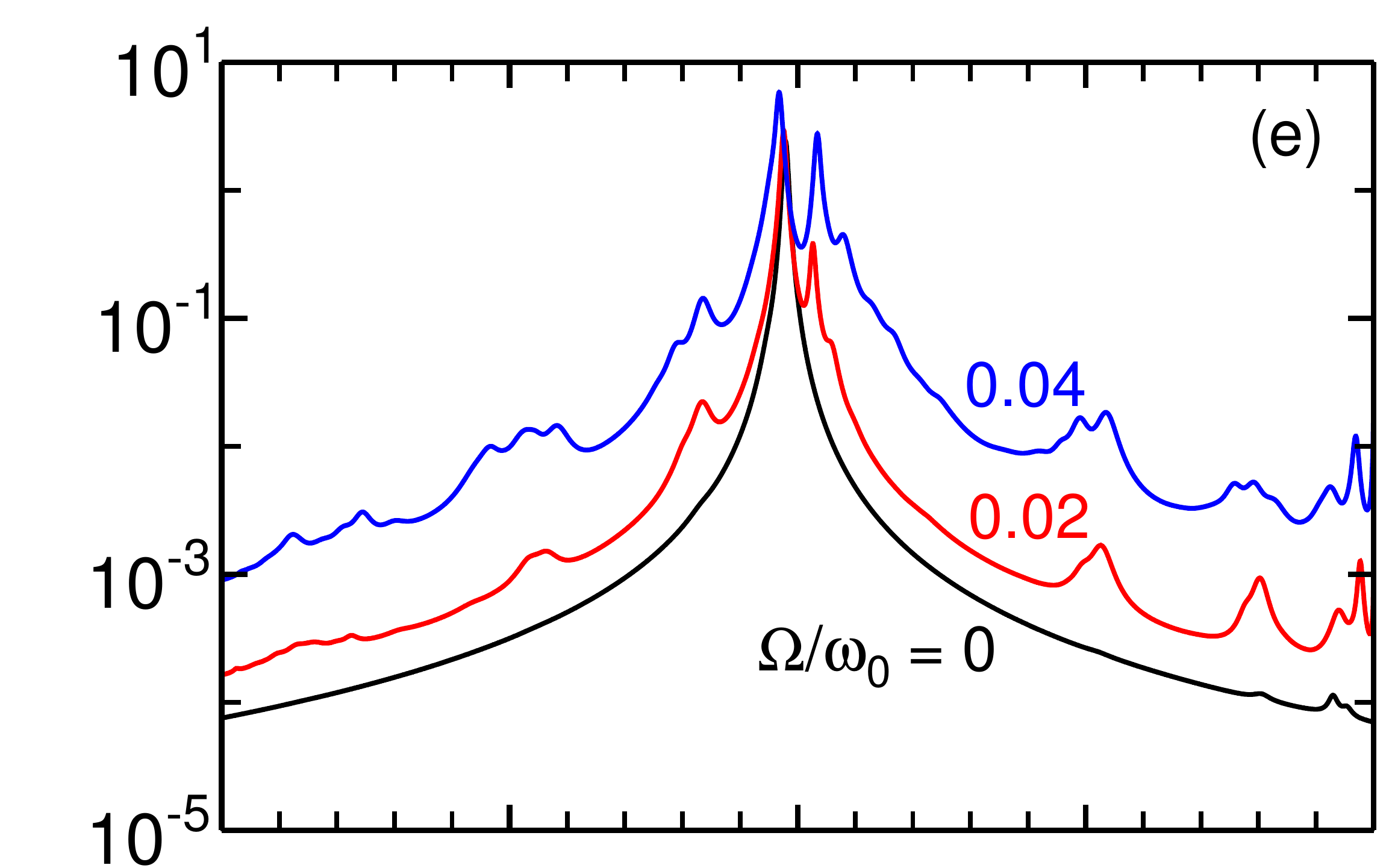}\\
  \includegraphics[width=0.49\linewidth]{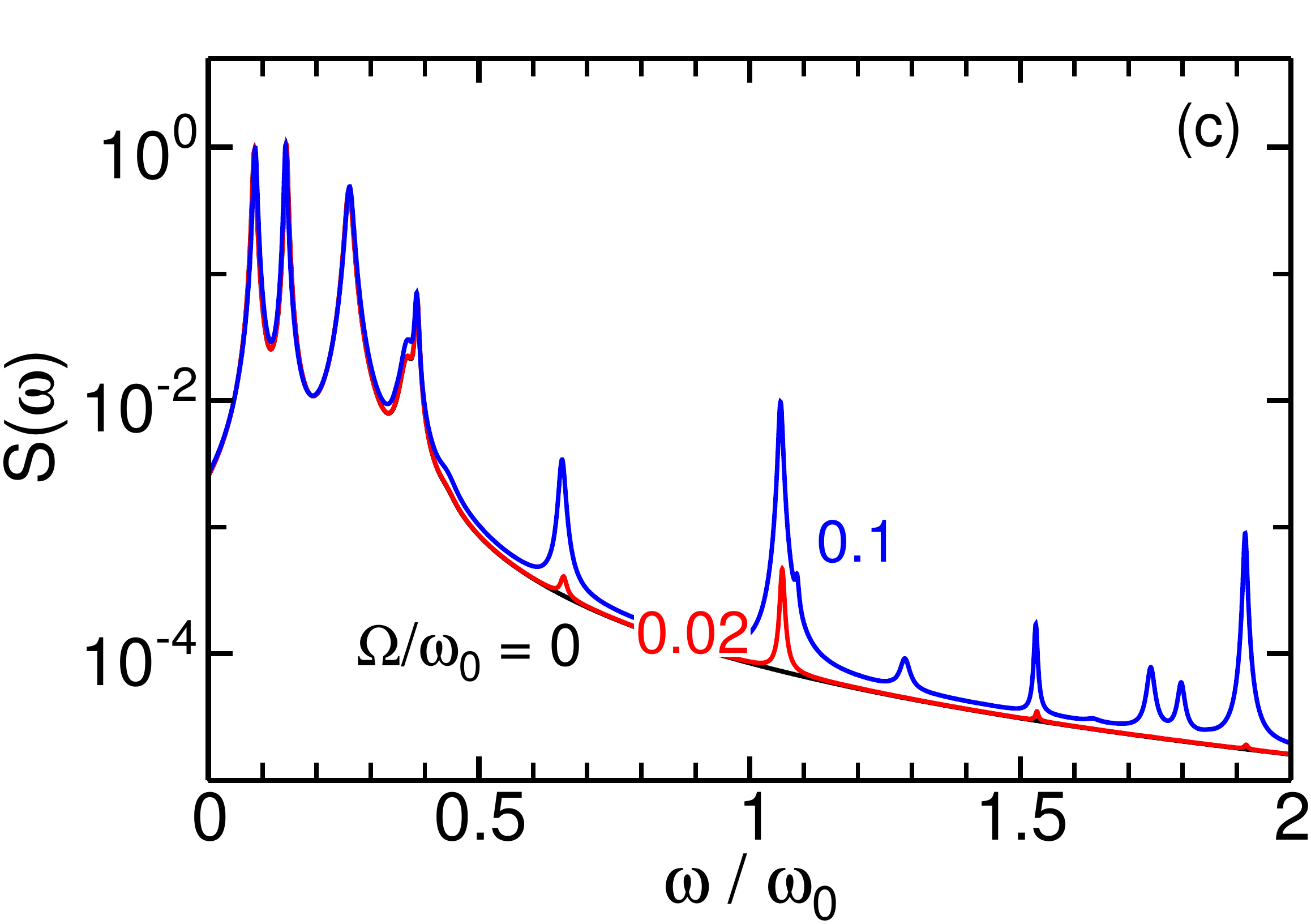}
  \includegraphics[width=0.49\linewidth]{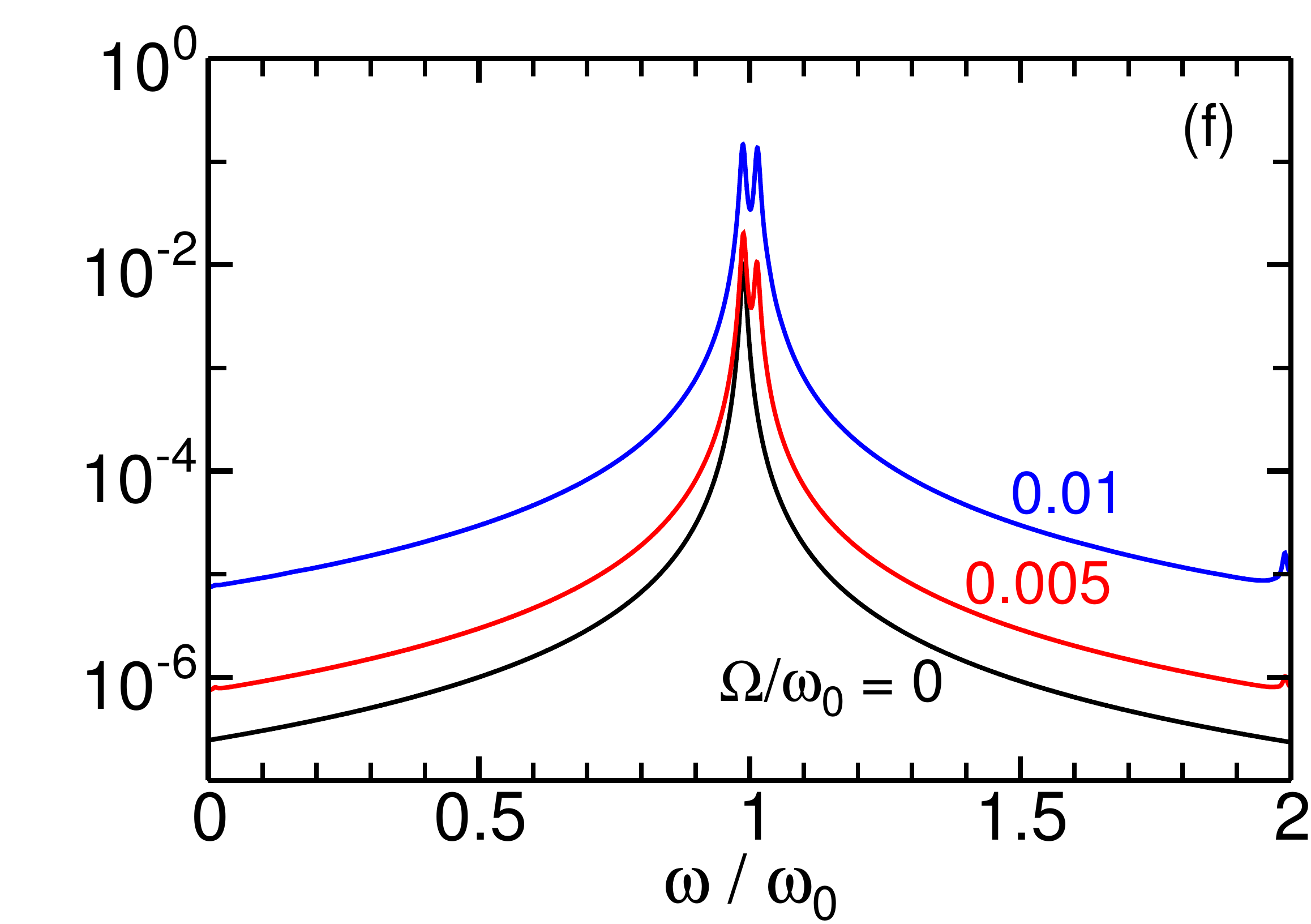}
  \caption{\label{fig:spct3}Emission spectra $S(\omega)$ for three emitters.
    The left (right) column depicts the results for $g' = \Omega' = 0$ ($g' = g$ and $\Omega' = \Omega$).
    The other parameters are the same as in Fig.~\ref{fig:spct1}.}
\end{figure}

The emission spectra for two and three emitters given in Figs.~\ref{fig:spct2} and~\ref{fig:spct3} are calculated with the formalism explained in Sec.~\ref{ssec:spectra}.
The results show the same behavior with increasing $\Omega$:
(i) The height of the peaks in Figs.~\ref{fig:spct2} and~\ref{fig:spct3} grows because the populations $\rho_{n,n}^\infty$ of higher excited states are enhanced,
(ii) new emission peaks appear due to the increased number of allowed transitions (e.g., between sidebands), and
(iii) the spectral lines shift.

Directly comparing Figs.~\ref{fig:spct2}(a)--\ref{fig:spct2}(c) with Figs.~\ref{fig:spct1}(a)--\ref{fig:spct1}(c), we notice that, within the rotating-wave approximation ($g' = 0$ and $\Omega' = 0$), the dominant spectral lines are shifted towards lower energies.
The trend continues when the number of emitters is increased from $N = 2$ to $N = 3$ in Figs.~\ref{fig:spct3}(a)--\ref{fig:spct3}(c).
This is a consequence of the quasienergy spectra given in Figs.~\ref{fig:energy}(a)--\ref{fig:energy}(c).
In a finite energy interval the number of states grows with the number of emitters.
Hence, the energies of transitions between the states become smaller.

Similar behavior cannot be observed if the counterrotating terms are included ($g' = g$ and $\Omega' = \Omega$), i.e., when Figs.~\ref{fig:spct2}(d)--\ref{fig:spct2}(f) and~\ref{fig:spct3}(d)--\ref{fig:spct3}(f) are compared with Fig.~\ref{fig:spct1}.
Instead, the energy of the dominant spectral line tends to converge to $\omega_0$.
This convergence is faster for larger emitter-cavity coupling $g$.
Again, this can be understood with the quasienergy spectra in Fig.~\ref{fig:energy}(d)--\ref{fig:energy}(e).
Increasing the number $N$ of emitters, pairs of system energies $E_n$ are very close to each other already at reduced emitter-cavity coupling $g$.
Importantly, the energy difference between neighboring pairs converges to $\omega_0$, both when $N$ is increased for fixed (but finite) $g$, and when $g$ is increased for fixed $N$.
This explains our observation.

\subsection{Dynamic Stark effect}
\begin{figure}
  \includegraphics[width=0.49\linewidth]{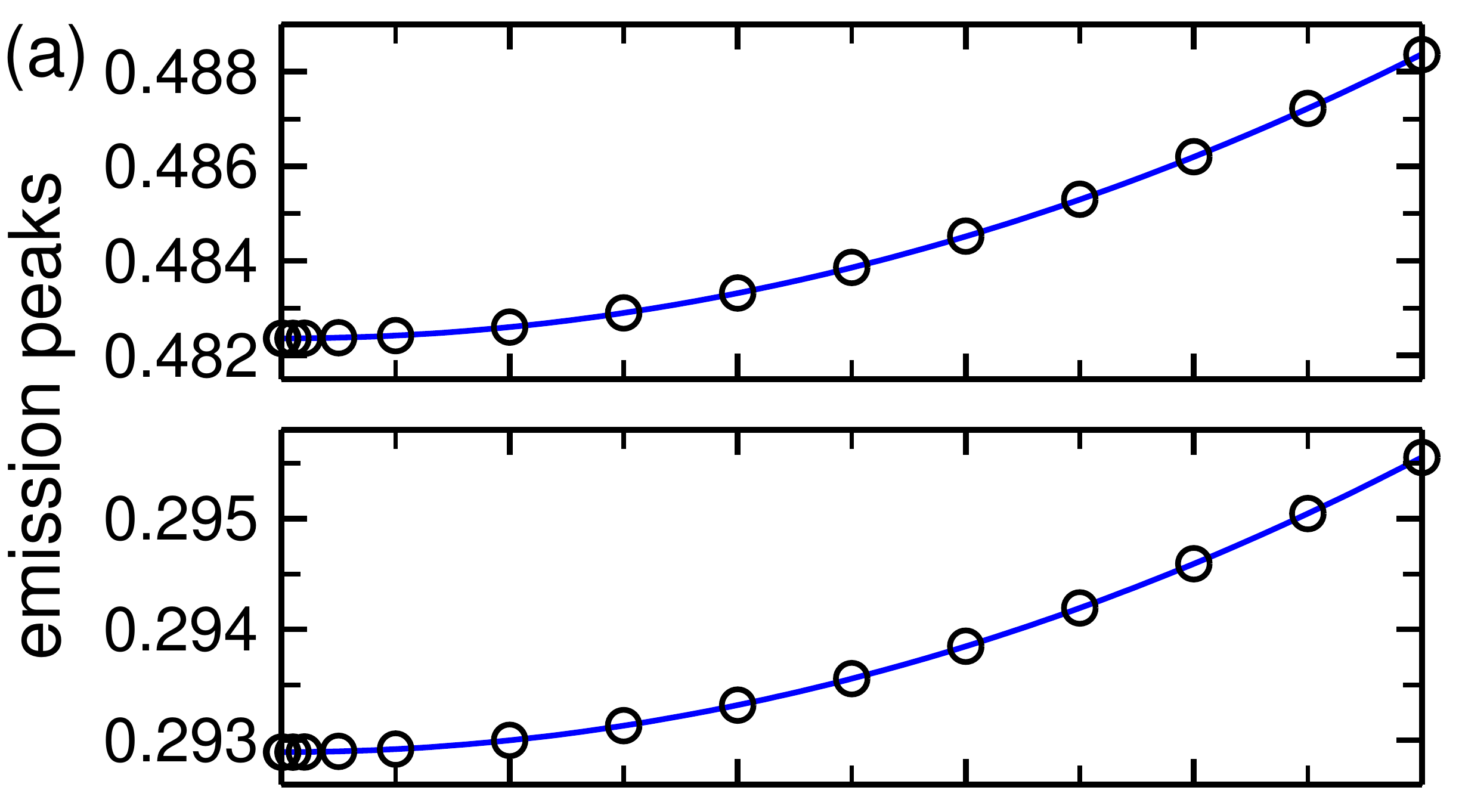}
  \includegraphics[width=0.49\linewidth]{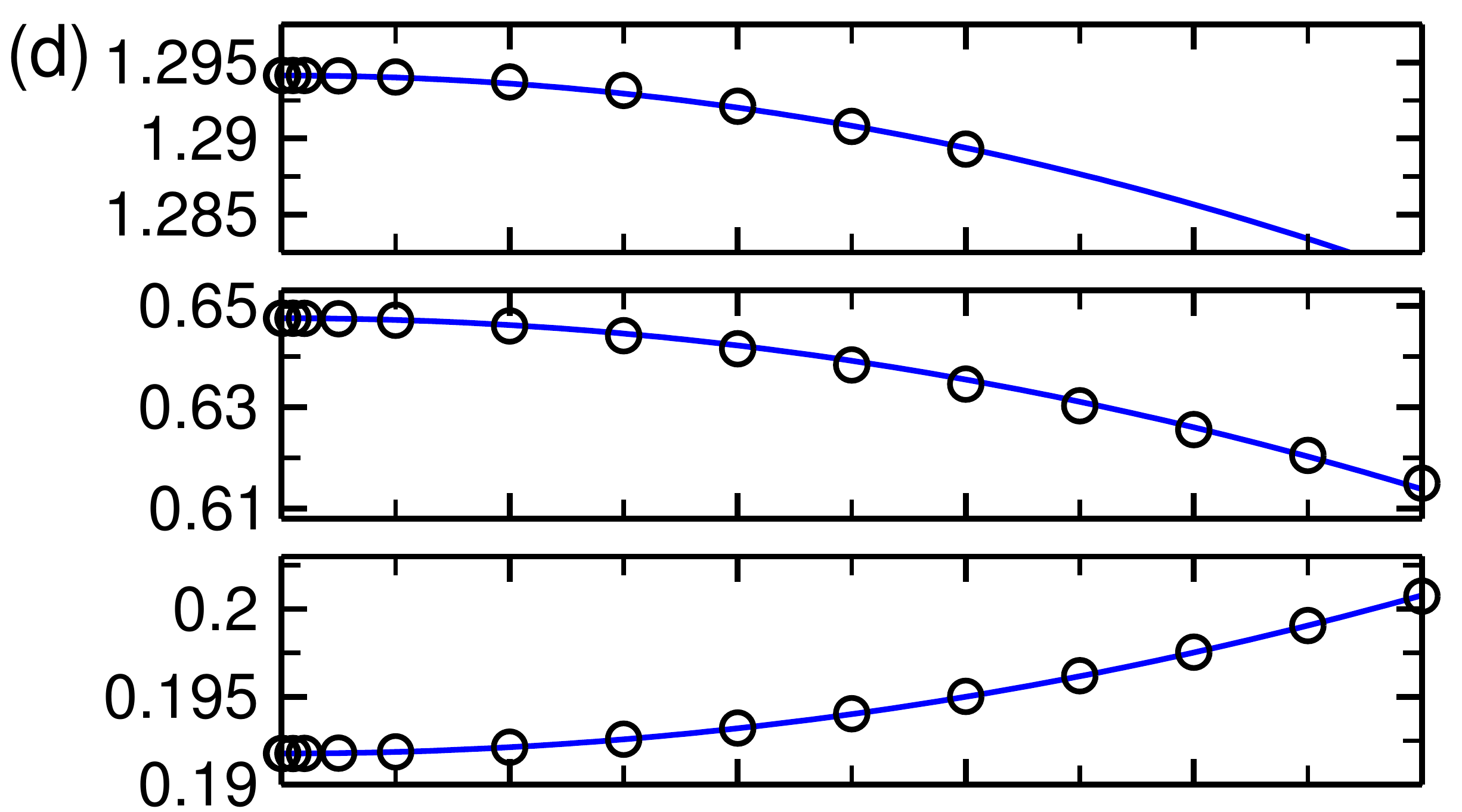}\\
  \includegraphics[width=0.49\linewidth]{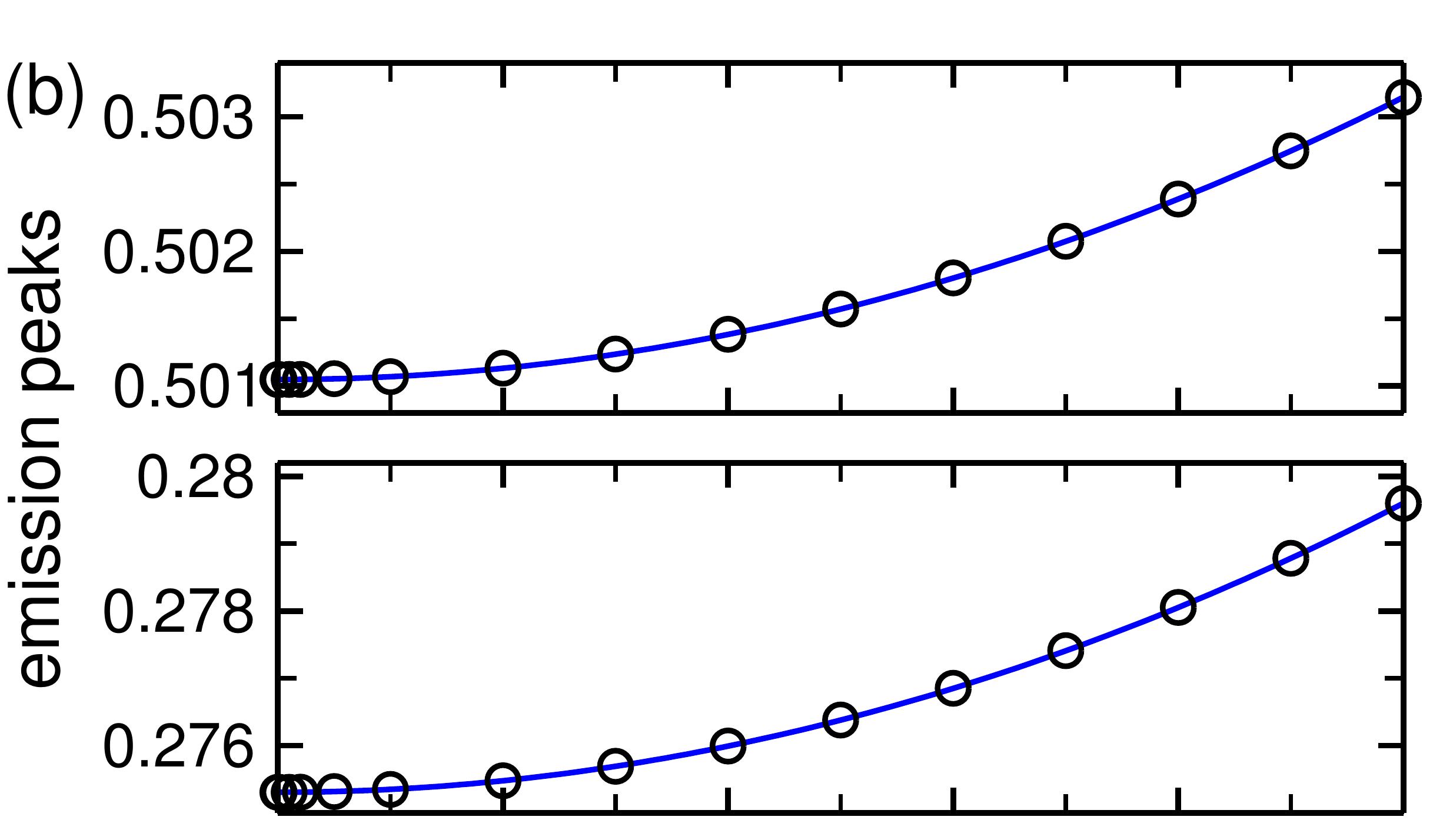}
  \includegraphics[width=0.49\linewidth]{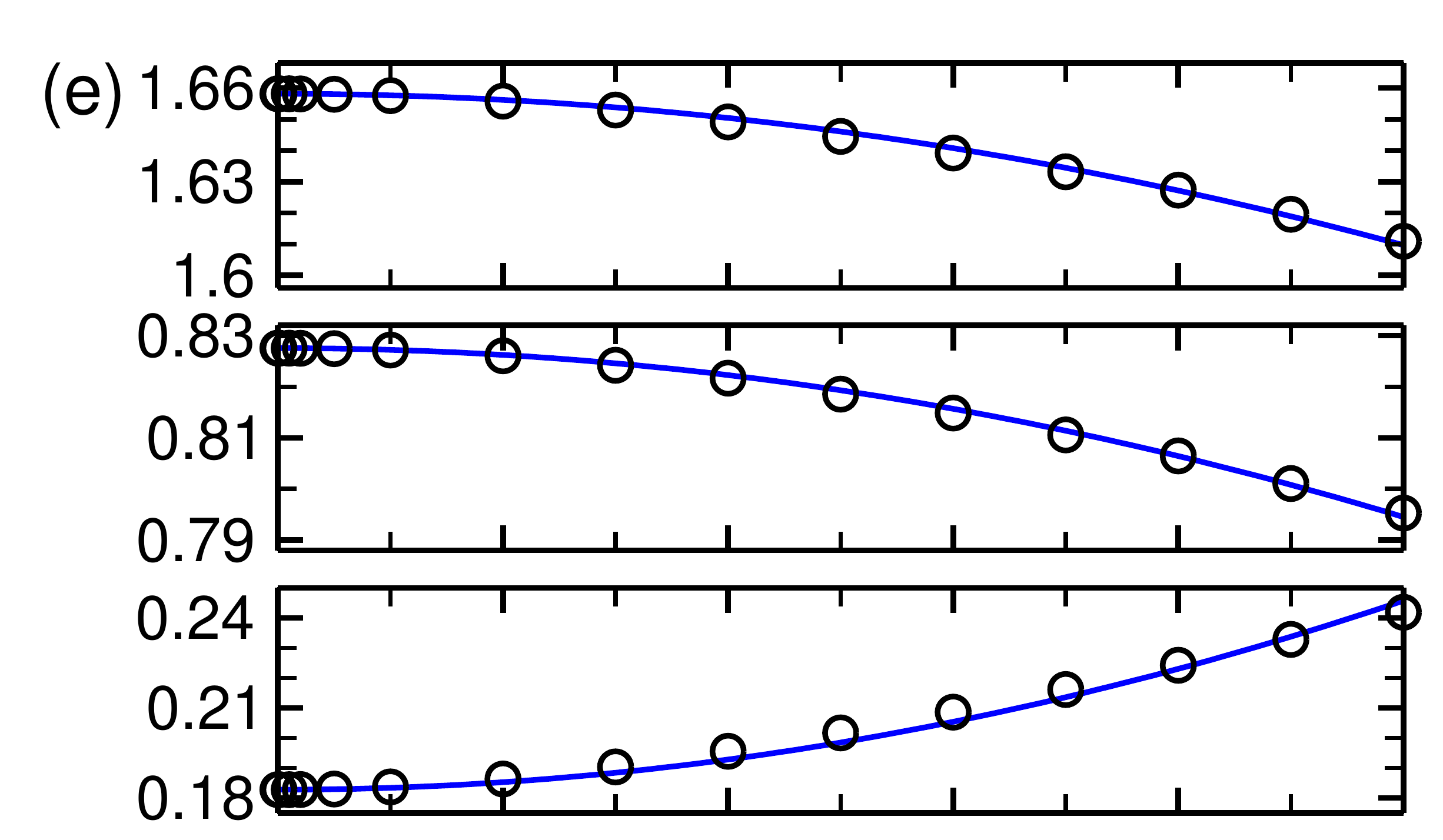}\\
  \includegraphics[width=0.49\linewidth]{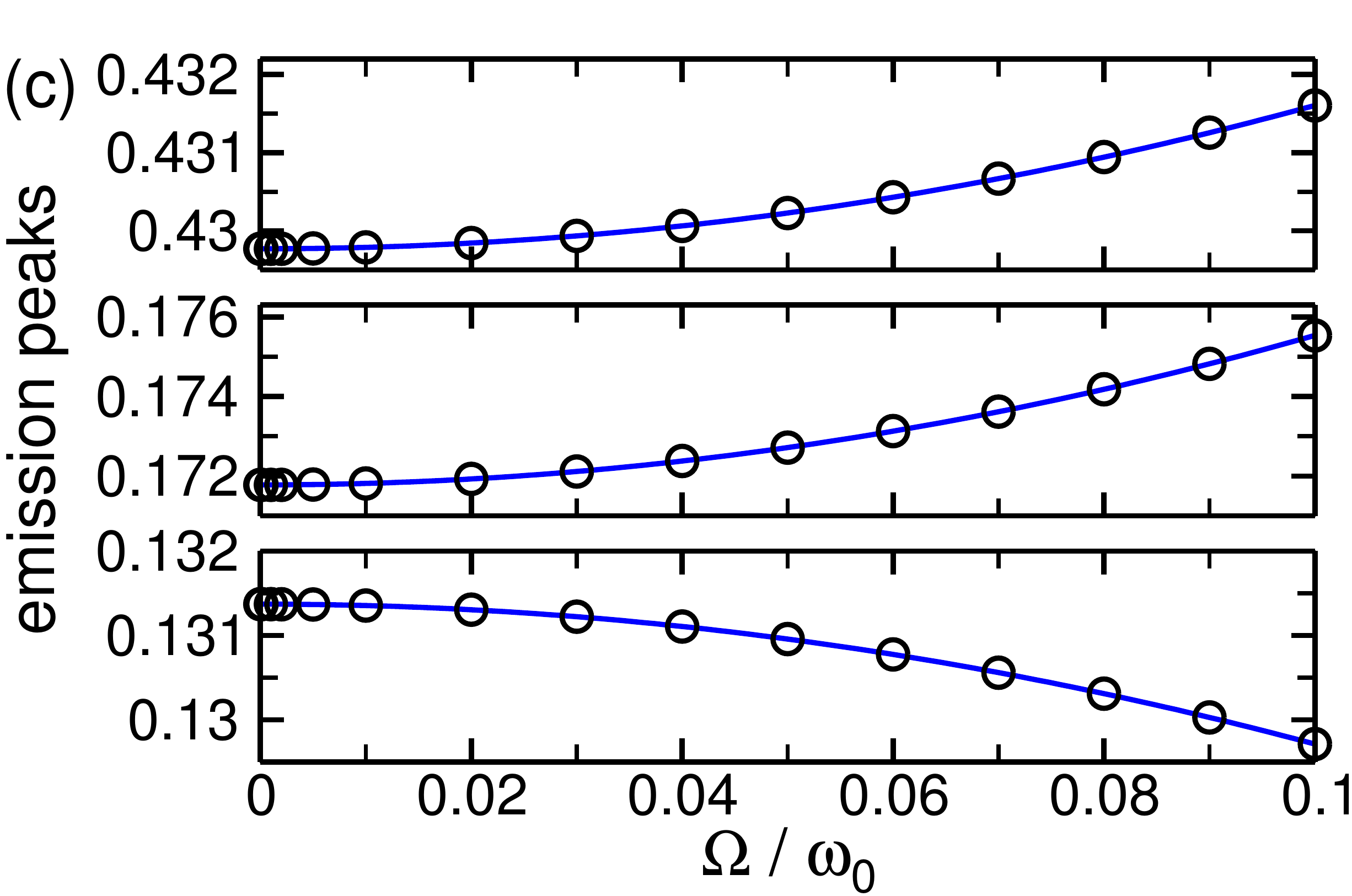}
  \includegraphics[width=0.49\linewidth]{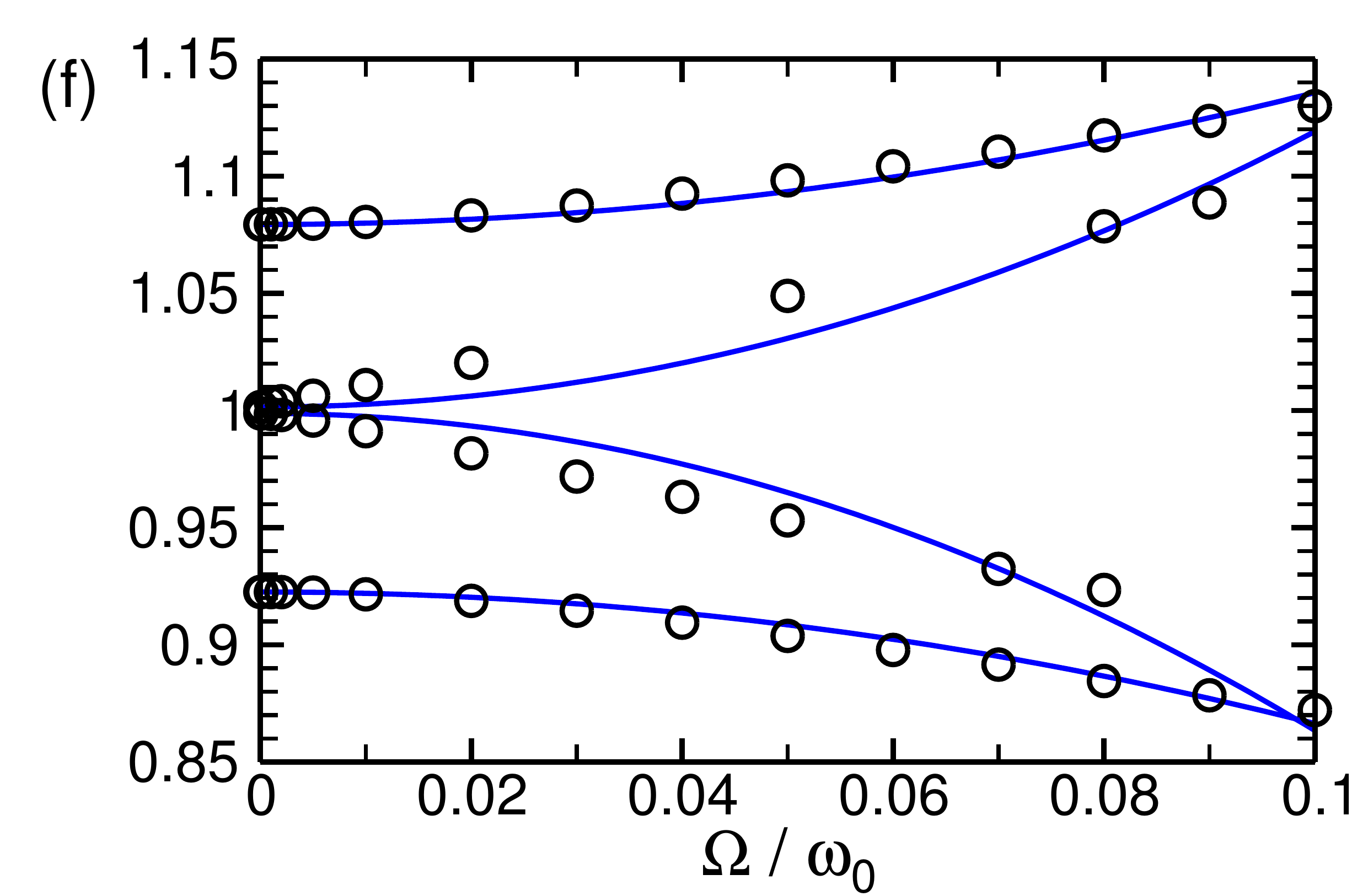}
  \caption{\label{fig:shft2}Shift of emission peaks as a function of the laser intensity $\Omega$ for two emitters.
    The left (right) column depicts the results for $g' = \Omega' = 0$ ($g' = g$ and $\Omega' = \Omega$).
    The other parameters are the same as in Fig.~\ref{fig:spct1}.
    The (blue) lines depict fitted values proportional to $[1 - (\Omega / g)^2]^{3/4}$.}
\end{figure}

\begin{figure}
  \includegraphics[width=0.49\linewidth]{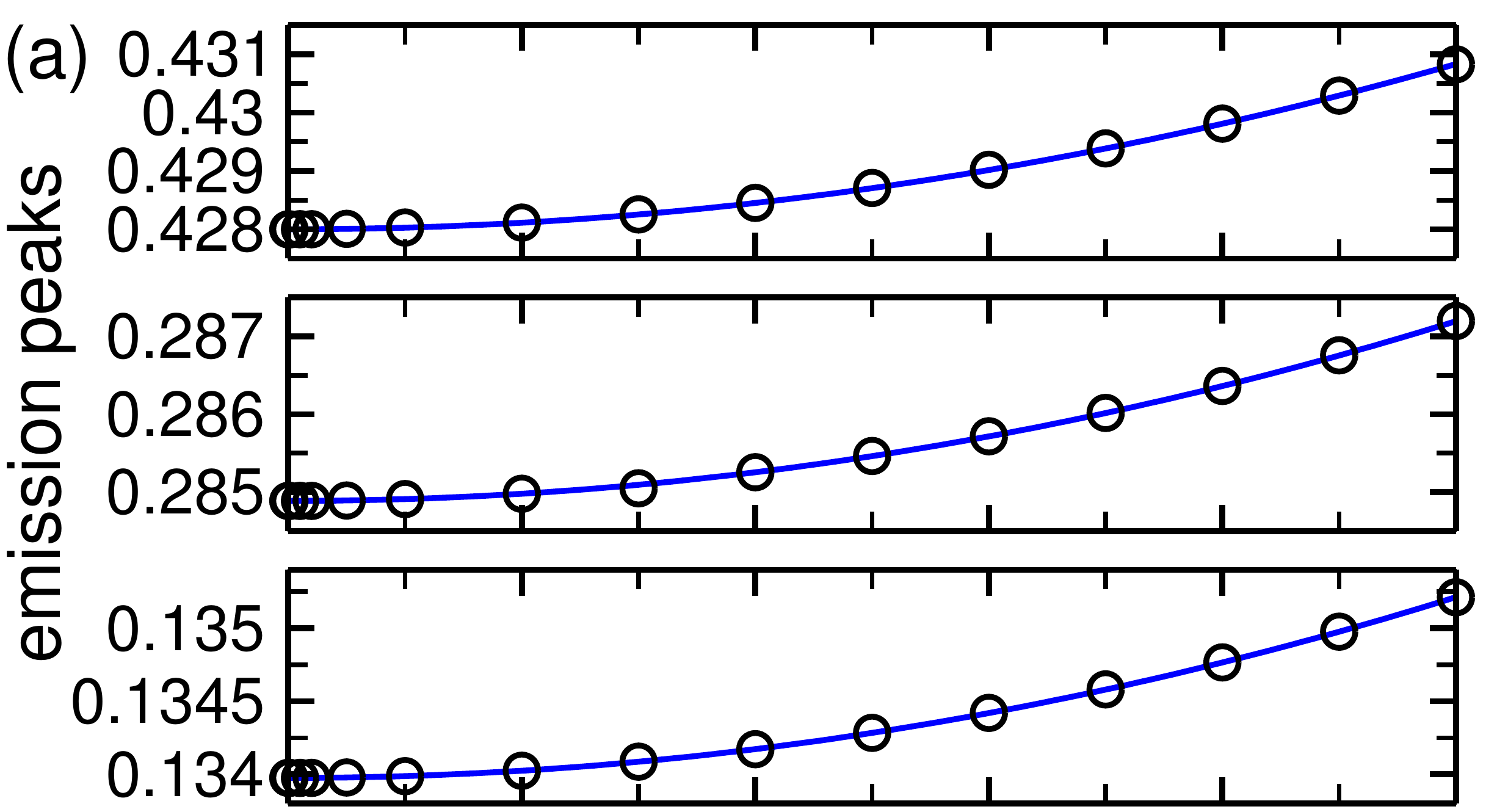}
  \includegraphics[width=0.49\linewidth]{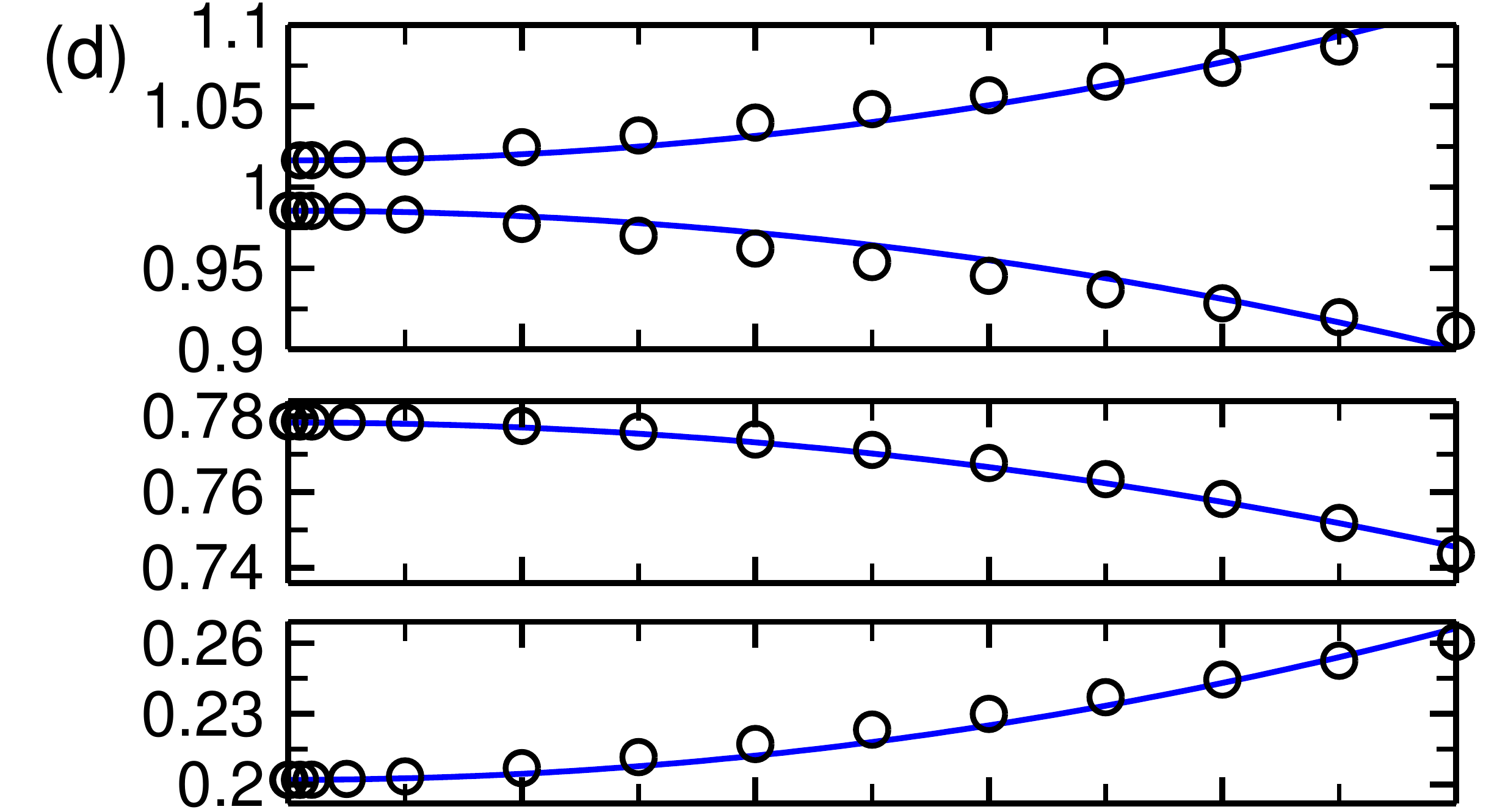}\\
  \includegraphics[width=0.49\linewidth]{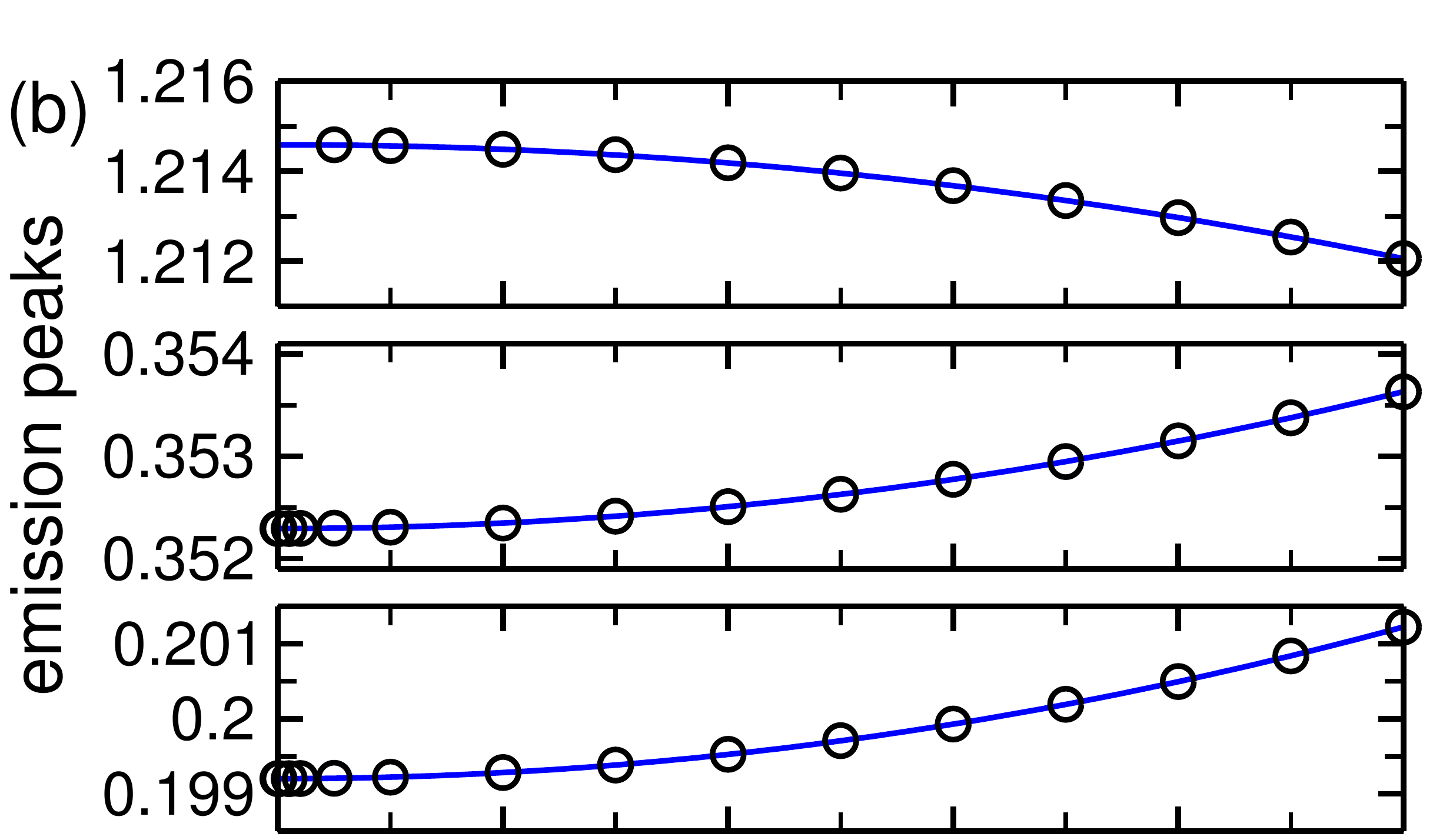}
  \includegraphics[width=0.49\linewidth]{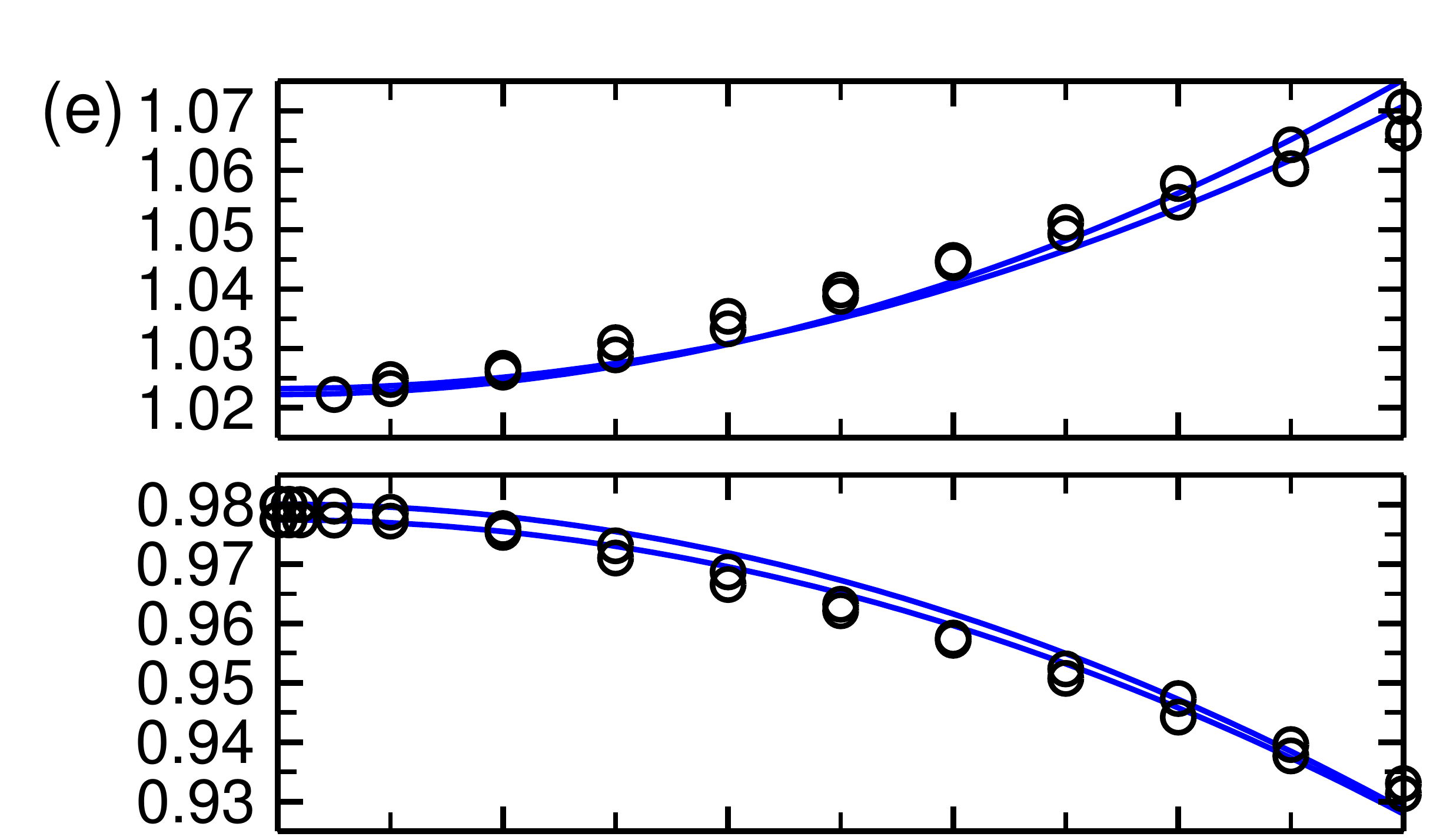}\\
  \includegraphics[width=0.49\linewidth]{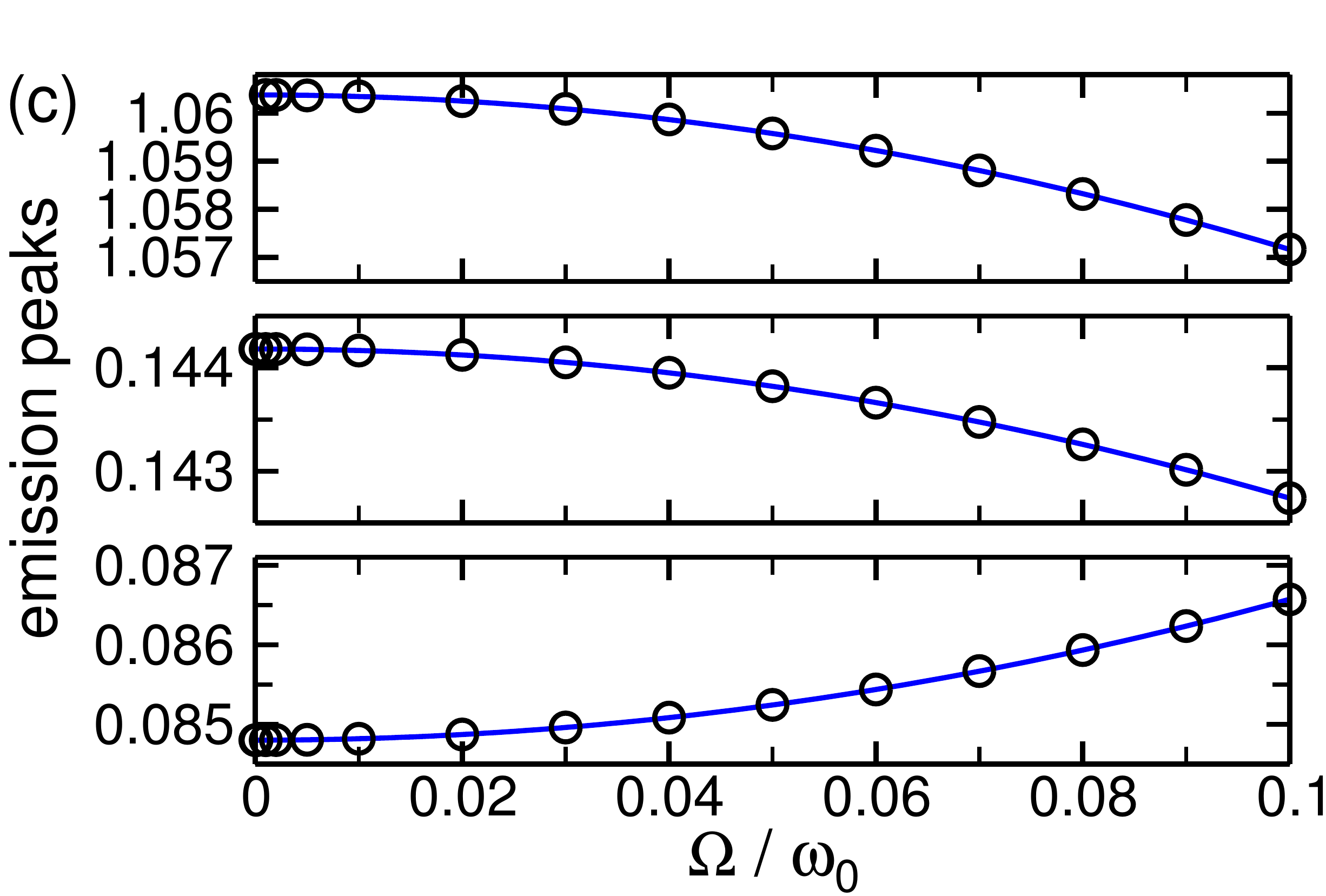}
  \includegraphics[width=0.49\linewidth]{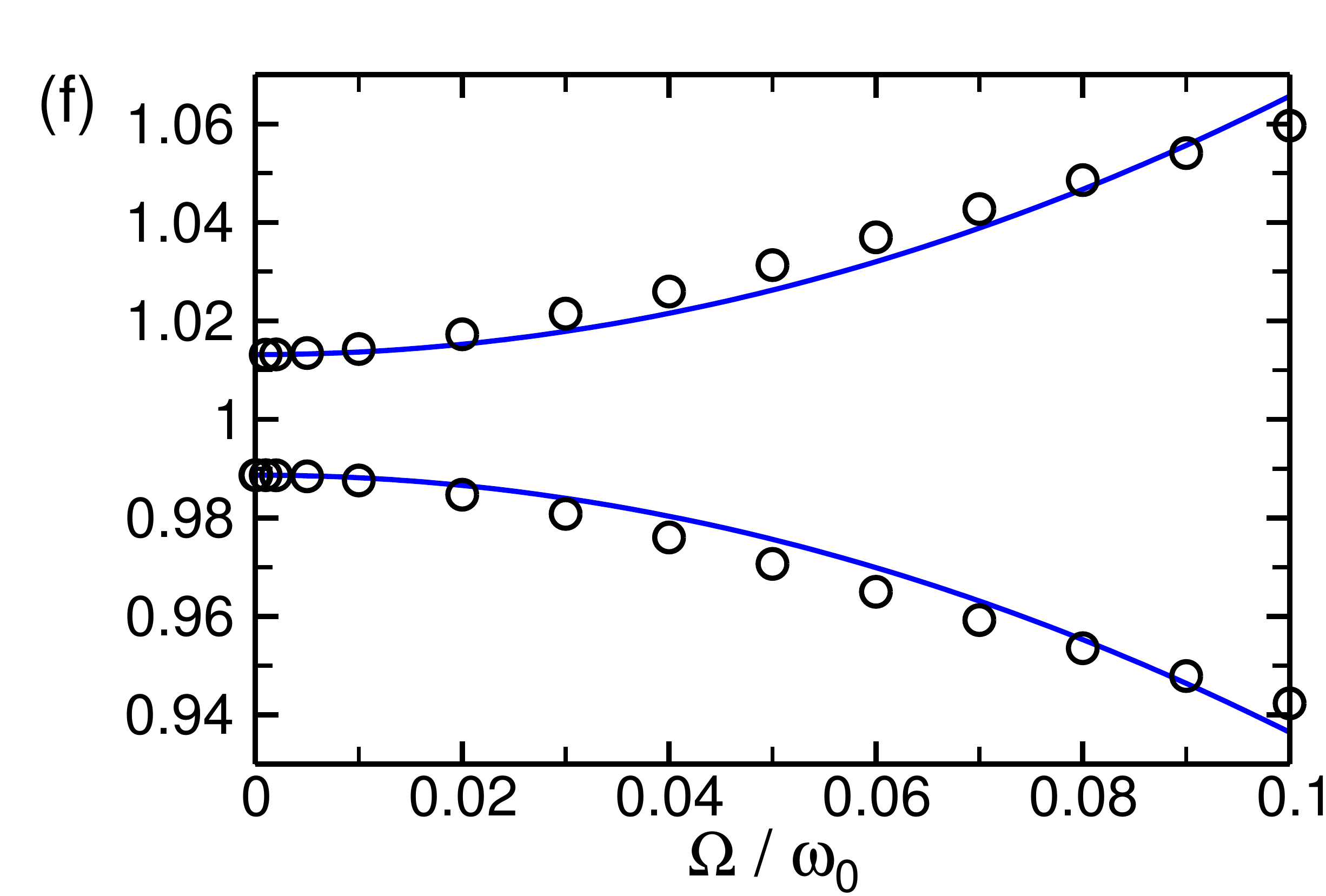}
  \caption{\label{fig:shft3}Shift of emission peaks as a function of the laser intensity $\Omega$ for three emitters.
    The left (right) column depicts the results for $g' = \Omega' = 0$ ($g' = g$ and $\Omega' = \Omega$).
    The other parameters are the same as in Fig.~\ref{fig:spct1}.
    The (blue) lines depict fitted values proportional to $[1 - (\Omega / g)^2]^{3/4}$.}
\end{figure}

\begin{figure}
  \includegraphics[scale=0.3]{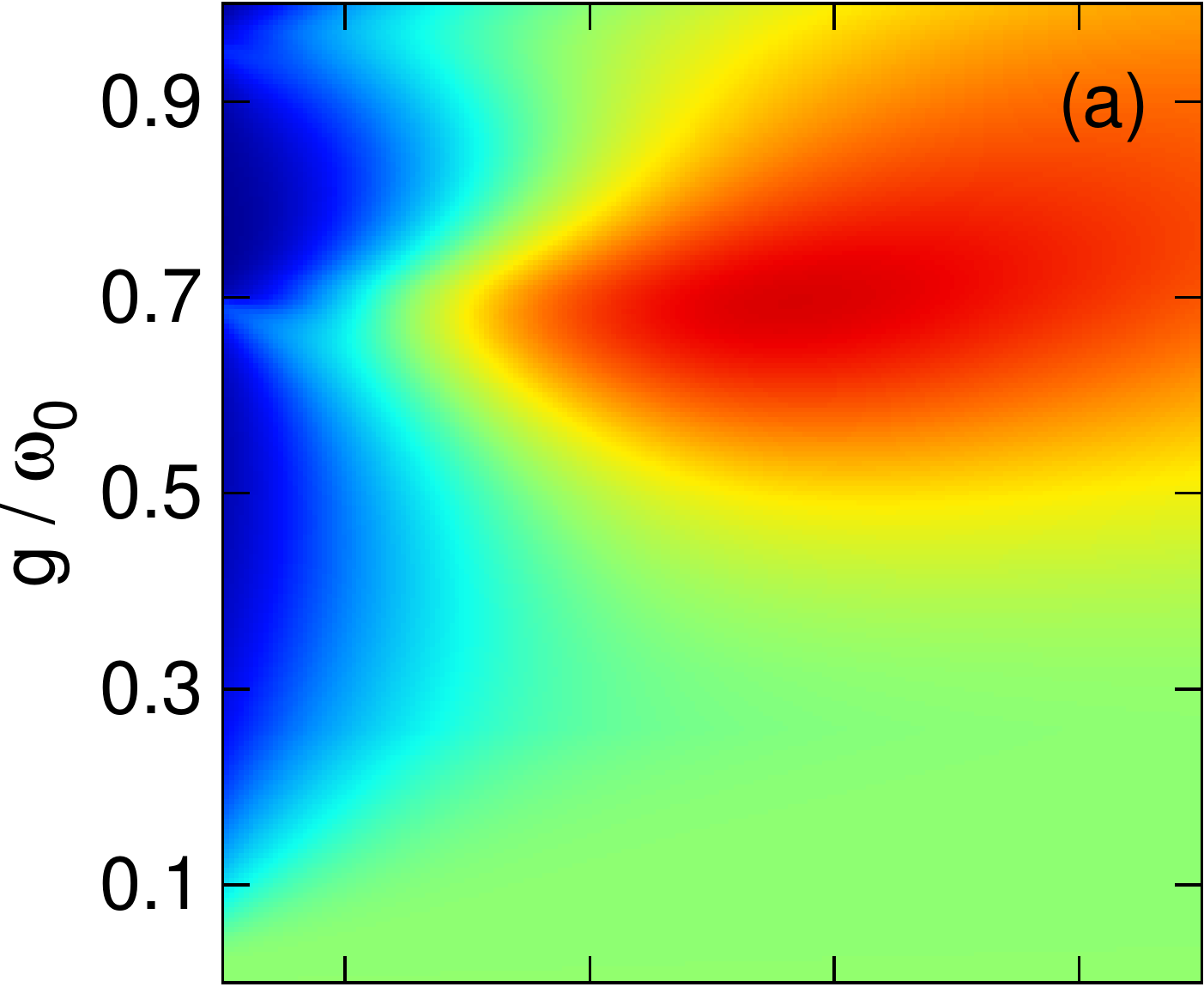}
  \includegraphics[scale=0.3]{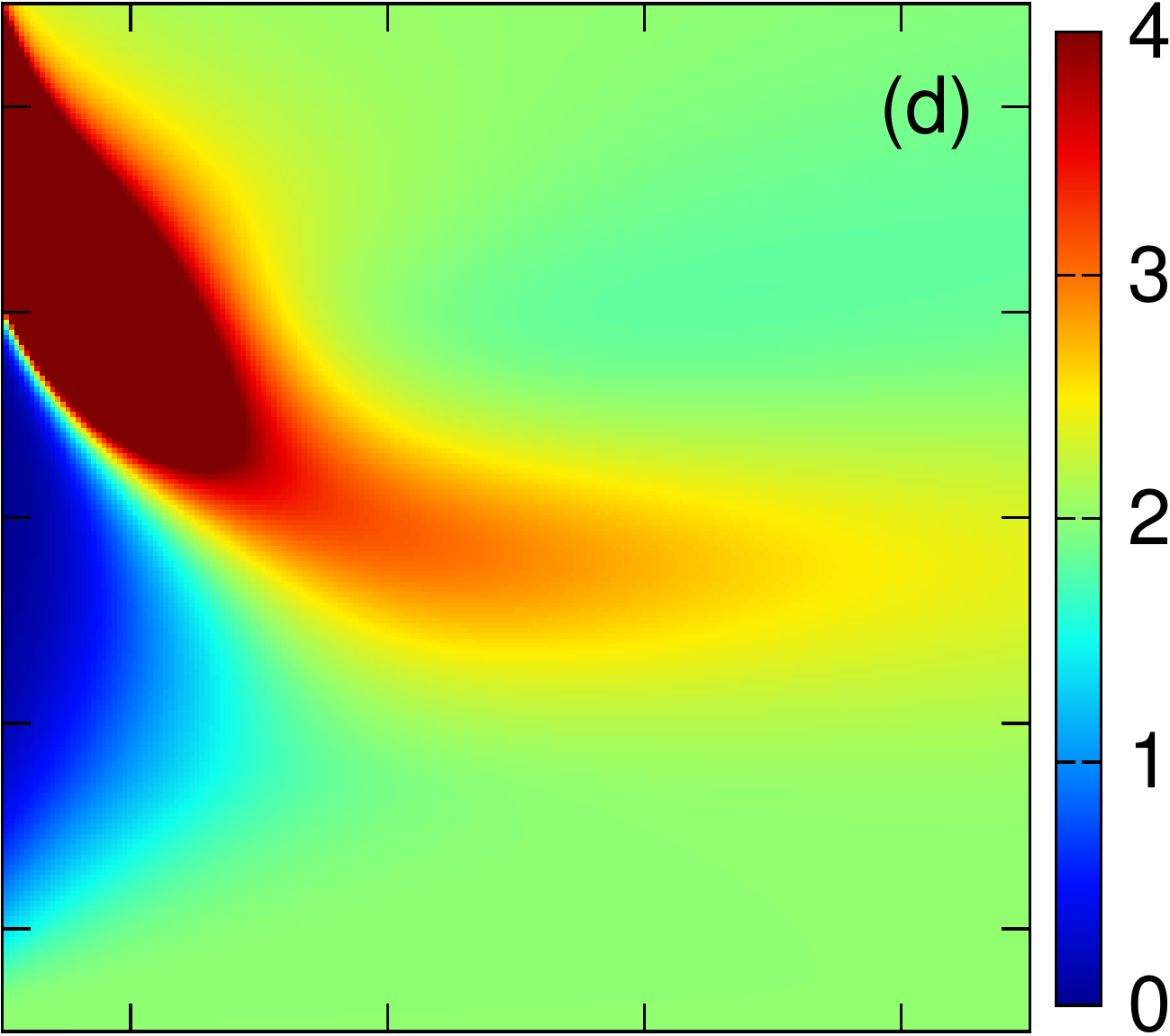}\\
  \includegraphics[scale=0.3]{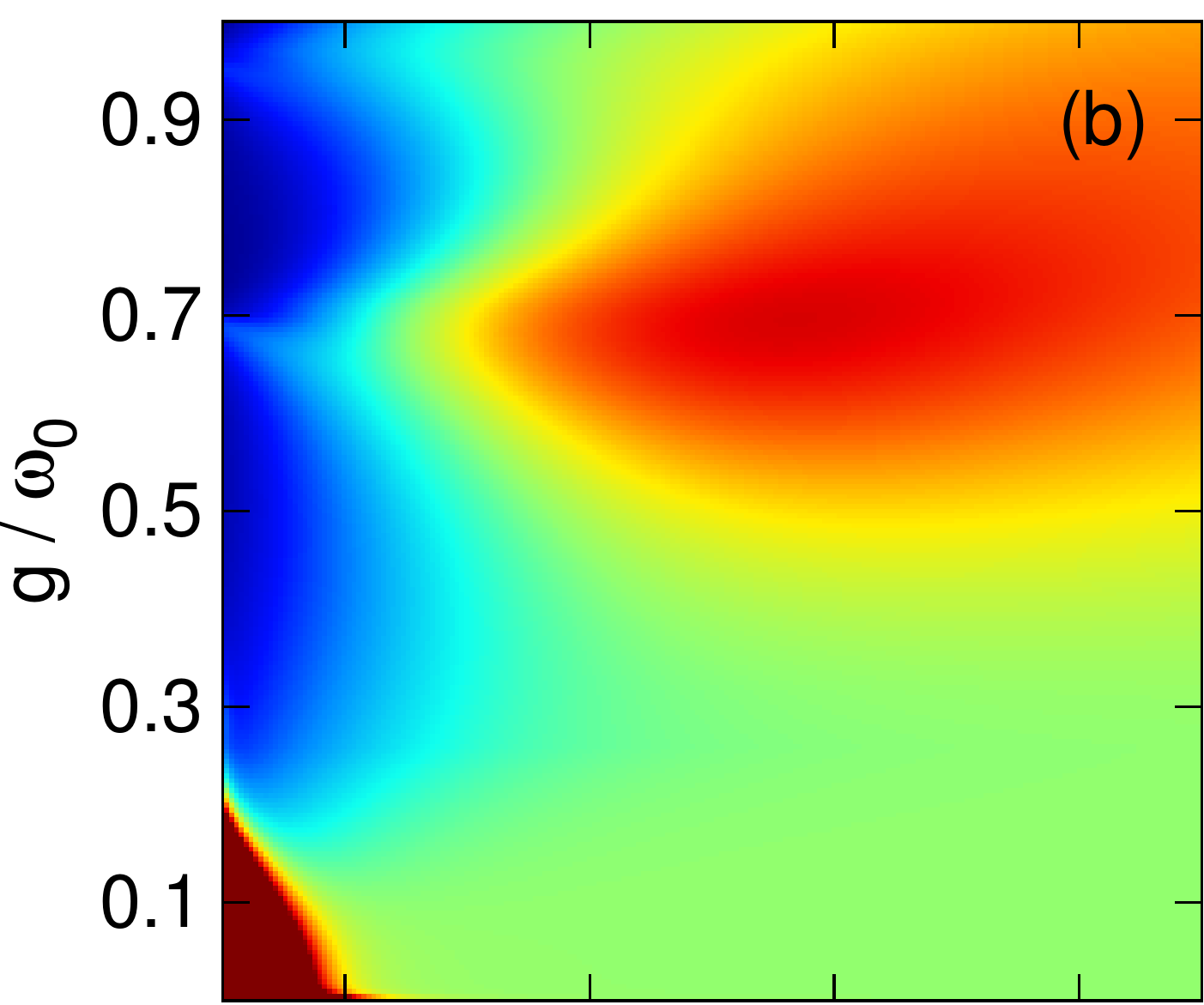}
  \includegraphics[scale=0.3]{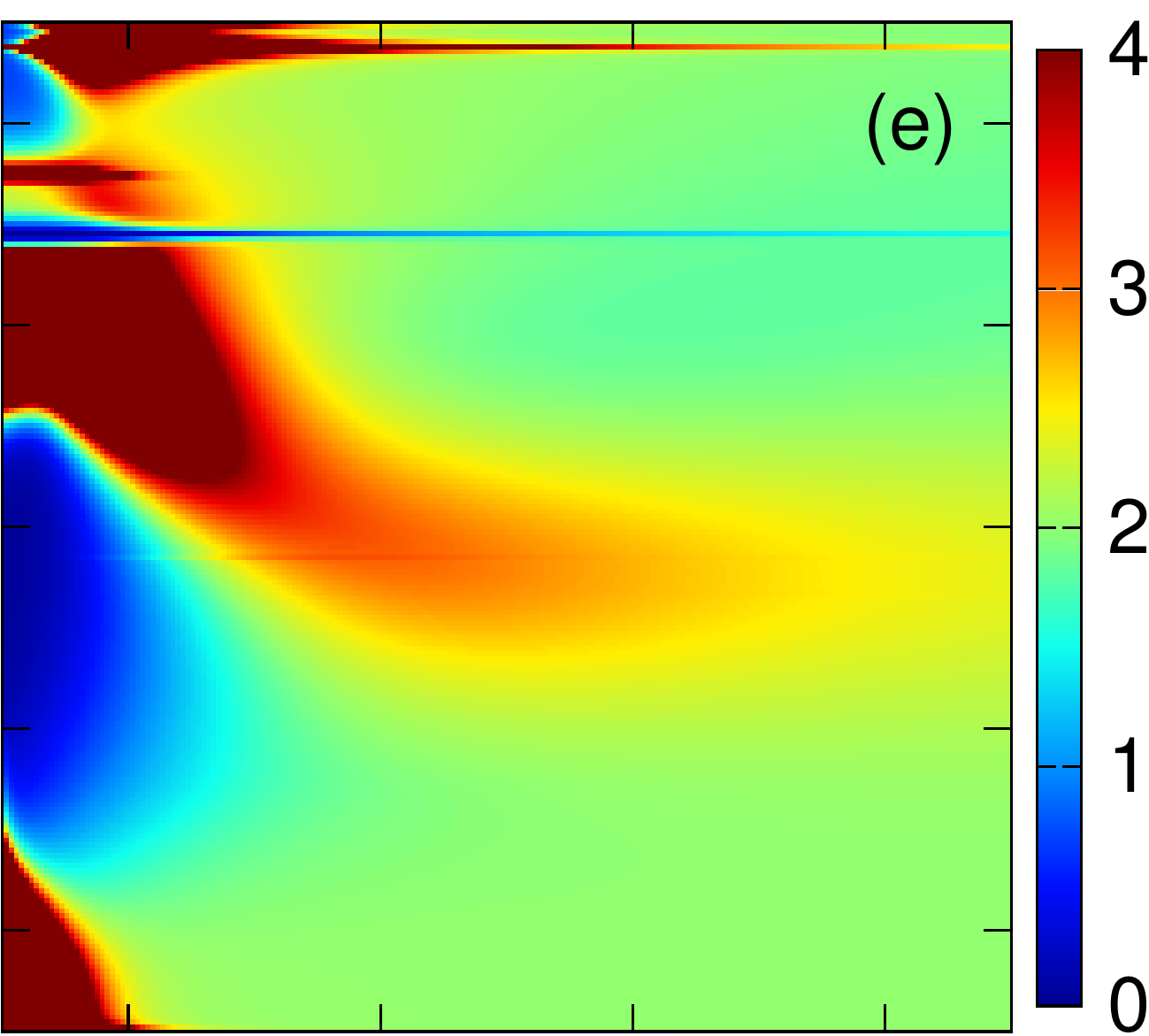}\\
  \includegraphics[scale=0.3]{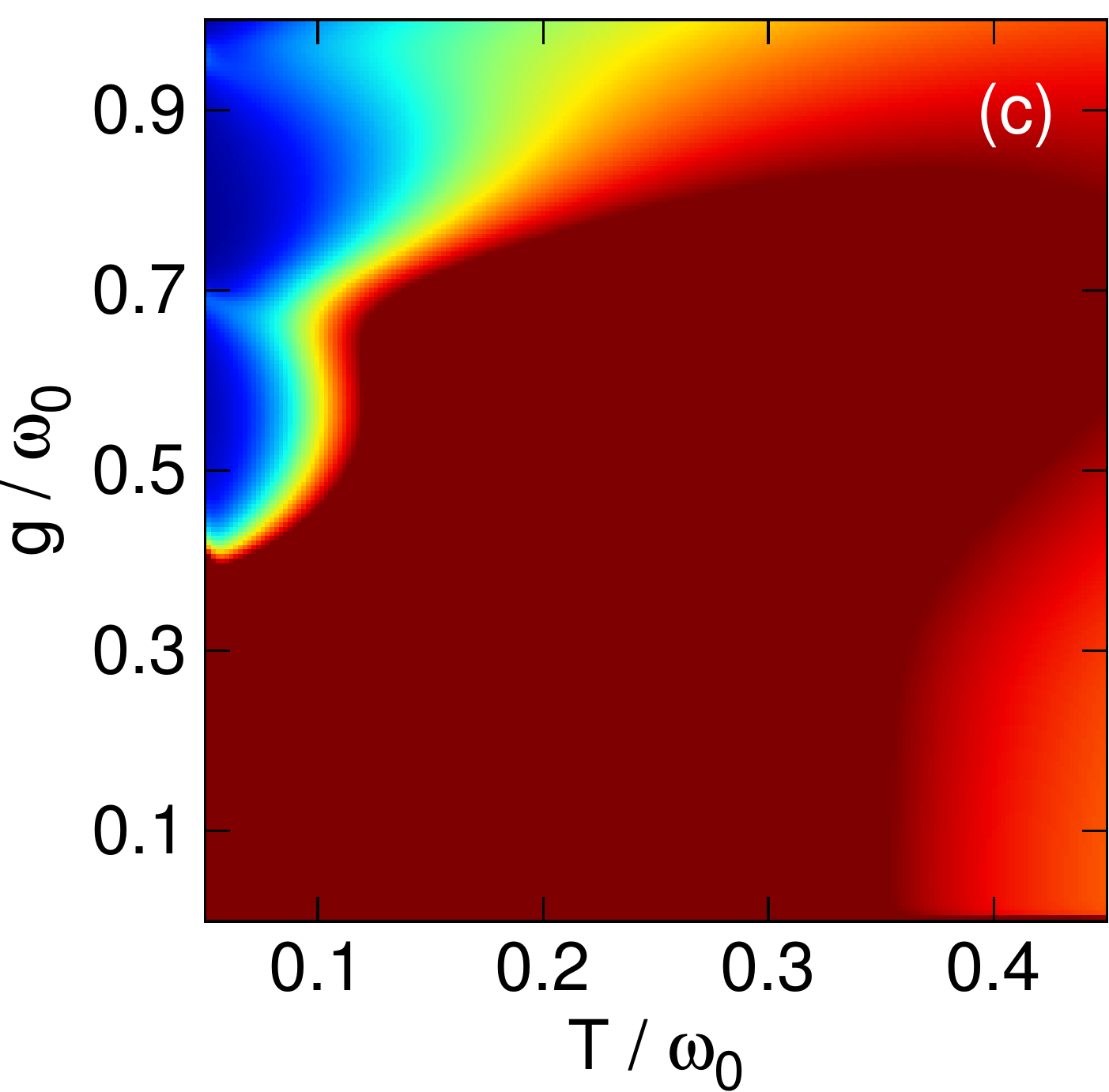}
  \includegraphics[scale=0.3]{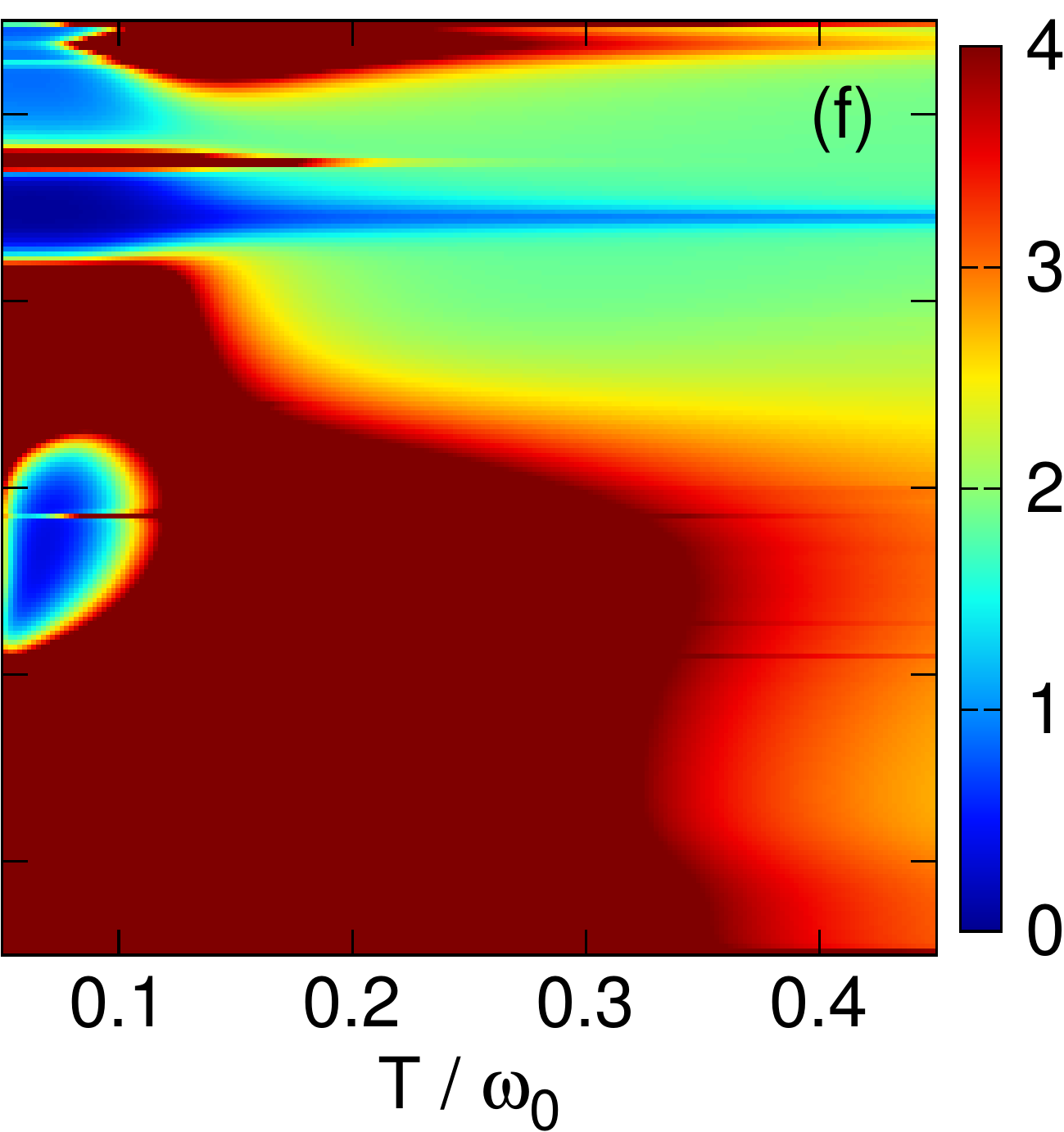}
  \caption{\label{fig:glauber2}Glauber function $g^{(2)}(0)$ for two emitters.
    The left (right) column depicts the results for $g' = \Omega' = 0$ ($g' = g$ and $\Omega' = \Omega$).
    The other parameters are the same as in Fig.~\ref{fig:glauber1}.}
\end{figure}

\begin{figure}
  \includegraphics[scale=0.3]{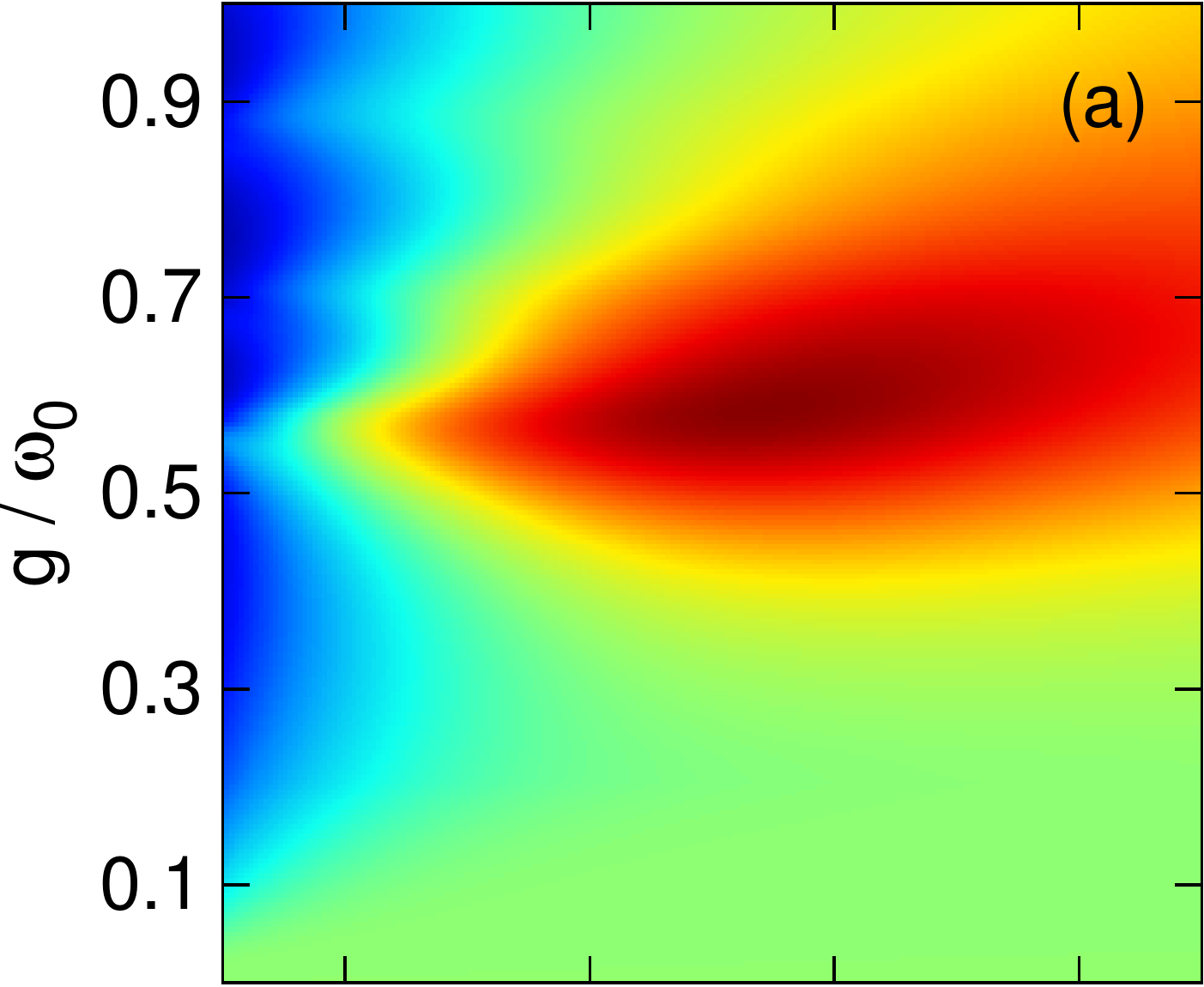}
  \includegraphics[scale=0.3]{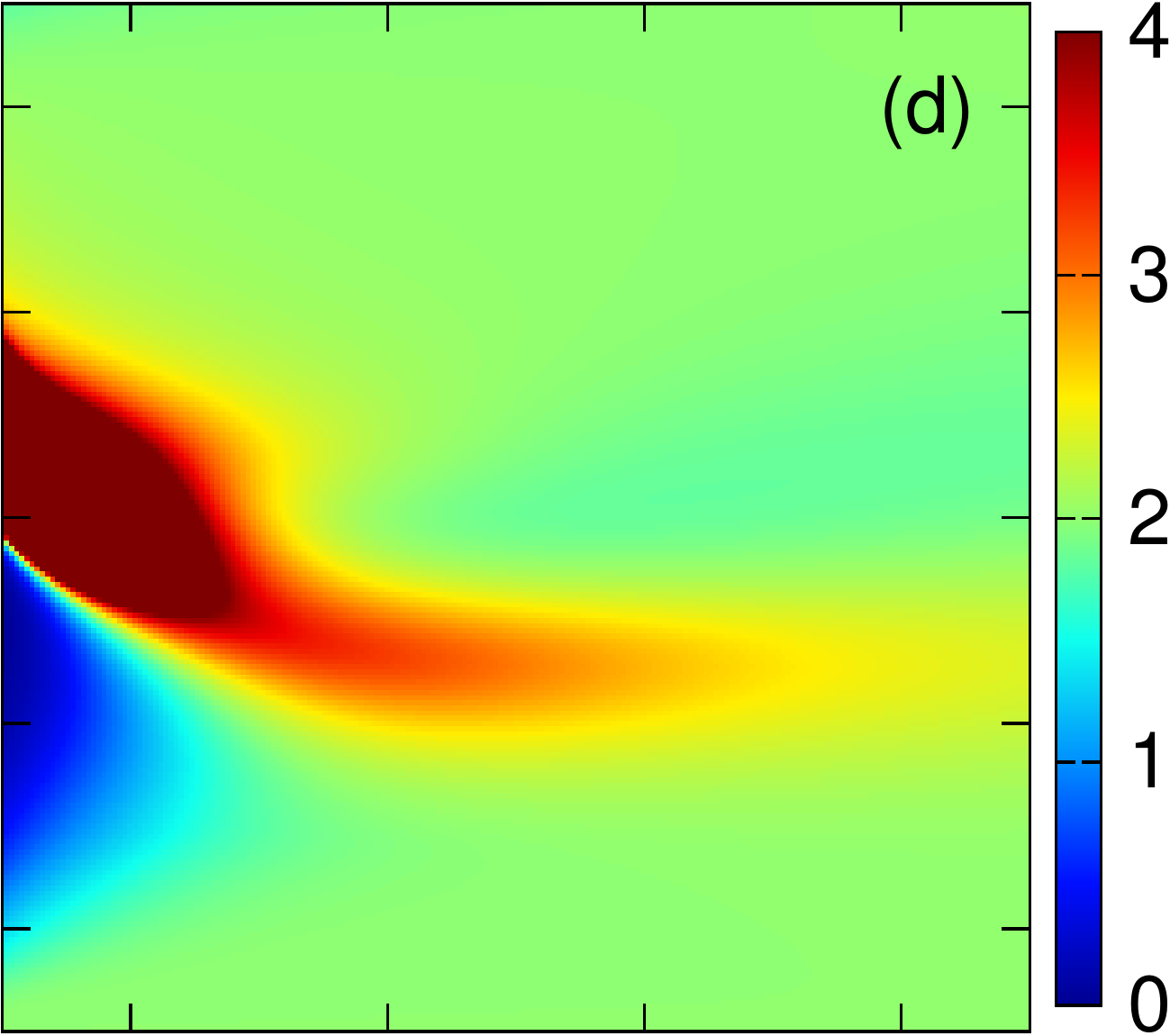}\\
  \includegraphics[scale=0.3]{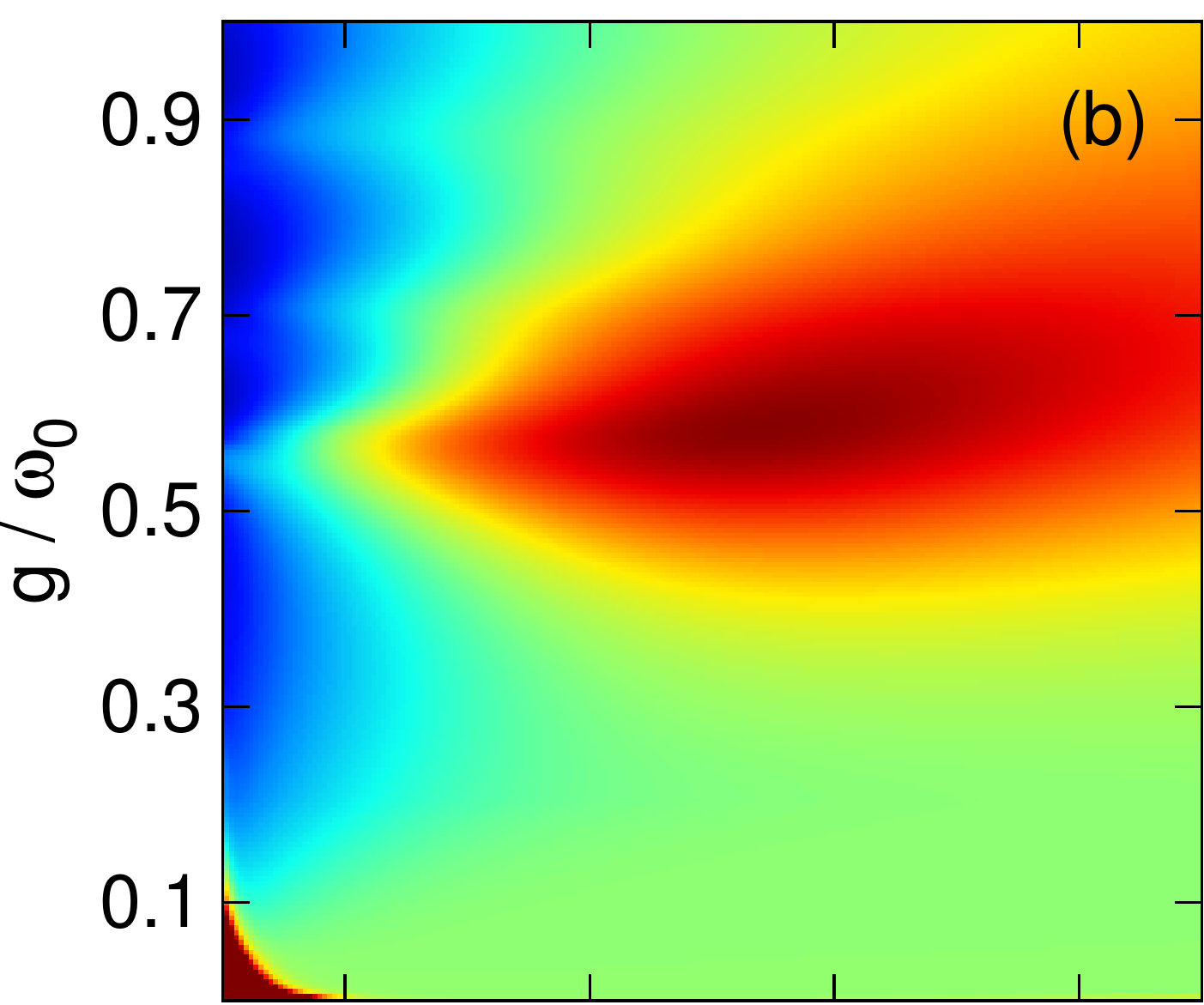}
  \includegraphics[scale=0.3]{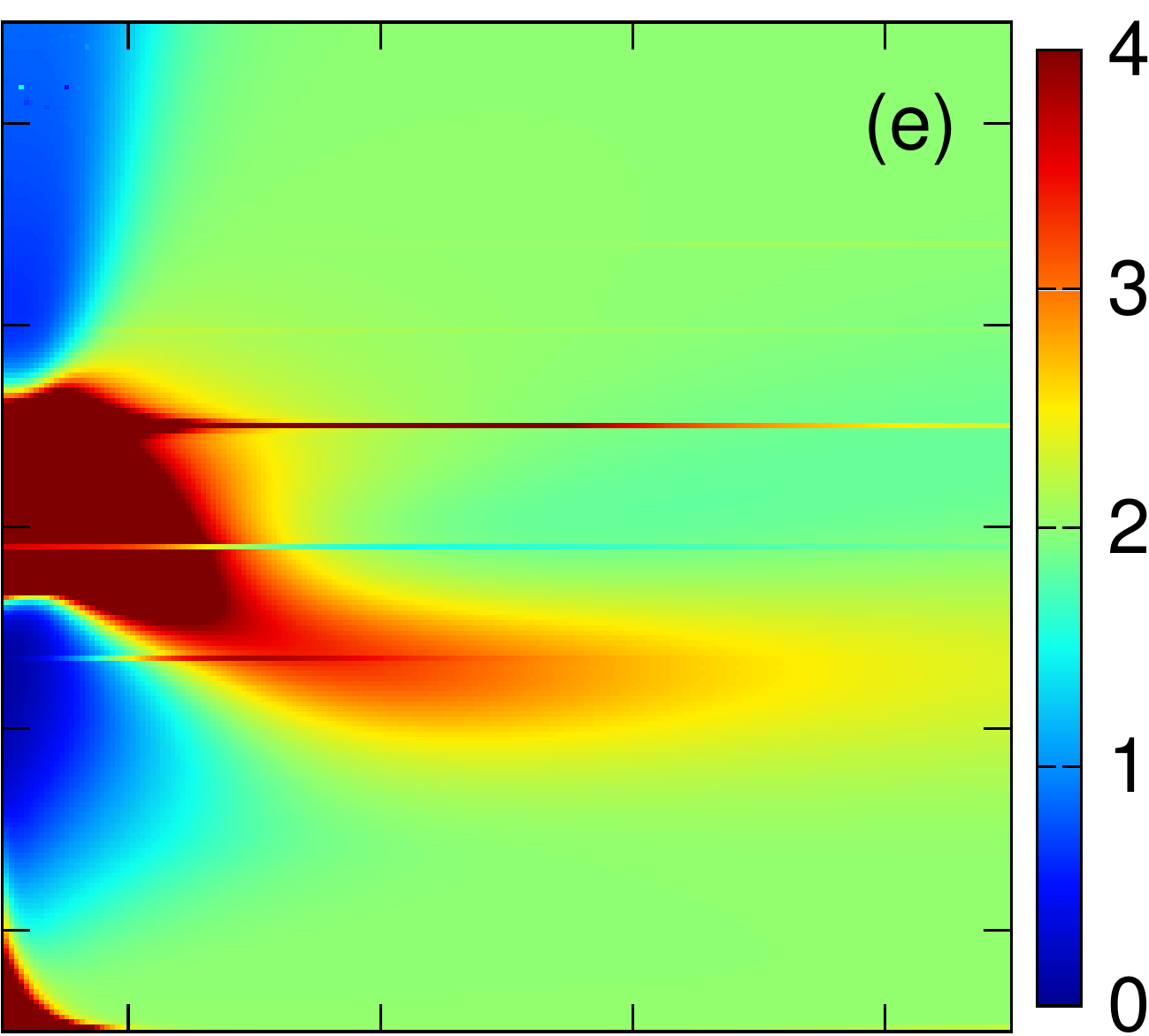}\\
  \includegraphics[scale=0.3]{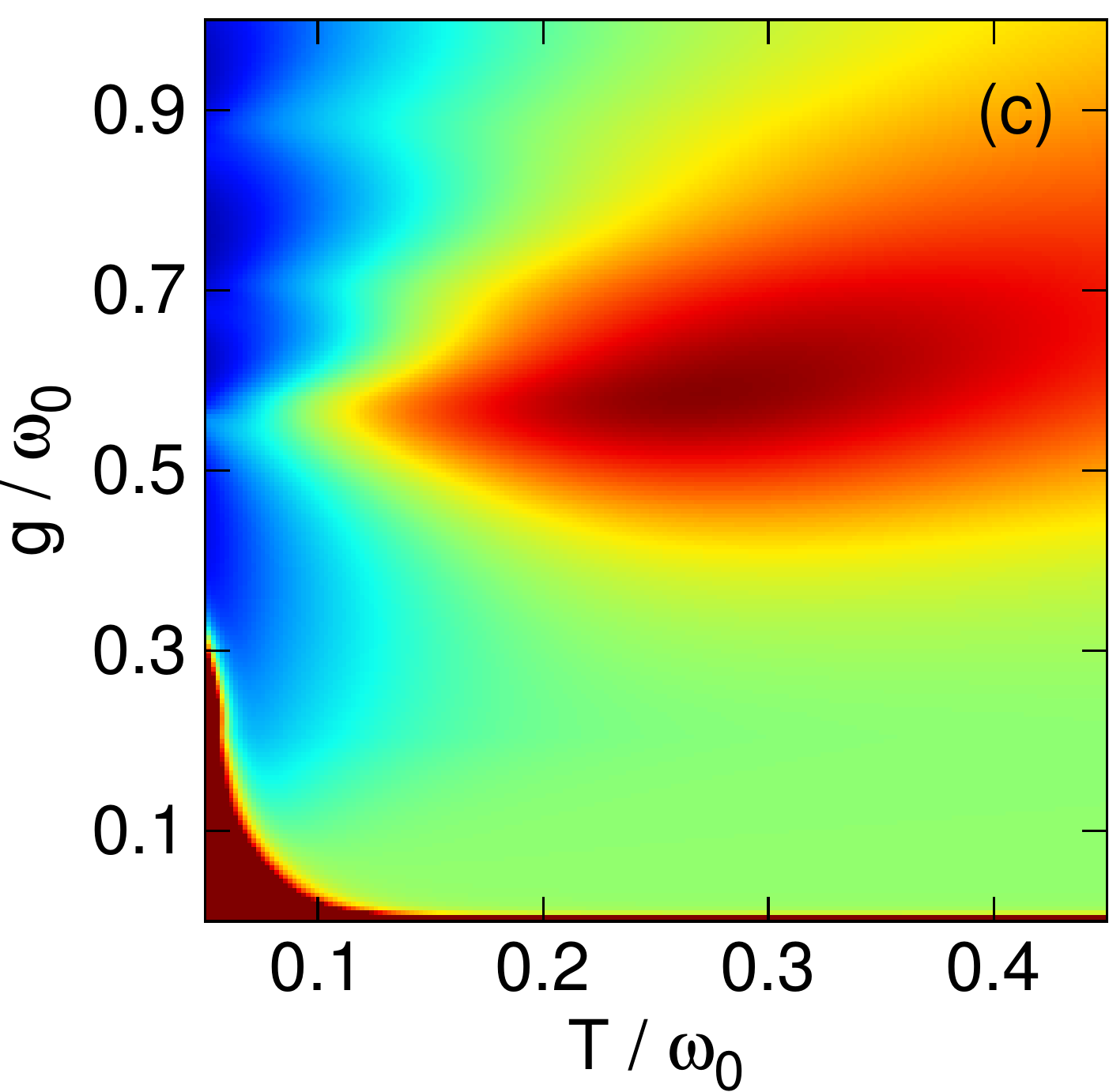}
  \includegraphics[scale=0.3]{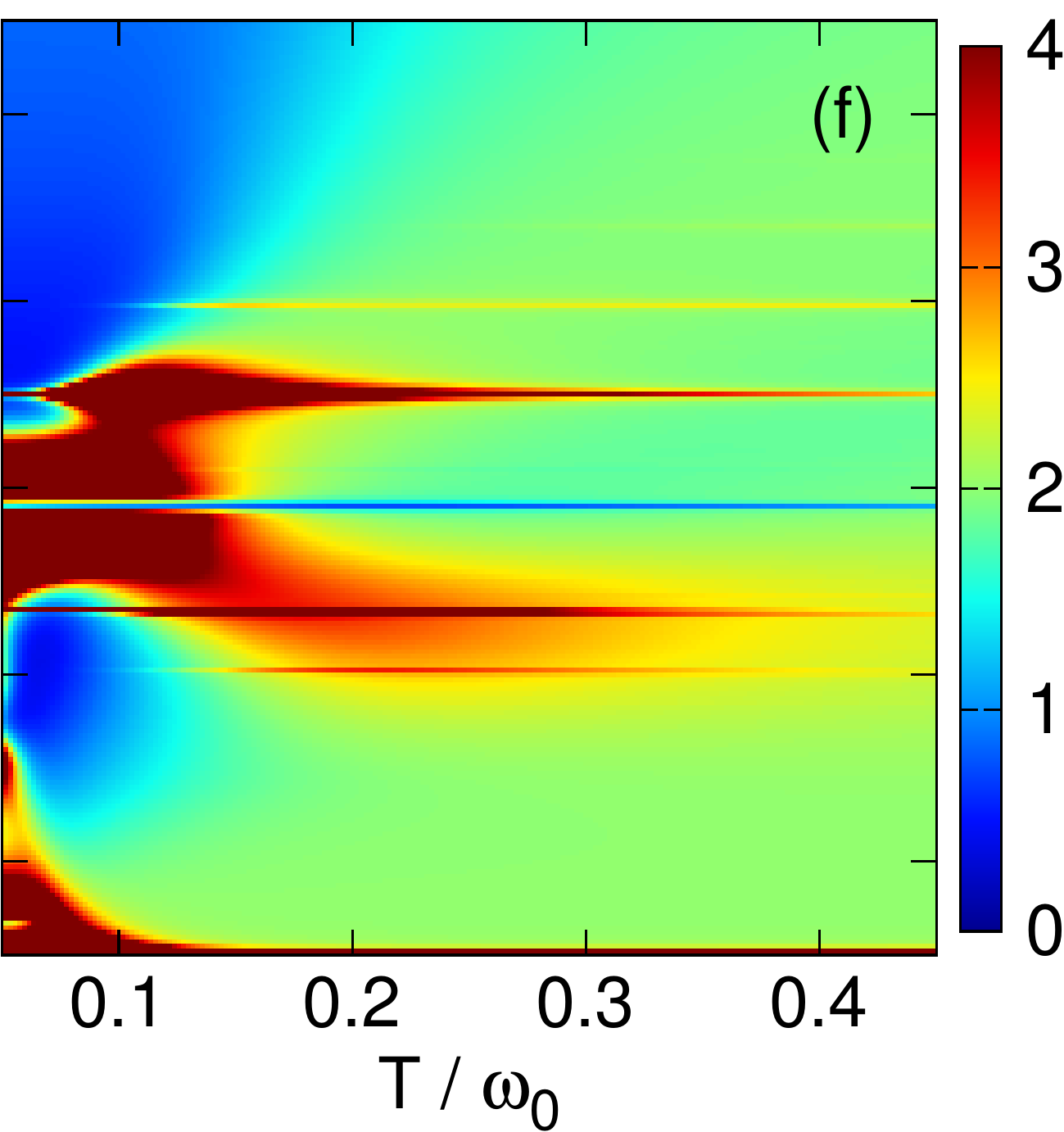}
  \caption{\label{fig:glauber3}Glauber function $g^{(2)}(0)$ for three emitters.
    The left (right) column depicts the results for $g' = \Omega' = 0$ ($g' = g$ and $\Omega' = \Omega$).
    The other parameters are the same as in Fig.~\ref{fig:glauber1}.}
\end{figure}

The shift of spectral lines in Figs.~\ref{fig:spct2} and~\ref{fig:spct3} is visualized in Figs.~\ref{fig:shft2} and~\ref{fig:shft3} in more detail.
Like in the corresponding Fig.~\ref{fig:shft1} for a single emitter, we include here circles marking the energy of spectral lines and solid (blue) lines depicting an $N$-independent fit proportional to $[1 - (\Omega / g)^2]^{3/4}$.
Again, we notice that the overall quality of the fit is very good, showing that the proportionality $\epsilon_n \propto [1 - (\Omega / g)^2]^{3/4}$ of the quasienergies is independent of the number of emitters.
This corroborates our conclusion from the main text that the dynamic Stark effect is quite universal in the driven Dicke system.
Nevertheless, because of avoided quasienergy crossings, the fit deviates from the numerical data in Figs.~\ref{fig:shft2}(f), \ref{fig:shft3}(e), and~\ref{fig:shft3}(f).
Thereby, anticrossings occur only in the case $g' = g$ and $\Omega' = \Omega$, which is due to the linearity of the system energies for $g' = 0$ and $\Omega' = 0$.
The number of avoided quasienergy crossings increases with the number of emitters because the quasienergy spectrum, upon projection into the Brillouin zone, becomes increasingly dense.
Thus, as is already evident from a comparison of Figs.~\ref{fig:shft2}(d)--\ref{fig:shft2}(f) with Figs.~\ref{fig:shft3}(d)--\ref{fig:shft3}(f), the general proportionality of the quasienergies will no longer be visible for many emitters $N \gg 3$.

\subsection{Glauber function}
The Glauber function for two and three emitters is given in Figs.~\ref{fig:glauber2} and~\ref{fig:glauber3}, respectively.
The three observations from Sec.~\ref{ssec:glauber-0} are recovered in each of the two plots.
In more detail, with increasing $\Omega$,
(i) a region of highly classical light emission [$g^{(2)}(0) > 2$] at low emitter-cavity-coupling strength and temperature appears,
(ii) a region of nonclassical light emission [$g^{(2)}(0) < 1$] emerges at ultrastrong coupling, and
(iii) additional horizontal lines with modified $g^{(2)}(0)$ appear at specific values of the emitter-cavity-coupling strength $g$.

For two emitters, the region of highly classical light emission for low emitter-cavity coupling $g$ and environment temperature $T$ in Figs.~\ref{fig:glauber2}(c) and~\ref{fig:glauber2}(f) is much greater than the corresponding region in Fig.~\ref{fig:glauber1} (Fig.~\ref{fig:glauber3}) for a single emitter (three emitters).
In addition, we observe that (again only for $N = 2$ but not for $N = 1, 3$) additional modifications of $g^{(2)}(0)$ in Fig.~\ref{fig:glauber2}(e) and~\ref{fig:glauber2}(f) are visible at ultrastrong coupling.
In contrast to the enlarged area of nonclassical light emission, these changes appear only if the counterrotating interaction terms are included in the Hamiltonian, i.e., for $g' = g$ and $\Omega' = \Omega$.
This increased sensitivity to the laser driving was already observed in the emission spectra for a single emitter in Fig.~\ref{fig:spct1} and we believe that the physics behind these observations is the same.

The appearance of horizontal lines in Figs.~\ref{fig:glauber2}(e), \ref{fig:glauber2}(f), \ref{fig:glauber3}(e), and~\ref{fig:glauber3}(f) follows (as for a single emitter) from the existence of resonant transition sequences.
They are marked in Fig.~\ref{fig:energy} with vertical (blue) lines. We immediately notice that all rules derived in Sec.~\ref{ssec:glauber-0} for a single emitter equally apply to $N = 2$ or $3$ emitters.

In Ref.~\cite{PAF15} we pointed out the existence of an approximate rule to relate the emission of a few emitters to the emission of a single emitter under appropriate scaling of the emitter-cavity-coupling strength $g$.
This rule follows from comparison of panel (a) or (d) in Figs.~\ref{fig:glauber1},~\ref{fig:glauber2}, and~\ref{fig:glauber3}.
As a consequence of the above modifications, the approximate rule for the dependence of the Glauber $g^{(2)}(0)$ function on the number of emitters $N$ does not hold for finite laser intensity $\Omega$.



%

\end{document}